\documentclass[aps,pre,reprint,superscriptaddress,nofootinbib,longbibliography,floatfix]{revtex4-2}

\usepackage{amsmath,amssymb,amsthm,mathtools,bm}
\usepackage{graphicx}
\usepackage{xcolor}
\usepackage{microtype}
\usepackage[colorlinks=true,linkcolor=red!55!black,citecolor=green!35!black,urlcolor=blue!55!black]{hyperref}

\setcitestyle{authoryear,round,aysep={,}}

\newtheorem{proposition}{Proposition}
\newtheorem{lemma}[proposition]{Lemma}

\newcommand{\R}{\mathbb{R}}
\newcommand{\avg}[1]{\langle #1\rangle}   
\newcommand{\Cov}{\operatorname{Cov}}
\newcommand{\Corr}{\operatorname{Corr}}
\newcommand{\Var}{\operatorname{Var}}
\newcommand{\diag}{\operatorname{diag}}
\newcommand{\bdiag}{\operatorname{blockdiag}}
\newcommand{\tr}{\operatorname{tr}}
\newcommand{\T}{\mathsf{T}}
\newcommand{\dd}{\mathrm{d}}
\newcommand{\Nc}{N_{\mathrm{c}}}  
\newcommand{\Bsrc}{\mathcal{B}}   
\newcommand{\Lop}{\mathcal{L}}    
\newcommand{\Sact}{\mathcal{S}}   
\newcommand{\Aph}{\breve{A}}      
\newcommand{\zmax}{z_{\max}}
\newcommand{\Kf}{\mathcal{K}}     
\newcommand{\Wf}{\mathcal{W}}     
\newcommand{\tcell}[1]{\parbox[t]{0.38\textwidth}{\raggedright #1\strut}}
\newcommand{\lcell}[1]{\parbox[t]{0.52\columnwidth}{\raggedright #1\strut}}
\newcommand{\cFR}{\mathrm{FR}}
\newcommand{\cIT}{\mathrm{IT}}

\begin{document}

\title{Bond yield dynamics: a multi-country multi-horizon string model}

\author{Paul Ruelloux}\email{paul.ruelloux@centralesupelec.fr}
\affiliation{Universit\'e Paris-Saclay, CentraleSup\'elec, MICS
}
\affiliation{Barclays Europe SE 
}


\begin{abstract}

This article extends the forward rate curve model of \citet{baaquie_bouchaud_2004} and \citet{lecoz_bouchaud_2024} to the multi-country setting of government bond markets. Cross-country blocks of correlation matrices of realized yield changes are found to display significant asymmetry between maturities. In the framework of the proposed model, this amounts to give to each country an idiosyncratic time scale. Each time scale describes how the perceived time between future dates depends on the dates themselves. The differences between these time scales structure the interactions between curves and reproduce the characteristic asymmetry of cross-country correlation matrices. The proposed model captures the full set of cross-curve correlations simultaneously at both daily and intra-day frequencies. When calibrated on government bonds issued by nine countries of the euro-zone, the typical reconstruction error is  1.7\%.

\end{abstract}

\maketitle

\section{Introduction}\label{sec:introduction}

\subsection{Motivation}\label{subsec:motivation}

A bond is a tradable promise of future cash flows, and a government bond is issued by a sovereign borrower. Given a bond $b$, let $CF_{b,j}$ be the $j^{th}$ cash flow paid at date $T_{b,j}$. Let $y_b(t)$ be its continuously compounded yield to maturity at date $t$, which is defined by the bond's price
\begin{equation}
  \mathcal P_b(t,y_b)=\sum_j CF_{b,j}\,
  \exp\!\left[-y_b(t)\,(T_{b,j}-t)\right].
  \label{eq:bond-price}
\end{equation}
If $T_b$ is the date of maturity of the bond, its tenor (or time-to-maturity) at date $t$ is $\theta_b \triangleq T_b-t$\footnote{Throughout the paper, I use $\triangleq$ to symbolize a definition.}. The empirical yield surface of a given issuer is defined by the collection of yields $y_b(t,\theta)$ across dates and tenors. The associated empirical yield curve, or curve, is the collection of yields at a given date $t$. Once deterministic carry is separated, the main first-order random contribution to the profit and loss over a period $\Delta t$ of a position in bond $b$ is given by the yield change $\Delta y_b$ over the same period,
\begin{equation}
  \Delta\mathcal P_b
  \simeq \mathrm{carry}_b-\mathrm{DV01}_b\,\Delta y_{b},
  \label{eq:dv01}
\end{equation}
where the $\Delta y_b$ has basis-point unit and $\mathrm{DV01}_b \triangleq -\partial\mathcal P_b/\partial y_{b}>0$ is the price loss produced by a one-basis-point rise in yield, a measure of rate sensitivity \citep{tuckman_serrat_2022}. 
To keep track of the issuing country of a specific bond, I denote by $Y_c(t,\theta)$ the yield to maturity of a bond $b$ issued by country $c$ that matures at $t+\theta$, that is to say:
\begin{equation}
    \label{eq:yield-surface}
    Y_c(t,\theta) \triangleq y_b(t).
\end{equation}
The risk of a portfolio of government bonds at date $t$ is therefore governed by the joint behavior of the yield changes $\Delta Y_c(t,\theta)$ across issuers $c$ and times to maturity $\theta$, which is the object of this paper. In the euro zone, the single currency and common monetary policy anchor national yield curves to shared rate expectations \citep{dunne_et_al_2007,ehrmann_et_al_2011}, which isolates cross-issuer co-movement from exchange-rate risk. Market integration nevertheless varies over time \citep{pozzi_wolswijk_2012}, and sovereign spreads respond to credit quality and liquidity in proportions that shift during periods of stress \citep{beber_et_al_2009,favero_et_al_2010,monfort_renne_2014}.

The co-variations of yields across maturities are commonly summarized by a few factors, such as level, slope, and curvature \citep{litterman_scheinkman_1991}, or by parsimonious parametric curves \citep{nelson_siegel_1987,svensson_1994,diebold_li_2006}, which \citet{diebold_li_yue_2008} and \citet{jotikasthira_et_al_2015} extend to several countries through global and country-specific factors. A second tradition models the entire curve as a random field indexed by maturity and driven by infinitely many correlated shocks \citep{kennedy_1994,goldstein_2000,santaclara_sornette_2001,filipovic_2001,cont_2005,carmona_tehranchi_2006}. Following \citet{bouchaud_et_al_1999}, the starting point of the present paper is to describe the yield curve as an elastic string: neighboring maturities resist moving independently and shocks propagate smoothly along the curve. \citet{baaquie_bouchaud_2004} cast this picture as a stiff field theory in a ``psychological" maturity coordinate that runs more slowly than calendar time at long horizons. This psychological maturity differs from calendar (physical) maturity in that the psychologically perceived time interval between two future events depends not only on their temporal separation but also on how far into the future they occur \citep[see also][]{baaquie_2009_book,baaquie_yang_2009}. \citet{lecoz_bouchaud_2024} later reformulated it as a relaxation equation for a correlated field driven by uncorrelated shocks, and \citet{lecoz_et_al_2025} related this elastic coupling to cross-maturity order flows and liquidity. On interest-rate futures, their model reproduces the cross-maturity correlation structure with a typical error of $1\%$. Because the field takes a finite time to relax, it also reproduces the build-up of correlations at short horizons known as the Epps effect \citep{epps_1979}.

I model the sovereign bond market as a set of elastic strings coupled on a graph at equal psychological, latent maturities that differ from observed, physical ones. These two spaces of maturities are mapped to one-another through ``clocks", whose name refers to the temporal aspect of the transform. Each country reads a shared latent maturity through its own clock. Unequal clocks bend the cross-country correlation ridges away from the equal-tenor diagonal and produce the ordered-block asymmetry found in the data. Calibrated on nine euro-zone sovereigns, a single spatial geometry, combined with one relaxation time and bond-level microstructure noise, reproduces the daily correlation surface and the intraday build-up of correlations.

Three empirical regularities of these markets guide the proposed model.

\begin{figure}[t]
  \centering
  \includegraphics[width=\columnwidth]{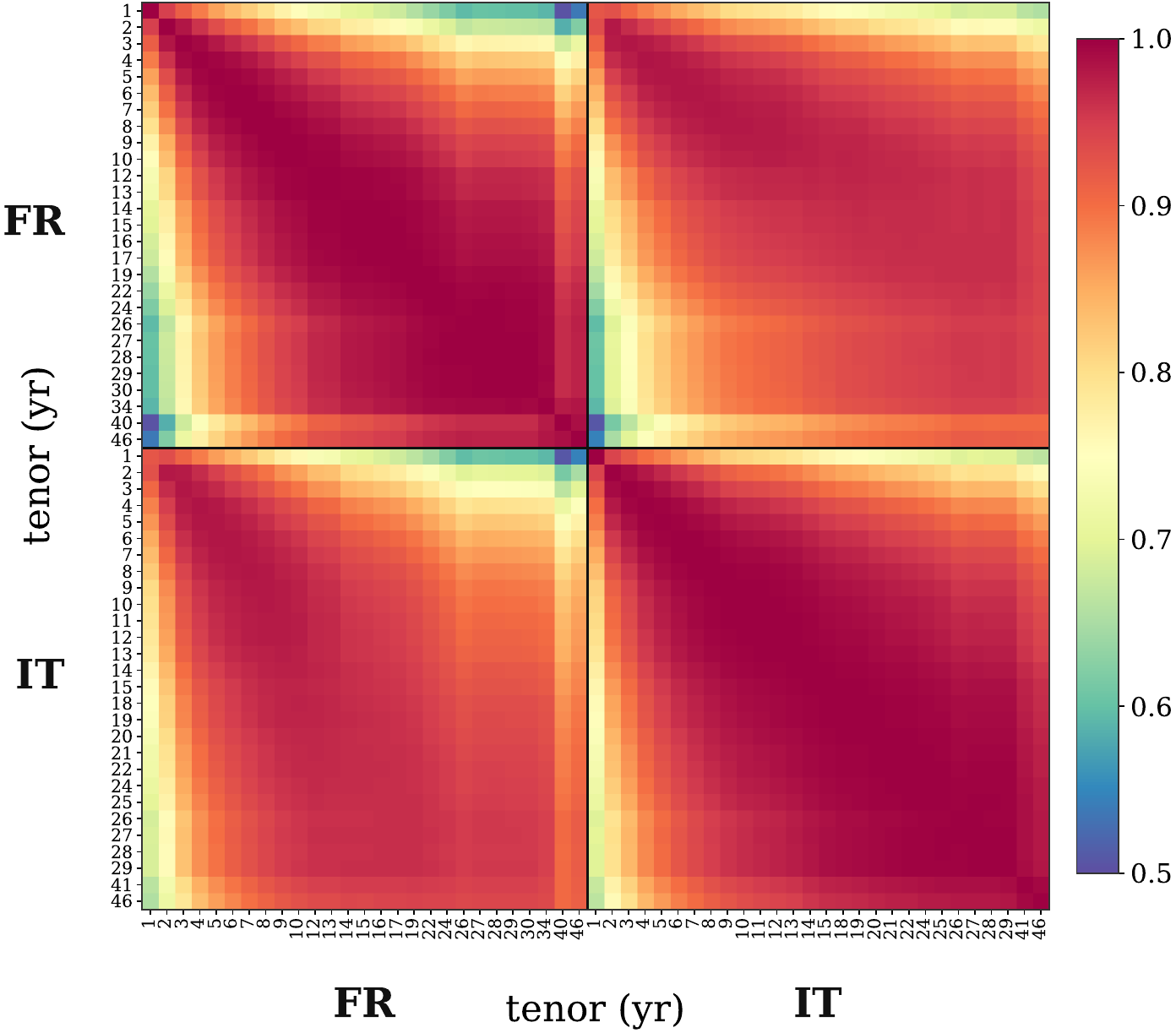}
  \caption{Daily yield-change correlation matrix for 27 French and 28 Italian government bonds of the daily sample (2024-2026), with tenors from 1 to 46 years in both countries, ordered by country and tenor. The diagonal blocks show within-country maturity dependence. The off-diagonal blocks are transposes of one another. Within the ordered France-Italy block, exchanging the two maturity indices changes the correlation over a broad region.}
  \label{fig:fr-it}
\end{figure}

\emph{Smooth, monotonic correlation surfaces.} Organized in blocks by issuer and ordered by maturity, daily yield changes of sovereign bonds form smooth correlation surfaces. Figure~\ref{fig:fr-it} illustrates this through the France-Italy cross-country realized Pearson correlation matrix. Within each diagonal block, correlations decay monotonically away from the diagonal, and the decay slows down at longer tenors so that the correlation ridge broadens with maturity. Cross-country blocks display the same geometry at a slightly lower overall level, a pattern that holds across all nine issuers (Sec.~\ref{subsec:fit}). This structure mirrors the fluctuations of forward-rate curves around their average shape \citep{bouchaud_et_al_1999,baaquie_bouchaud_2004,lecoz_bouchaud_2024}.

\emph{Ordered-block asymmetry.} Let $\rho_{cd}(\theta_i,\theta_j)$ be the Pearson correlation coefficient between the yield changes of country $c$ at tenor $\theta_i$ and of country $d$ at tenor $\theta_j$. In general:
\begin{equation}
  \rho_{cd}(\theta_i,\theta_j)\neq \rho_{cd}(\theta_j,\theta_i)
  .
  \label{eq:block-asymmetry}
\end{equation}
This ordered-block asymmetry under exchange of the two maturity indices within a single off-diagonal block is qualitatively clear in Fig.~\ref{fig:fr-it}. To quantitatively test the asymmetry coefficient by coefficient, I retain the 22 tenors between 1 and 46 years at which both France and Italy have a bond in Fig.~\ref{fig:fr-it}, forming 231 distinct maturity pairs. Let $\Delta Y_{\cFR,i}$ denote the daily yield change of the French bond at tenor $\theta_i$, and similarly for Italy. For two distinct maturities $i\neq j$, compare
\begin{equation}
  \rho(\Delta Y_{\cFR,i},\Delta Y_{\cIT,j})
  \quad\text{with}\quad
  \rho(\Delta Y_{\cFR,j},\Delta Y_{\cIT,i}).
  \label{eq:steiger-pair}
\end{equation}
A joint block-bootstrap test of exact maturity-index symmetry across the 231 cells rejects the null with $p$-value $p<1\times10^{-4}$, and 170 pairs remain significant at a $5\%$ familywise error rate under the Romano-Wolf step-down procedure in a single coherent region (Fig.~\ref{fig:asymmetry}; protocol, exact-maturity alignment, and bond-substitution checks in Appendix~\ref{app:statistics}). 

\begin{figure}[t]
  \centering
  \includegraphics[width=\columnwidth]{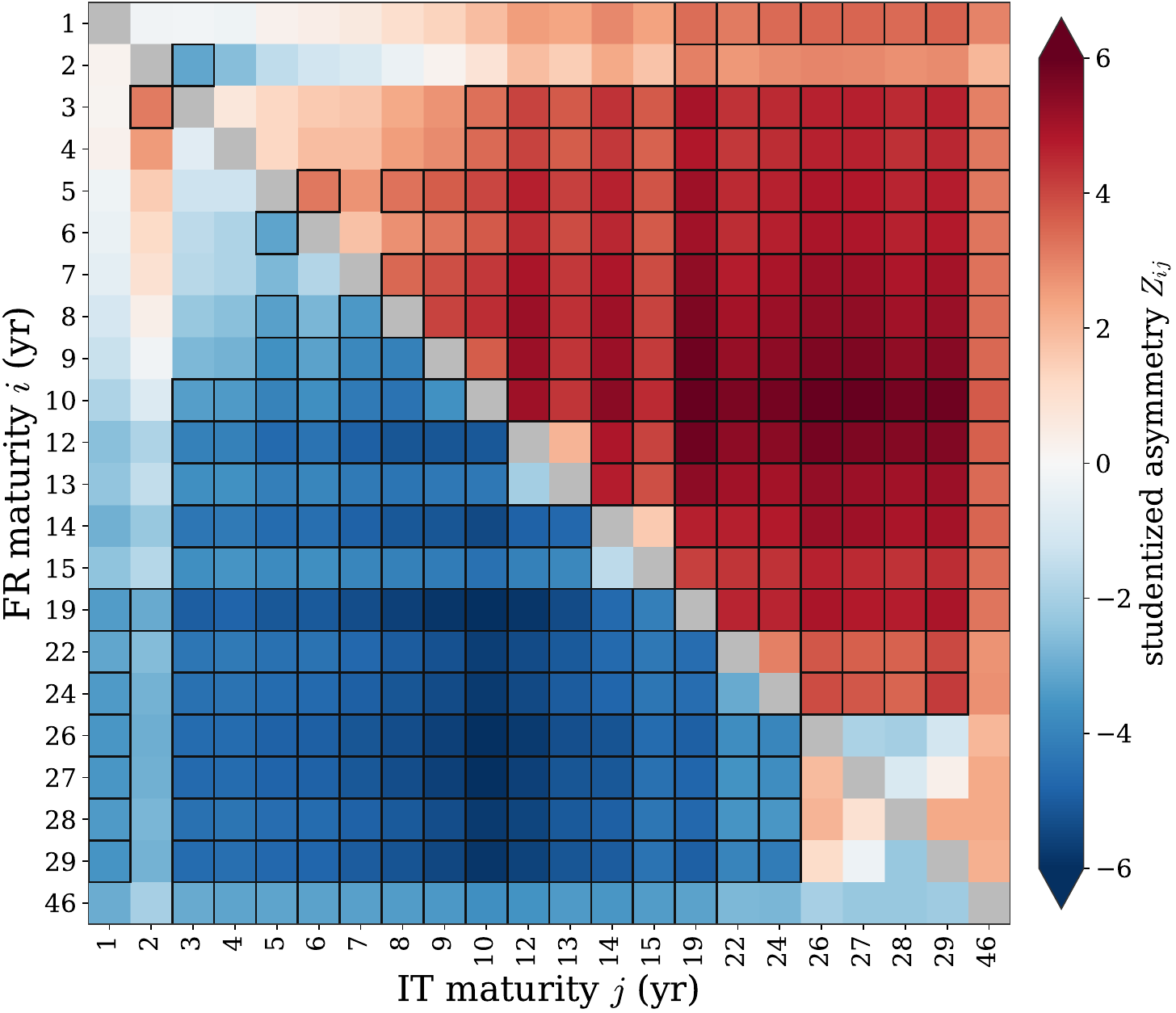}
  \caption{Ordered-block asymmetry of the France-Italy cross block (daily sample). Each cell $(i,j)$ shows the antisymmetric studentized block-bootstrap statistic $Z_{ij}$ of $\operatorname{atanh}\rho(\Delta Y_{\cFR,i},\Delta Y_{\cIT,j})-\operatorname{atanh}\rho(\Delta Y_{\cFR,j},\Delta Y_{\cIT,i})$. Outlined cells are significant at $5\%$ familywise error (Romano-Wolf), and the global maturity-index symmetry test gives $p<1\times10^{-4}$.}
  \label{fig:asymmetry}
\end{figure}

\begin{figure*}[t]
  \centering
  \includegraphics[width=\textwidth]{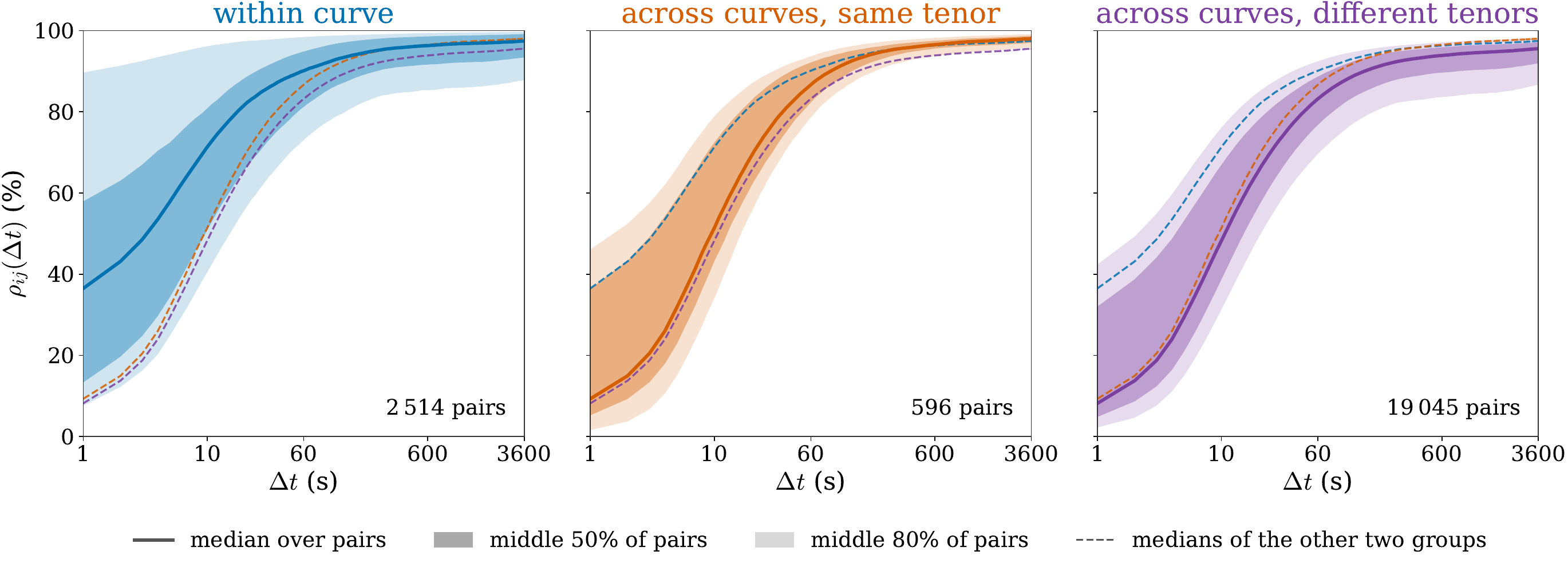}
  \caption{Epps effect in raw data (intraday sample): Pearson correlation $\rho_{ij}(\Delta t)$ of yield changes measured over a horizon $\Delta t$ from one second to one hour, partitioned across all bond pairs into within-curve pairs ($2{,}514$ pairs, left), same-tenor cross-country pairs ($596$ pairs, center), and cross-tenor cross-country pairs ($19{,}045$ pairs, right). Solid lines show the median correlation within each group, dark and light shaded bands indicate the middle $50\%$ and $80\%$ of pairs, and dashed lines reproduce the medians of the other two groups for comparison.}
  \label{fig:epps-empirical}
\end{figure*}

\emph{The Epps effect.} At intraday horizons, the Pearson correlation of two yield changes depends on the horizon $\Delta t$ over which they are measured, as reported in Fig.~\ref{fig:epps-empirical}. Most of the correlation build-up takes place within the first minute. This dependence on the sampling horizon is the Epps effect \citep{epps_1979}. It has been documented in many markets \citep{reno_2003,toth_kertesz_2009,munnix_et_al_2010,mastromatteo_et_al_2011}, and is a primary concern of high-frequency covariance estimation from asynchronous and noisy prices \citep{hayashi_yoshida_2005,zhang_2011,bacry_et_al_2013,hoffmann_et_al_2013,huth_abergel_2014}.

A convincing model would have to produce all three stylized facts described above. The model I develop below produces the first two facts through elastic string coupling along and across curves at equal latent maturity. A finite relaxation time combined with bond-level microstructure noise explains the Epps effect, as in \citet{lecoz_bouchaud_2024}.

\subsection{Organization of the paper}\label{subsec:organization}

Section~\ref{sec:data} describes the daily and intraday yield datasets. Section~\ref{sec:model} develops the theory: Sec.~\ref{subsec:baseline} recalls the single-curve model of \citet{lecoz_bouchaud_2024}, Sec.~\ref{subsec:clocks} couples the curves on a graph at equal latent maturity through country clocks, and Sec.~\ref{subsec:discretization} discretizes the dynamical operators and obtains the covariance of yield changes over any window. Section~\ref{sec:daily} calibrates the spatial geometry on daily yield changes, and Section~\ref{sec:epps} estimates the relaxation time and microstructure noise at intraday horizons.

\section{Data}\label{sec:data}

The application focuses on nominal fixed-coupon government bonds issued by nine euro-area sovereigns: Austria (AT), Belgium (BE), Germany (DE), Spain (ES), France (FR), Ireland (IE), Italy (IT), the Netherlands (NL), and Portugal (PT). I retain the sovereigns whose curves are sufficiently liquid and densely quoted across maturities. Smaller or emerging issuers are left out for clarity of exposition. I likewise exclude STRIPS, inflation-linked bonds, and other securities whose pricing depends on additional risk factors. In this model, such instruments can be incorporated by treating each distinct instrument class as a separate curve, that is, by adding a new vertex to the graph (for instance, an Italian STRIPS curve) coupled to the nominal curves.

Each time series samples the empirical yield surface $Y_c(t,\theta)$ of Eq.~\eqref{eq:yield-surface} at a fixed integer tenor $\theta$ (in years): at any date $t$, it tracks the benchmark instrument of issuer $c$ maturing $\theta$ years ahead, rolling from one ISIN to the next as the calendar advances. Following standard market convention, as in DE10Y for the German ten-year yield, I refer to each constant-tenor series simply as a bond, indexed by its issuer and tenor. Throughout the paper, $c,d$ index the $\Nc=9$ countries and $i,j$ index observed bonds.

I use two datasets. The first consists of daily close yields across all active trading days from 2 January 2024 to 31 August 2026. The full nine-country calibration uses 196 bonds with tenors ranging from 1 to 95 years, listed by country in the caption of Fig.~\ref{fig:daily-surface}, and their realized volatilities are reported in Appendix~\ref{app:vol}. The second dataset records intraday mid-yields on a one-second grid from May to August 2026 across 211 bonds, which I use to study the horizon dependence of correlations.

The number of bonds varies slightly across samples, for two practical reasons. The active universe evolves over time as old issues mature and new benchmarks are issued, which accounts for the difference between the 196 bonds of the multi-year daily panel and the 211 bonds of the 2026 intraday window. In addition, to keep the empirical correlation surfaces free of stale-quote artifacts, I screen out insufficiently liquid off-the-run tenors when constructing a synchronous multi-country panel.

\section{Theory}\label{sec:model}

This section extends the single-curve elastic string framework to a multi-issuer market and derives the covariance of yield changes over any observation window. I first formulate a continuous field theory whose action governs how shocks propagate across maturities and issuers, and then discretize the field onto a finite maturity grid, where specifying the measure of the primitive white noise selects a stochastic realization. Section~\ref{subsec:baseline} recalls how these two steps work for a single curve in the model of \citet{lecoz_bouchaud_2024}. Section~\ref{subsec:clocks} extends the continuous geometry to a graph of curves. I hereby show that naively coupling countries at equal quoted tenors leaves every cross-country covariance block symmetric under exchange of its maturity indices, hence a naive extension of the single-curve framework would fail to explain the cross-country covariance asymmetry. Section~\ref{subsec:clocks} resolves this problem by proposing interaction of curves at equal-latent maturity, which maps back to physically quoted tenors via country-dependent clocks. Section~\ref{subsec:discretization} then discretizes the curves and derives the exact covariance of yield changes in the general case.

\subsection{The single-curve framework}\label{subsec:baseline}

Dropping the country index of Eq.~\eqref{eq:yield-surface}, consider a single curve whose yield $Y(t,\theta)$ is scalar-valued, and decompose its equation of motion at date $t$ and tenor $\theta$ as
\begin{equation}
  \frac{\partial Y}{\partial t}(t,\theta)
  =\gamma(t,\theta)+\frac{\sigma(t,\theta)}{\sigma_A(\theta)}\,A(t,\theta),
  \label{eq:yield-dynamics}
\end{equation}
where $\gamma$ is a deterministic drift, $\sigma(t,\theta)>0$ is the univariate yield volatility at tenor $\theta$, and $A(t,\theta)$, which I call the velocity field, is the raw, driftless scalar field that drives the deformation of the curve. Parameters and amplitudes are held fixed within each calibration window. Following \citet{lecoz_bouchaud_2024}, the amplitude $\sigma_A(\theta)>0$ normalizes the velocity field by its integrated variance per unit horizon, here evaluated at a fixed reference horizon $\Delta t_{\mathrm{ref}}$ taken as one trading day:
\begin{equation}
  \sigma_A(\theta)^2\triangleq\frac{1}{\Delta t_{\mathrm{ref}}}\Var\!\left[\int_t^{t+\Delta t_{\mathrm{ref}}}A(u,\theta)\,\dd u\right].
  \label{eq:sigma-A}
\end{equation}
The field-driven yield change over $\Delta t_{\mathrm{ref}}$ thus has variance $\Delta t_{\mathrm{ref}}\,\sigma(\theta)^2$. Determined by the covariance of $A$, the amplitude $\sigma_A$ adds no fitted parameter and differs from the equal-time standard deviation of the velocity field $A$.

The drift $\gamma$ collects the no-arbitrage drift of \citet{hjm_1992}, the roll-down of a bond along the curve, and risk premia under the empirical measure. In the field theory of \citet{baaquie_2001,baaquie_2004_book}, the no-arbitrage component is expressed through the volatility and correlation kernel of the velocity field. Because $\gamma$ is numerically negligible compared to daily and intraday yield fluctuations, I drop it henceforth (see detailed claim in Appendix~\ref{app:drift}). Within each calibration window, the volatility scale $\sigma$ is held fixed and estimated as the sample standard deviation of yield changes over the calibration window. Henceforth, I factor out the positive amplitude $\sigma/\sigma_A$ and express yield changes in the arbitrary model units, re-using the yield $Y$ notation. Their Pearson correlations are identical to those of the corresponding unscaled yield changes. Over a horizon $\Delta t$, they are therefore
\begin{equation}
  \Delta Y(t,\theta)=\int_t^{t+\Delta t}A(u,\theta)\,\dd u .
  \label{eq:centered-increment}
\end{equation}
This integral runs over calendar dates $u$ at a fixed tenor $\theta$. For a held bond, the aging of its tenor over $\Delta t$ enters the roll-down term of the drift $\gamma$ and is negligible at daily and intraday horizons (Appendix~\ref{app:drift}).

The first ingredient of the model is the dynamics of the scalar velocity field $A$. Random shocks $\eta$ drive $A$, which relaxes over a characteristic time $\tau$ towards configurations favored by a quadratic action $\Sact$ that encodes interactions across maturities:
\begin{equation}
  \begin{aligned}
  \tau\,\frac{\partial A}{\partial t}(t,\theta)
  &=\frac{\delta\Sact}{\delta A}[A]\bigg|_{t,\theta}+\eta(t,\theta),\\[0.2em]
  \avg{\eta(t,\theta)\,\eta(t',\theta')}
  &=2D\,\delta(t-t')\,\delta(\theta-\theta').
  \end{aligned}
  \label{eq:primitive-noise}
\end{equation}
The shocks $\eta$ are Gaussian and independent across dates and tenors with amplitude $D>0$, and $\tau$ is the relaxation time. The specification of the action fundamentally structures the resulting joint behavior of model-implied yield changes. The action of \citet{lecoz_bouchaud_2024} contains three scalar terms penalizing the amplitude of the velocity field, its maturity gradient, and its curvature $(\partial_\theta^2A)^2/\nu^4$. Following these authors, I retain here only the first two terms\footnote{\citet{lecoz_bouchaud_2024} find that the third quartic term is not needed empirically, because the discretization step regularizes the resulting correlation structures by acting as a stiffness term.},
\begin{equation}
  \Sact[A]\triangleq-\frac12\int\dd t\int_0^\infty\dd\theta
  \left[A^2+\frac{(\partial_\theta A)^2}{\mu^2}\right].
  \label{eq:action-single}
\end{equation}
The first term controls the amplitude of the velocity field, while the second, weighted by the line tension $\mu^{-2}$, penalizes relative motion between neighboring maturities. Reflecting Neumann boundary conditions are further imposed at zero maturity,
\begin{equation}
  \partial_\theta A(t,\theta)\big|_{\theta=0}=0.
  \label{eq:neumann-main}
\end{equation}

Taking the functional derivative of the action yields the stochastic differential equation of motion
\begin{align}
  \tau\,\frac{\partial A}{\partial t}(t,\theta)
  =-\Lop[A](t,\theta)+\eta(t,\theta),
  \label{eq:field-relaxation}
\end{align}
where
\begin{equation}
    \Lop[A] \triangleq - \frac{\delta\Sact}{\delta A}[A]|_{t,\theta}  = A-\frac{\partial_\theta^2A}{\mu^2}
\end{equation}
The linear operator $\Lop$ transmits and smooths shocks across neighboring maturities, so a local shock on the curve produces a smooth deformation along its full length. \citet{lecoz_bouchaud_2024} interpret this coupling as the action of market participants who price each maturity by interpolating its neighbors. With white forcing and a self-adjoint operator $\Lop$, the stationary law of the velocity field is Gaussian with density proportional to $\exp\{-(\tau/2D)\int\dd\theta\,[A^2+(\partial_\theta A)^2/\mu^2]\}$. Thus, the action alone governs equal-time correlations, whereas $\tau$ sets the rate at which they are established (Appendix~\ref{app:variational}).

The second ingredient is spatial discretization. Because traded maturities are finite in number and discretely quoted in nature, one has to evaluate the equations on a discrete grid of tenors. Collecting the nodal values in the discretized velocity field $a_t$ turns the relaxation equation into a multivariate Ornstein-Uhlenbeck process,
\begin{equation}
  \tau\,\dd a_t=-Ma_t\,\dd t+\sqrt{2D}\,J\,\dd W_t,
  \label{eq:generic-ou}
\end{equation}
where $W_t$ is a standard Brownian motion with one independent component per node. The discrete drift operator is
\begin{equation}
  M=I_m+\mu^{-2}L_\theta,
  \label{eq:single-grid-drift}
\end{equation}
where $I_m$ is the $m\times m$ identity matrix and $L_\theta$ discretizes $-\partial_\theta^2$ with Neumann boundary conditions. On a uniform physical-tenor grid of spacing $\delta\theta$, its interior rows act as $(L_\theta a)_n=(2a_n-a_{n-1} -a_{n+1})/(\delta\theta)^2$. Following \citet{lecoz_bouchaud_2024}, the source loading matrix is $J=\diag(2,1,\ldots,1)$, with source covariance $\Bsrc\triangleq JJ^{\T}$ (see details in Appendix~\ref{app:galerkin}).

This single-curve model already accounts for two of the three stylized facts. Smooth correlation surfaces follow from the string coupling: shocks propagate across neighboring maturities through $\Lop$, so the correlation $\rho_{\theta\theta'}$ between yield changes at tenors $\theta$ and $\theta'$ varies smoothly and decays as the two tenors move apart. The Epps effect follows from the finite relaxation time $\tau$, since over windows much shorter than $\tau$ yield changes inherit the correlations of the instantaneous velocity field, which are lower than those accumulated over long horizons. At the shortest horizons, adding  an independent microstructure noise of intensity $\epsilon$ lowers correlations further, as in \citet{lecoz_bouchaud_2024} (Sec.~\ref{sec:epps}).

The final element of the single-curve model, first introduced by \citet{baaquie_bouchaud_2004}, is the maturity coordinate in which the string lives. Market participants do not perceive maturity linearly: a one-month shift at a one-year tenor weighs more than a one-month shift at a ten-year tenor. The action is then formulated in a psychological time that I denote $z$, linked to physical tenor by a strictly increasing map $z(\theta)$ that I call a clock. Following \citet{lecoz_bouchaud_2024}, I consider the logarithmic clock
\begin{equation}
  z(\theta)\triangleq\psi\log\!\left(1+\frac{\theta}{\psi}\right),
  \qquad \psi>0.
  \label{eq:clock-single}
\end{equation}
This map is approximately linear for $\theta\ll\psi$ and logarithmic for $\theta\gg\psi$. An exponential discount factor in psychological time corresponds to a power of $1+\theta/\psi$, and hence to hyperbolic discounting in physical time, consistent with \citet{farmer_geanakoplos_2009}, who show that interest-rate uncertainty induces hyperbolic discounting at long horizons. In this coordinate, the velocity field $A$ behaves as a string in $z$ governed by Eqs.~\eqref{eq:primitive-noise}--\eqref{eq:field-relaxation} with $\theta$ replaced by $z$, and the yield curve is evaluated at its quoted tenors through $z(\theta)$. This change of variable changes the form of the Laplacian $\Lop$ and its discrete counterpart $M$.

The logarithmic clock explains the broadening of the correlation ridge noted in Sec.~\ref{subsec:motivation}. For tenors beyond $\psi$, $z(\theta')-z(\theta)\simeq\psi\log(\theta'/\theta)$, so a fixed tenor gap shrinks in psychological time as maturity increases. Two bonds separated by one year are therefore closer in $z$ at long tenors than at short tenors, making their correlation higher and slowing the off-diagonal decay at long maturities.

\subsection{A graph of curves: country clocks and a common latent maturity}\label{subsec:clocks}

\subsubsection{The naive extension}\label{subsubsec:naive-extension}

A direct extension of the single-curve action couples the
$\Nc$ velocity fields at equal physical tenor. Writing
$A=(A_1,\ldots,A_{\Nc})^{\T}$, consider
\begin{equation}
  \Sact_0[A]\triangleq
  -\frac12\int\dd t\int_0^\infty\dd\theta\,
  \Bigl[
    A^{\T}QA+
    (\partial_\theta A)^{\T}K(\partial_\theta A)
  \Bigr],
  \label{eq:naive-action}
\end{equation}
where $Q=Q^{\T}\succ0$ couples countries and
$K=\diag(k_1,\ldots,k_{\Nc})\succ0$ collects their
line tensions. With Neumann boundary conditions, the
associated drift Laplacian is
\begin{equation}
  (\Lop_0A)_c
  =\sum_d Q_{cd}A_d-k_c\partial_\theta^2 A_c.
  \label{eq:naive-physical-operator}
\end{equation}
Thus each country interacts with the others at the same
quoted maturity, while its tension couples neighboring
maturities along its own curve.

Following the single-curve construction, one can instead
postulate this same string action in a common psychological
coordinate $z=z(\theta)$ given by Eq.~\eqref{eq:clock-single}.
This replaces the physical maturity geometry by a latent
one shared by all countries. Writing
$\Aph_c(t,\theta)=A_c(t,z(\theta))$ and
$h(\theta)\triangleq(\dd z/\dd\theta)^{-1}
=1+\theta/\psi$, the resulting operator expressed in
physical tenor is
\begin{equation}
  \begin{aligned}
  (\Lop_{\mathrm{naive},\theta}\Aph)_c
  ={}&\sum_dQ_{cd}\Aph_d\\
     &-k_c\left[
       h^2\partial_\theta^2\Aph_c
       +hh'\partial_\theta\Aph_c
     \right],
  \end{aligned}
  \label{eq:naive-clocked-operator}
\end{equation}
where $h'=\dd h/\dd\theta$. The common clock changes
maturity distances but preserves equal-tenor coupling:
$z(\theta)=z(\theta')$ still implies $\theta=\theta'$.

The discretization of the maturity axis promotes the Laplacian into the following matrix-valued drift operator:
\begin{equation}
  M_{\mathrm{naive}}
  =Q\otimes I_m+K\otimes L_{\mathrm{grid}},
  \label{eq:naive-grid-drift}
\end{equation}
where $L_{\mathrm{grid}}$ discretizes $-\partial_z^2$
with Neumann boundary conditions, as specified in
Eq.~\eqref{eq:latent-grid-laplacian}.
Here $\otimes$ denotes the Kronecker product:
the $(c,d)$ block of $Q\otimes I_m$ is $Q_{cd}I_m$,
while $K\otimes L_{\mathrm{grid}}$ is block diagonal
with blocks $k_cL_{\mathrm{grid}}$.
Every country therefore shares the same maturity
operator, and every cross-country coupling connects
matching maturity nodes.

The limitation follows by resolving this common maturity
operator into its eigenmodes. In its symmetric
representation, let $u_q$ denote a maturity eigenvector
with eigenvalue $\ell_q$. The country dynamics within
that mode are governed by $Q+\ell_qK$.
If the source covariance does not mix different maturity
modes, the covariance block between countries $c$ and $d$
over any common horizon has the form
\begin{equation}
  C_{cd}
  =\sum_q(\Gamma_q)_{cd}\,u_qu_q^{\T}
  =C_{cd}^{\T},
  \label{eq:naive-block-symmetry}
\end{equation}
where $\Gamma_q$ is the country covariance carried by
mode $q$. The weighted representation needed for the
Neumann discretization, and the source assumptions,
are detailed in Proposition~\ref{prop:common-grid}
of Appendix~\ref{app:naive}. Evaluating both countries through the same clock preserves this maturity-index symmetry.\footnote{ Correlation blocks can nevertheless be asymmetric through maturity-dependent variance ratios, as the fields are standardized before identifying the Pearson correlation matrix. The finite-grid source convention introduces an additional boundary-generated correction, which need not be confined to boundary nodes. Both effects are discussed in Appendix~\ref{app:naive}.}

This construction therefore provides no country-dependent
matching of physical maturities. Its remaining sources
of correlation asymmetry do not adequately reproduce
the observed pattern in the common-clock fits reported
in Appendix~\ref{app:naive}. Those observations motivate the need for a different mechanism.

\subsubsection{The proposed model}
\label{subsubsec:proposed-mechanism}

I postulate that the sovereign curves interact
at equal latent maturity $z$, with each country $c$
mapping physical tenor onto that coordinate through
its own clock $z_c(\theta)$. Equal latent maturities can
then correspond to different quoted tenors across
issuers, allowing the cross-country correlation ridges
to move away from the equal-tenor diagonal.

The vector velocity field $A(t,z)=(A_1,\ldots,A_{\Nc})^{\T}$ is defined over a shared latent maturity coordinate $z\in I=[0,\zmax]$, replacing the scalar integrand of Eq.~\eqref{eq:action-single} with the matrix-coupled action
\begin{equation}
  \Sact[A]\triangleq-\frac12\int\dd t\int_I\dd z
  \Bigl[A^{\T}QA+(\partial_zA)^{\T}K(\partial_zA)\Bigr].
  \label{eq:action}
\end{equation}
The symmetric positive-definite matrix $Q=Q^{\T}\succ0$ couples countries at equal latent maturity. Its diagonal entries set the level stiffness of each curve. An off-diagonal entry $Q_{cd}<0$ lowers the quadratic form $A^{\T}QA$ for same-sign deviations of countries $c$ and $d$, raising the action and favoring co-movement, whereas $Q_{cd}>0$ penalizes co-movement. Because $Q$ enters a Gaussian action, it is a precision matrix, as in covariance selection and Gaussian graphical models \citep{dempster_1972,lauritzen_1996,anufriev_panchenko_2015,millington_niranjan_2020}: its entries represent direct couplings conditional on all other countries (Sec.~\ref{subsec:coupling}). The country tensions $k_c$ penalize relative motion between neighboring latent maturities along each curve and generalize the single curve tension $\mu^{-2}$. A diagonal $K$ suffices to reproduce all three stylized facts. Taking $K$ as dense would also couple maturity gradients across curves, which does not seem inherently needed to explain the observed correlation structure.
A common rescaling of both $Q$,$K$ and $\frac{1}{\tau}$ leaves the scaled action invariant, as further discussed in Sec.\ref{subsubsec:scale-invar}. I therefore fix this scale by imposing 
\begin{equation}
    \operatorname{tr}Q=N_c, 
\end{equation}
which, for $ N_c=1$, recovers the single-curve model convention.

Each country receives primitive shocks of common amplitude $D$ as in Eq.~\eqref{eq:primitive-noise}, with the measure in which they are white determined by the discretization (Sec.~\ref{subsec:discretization}). Since $(\delta\Sact/\delta A)[A]|_{t,z}=-(\Lop_zA)(t,z)$, the relaxation equation~\eqref{eq:primitive-noise} becomes $\tau\,\partial_tA=-\Lop_zA+\eta$, with latent operator
\begin{equation}
  (\Lop_zA)_c=\sum_dQ_{cd}A_d-k_c\,\partial_z^2A_c
  \label{eq:latent-operator}
\end{equation}
and free-endpoint Neumann boundary conditions $\partial_zA_c=0$ on $\partial I$ (Appendix~\ref{app:variational}).

For a single curve, the clock of Sec.~\ref{subsec:baseline} shapes correlations along that curve. On a graph of curves, it also determines which physical maturities of different countries face each other through the coupling $Q$. Because equal physical tenors need not occupy the same position on different sovereign curves, each country $c$ maps physical tenor to latent maturity through its own logarithmic clock of the form~\eqref{eq:clock-single},
\begin{equation}
  z_c(\theta)\triangleq\psi_c\log\!\left(1+\frac{\theta}{\psi_c}\right),
  \label{eq:log-clock}
\end{equation}
with inverse $z_c^{-1}(z)$ returning the physical tenor of country $c$ at latent maturity $z$. The latent coordinate $z$ is thus a shared psychological time across issuers, while $\psi_c$ controls how strongly country $c$ compresses long tenors. I write $\psi=(\psi_1,\ldots,\psi_{\Nc})$. With the drift omitted, the yield velocity of country $c$ at tenor $\theta$, in the factored model units, is $\partial_tY_c(t,\theta)=A_c(t,z_c(\theta))$. Two countries $c$ and $d$ therefore interact through $Q$ along the equal-latent locus $z_c(\theta)=z_d(\theta')$.

Over a common horizon, let $C^{\mathrm{lat}}_{cd}(z,z')$ denote the covariance between the integrated raw velocity fields of country $c$ at latent maturity $z$ and country $d$ at $z'$, and let $C_{cd}(\theta,\theta')$ denote the covariance between the yield changes of country $c$ at tenor $\theta$ and country $d$ at tenor $\theta'$, before the positive observation amplitudes $\sigma/\sigma_A$ are restored. Proposition~\ref{prop:common-grid} identifies the conditions under which $C^{\mathrm{lat}}_{cd}(z,z')$ is symmetric under $z\leftrightarrow z'$. The finite-grid correction to this symmetry is discussed in Appendix~\ref{app:naive}. Independently of that symmetry, evaluation through the country clocks gives
\begin{equation}
  \begin{aligned}
    C_{cd}(\theta,\theta')
      &=C^{\mathrm{lat}}_{cd}
        \bigl(z_c(\theta),z_d(\theta')\bigr),\\
    C_{cd}(\theta',\theta)
      &=C^{\mathrm{lat}}_{cd}
        \bigl(z_c(\theta'),z_d(\theta)\bigr).
  \end{aligned}
  \label{eq:clock-asymmetry}
\end{equation}
These expressions generally differ when the country clocks differ, even for an exchange-symmetric latent covariance, allowing the correlation asymmetry of Eq.~\eqref{eq:block-asymmetry}. In physical coordinates, the ridge of a cross-country block moves away from the equal-tenor diagonal $\theta=\theta'$ onto the equal-latent locus $z_c(\theta)=z_d(\theta')$. Unequal clocks $z_c$ displace and bend the cross-country ridges, which generates the ordered-block asymmetry of Fig.~\ref{fig:asymmetry}, while the coupling $Q$ and the tensions $k_c$ set the width of the ridge and determine how strongly that displacement translates into correlation asymmetry. Section~\ref{subsec:fit} tests the predicted ridge displacement and the sign of the asymmetry against the data. Factor models with country-specific maturity loadings can also break maturity-index symmetry, but only by introducing free loadings at every maturity. The proposed model instead controls the lopsidedness of cross-country correlations through a small number of identified parameters, as discussed in Sec.~\ref{subsec:fit}.

\subsection{Discretization}\label{subsec:discretization}

The discretization of the continuous theory follows that of \citet{lecoz_bouchaud_2024}. I however emphasize here that it involves choosing the maturity grid and the measure in which the primitive shocks $\eta$ are white. Indeed, one either postulates that the shocks are white per unit of physical maturity or per unit of latent maturity, leading to two different stochastic processes.

\subsubsection{Two native-white realizations}\label{subsubsec:native}
The discretization of the continuous dynamics associated with Eq.~\eqref{eq:latent-operator} proceeds by representing the velocity field on a finite maturity grid. Stacking its nodal values across countries gives a vector $a_t$, and the spatial operator $\Lop_z$ is replaced by a matrix acting on this vector. The resulting dynamics have the Ornstein-Uhlenbeck form of Eq.~\eqref{eq:generic-ou},
\begin{equation}
\label{eq:general_dynamics}
    \tau\, da_t
    = -M a_t\,dt + \sqrt{2D}\,J\,dW_t.
\end{equation}
$J$ specifies the loading of the primitive shocks. $M$ represents the discrete, deterministic, drift operator of the equation of motion of $a_t$. The construction of $M$ is first based on replacing the spatial derivatives using finite-difference Laplacian matrices. However, encoding the coupling of velocity fields at latent coordinates requires careful examination.  Indeed, one needs to choose both the maturity grid and the measure in which the shocks are white in order to properly define the discretization. Two choices of discretization lead to two different matrix operators $M$. The first choice is to discretize the latent maturity $z$ axis into a regular grid, leading to the dynamics-governing operator $M \equiv M_z$. The second is to discretize the physical maturity $\theta$ axis into a regular grid, for which the drift operator becomes $M \equiv M_{\theta}$. The latter choice corresponds to the design of \citet{lecoz_bouchaud_2024}. 

The dynamics of Eq.~\eqref{eq:general_dynamics} are, however, fully defined only once one specifies in which space the forcing $dW_t$ is white. I call native-coordinates the coordinates in which the random shocks are white. One can then define two different stochastic processes by declaring on which regular grid the random shocks primarily occur. I call Model A the model for which the dynamics is discretized on a regular grid in latent space, for which the random forcing is latent-native, \textit{id est} white on the latent regular grid. Conversely, adopting a regular discrete grid in physical space with physical-native white forcing leads to Model B.  

In Model A, the primitive shocks are white per unit of latent maturity. All countries share a regular grid $\{z_n\}_{n=1}^{m}$, with nodal values
$a_{c,n}(t)=A_c(t,z_n)$. On the regular latent grid with spacing
$\delta z \triangleq z_{\max}/(m-1)$, the discrete Neumann
Laplacian is defined by
\begin{equation}
\label{eq:latent-grid-laplacian}
    L_{\mathrm{grid}} \triangleq
    \frac{1}{(\delta z)^2}
    \begin{pmatrix}
         2 & -2 & 0 & \cdots & 0 \\
        -1 &  2 & -1 & \ddots & \vdots \\
         0 & \ddots & \ddots & \ddots & 0 \\
        \vdots & \ddots & -1 & 2 & -1 \\
         0 & \cdots & 0 & -2 & 2
    \end{pmatrix},
\end{equation}
The first and last rows impose the Neumann conditions at $z=0$ and $z=z_{\max}$ by reflecting the field across each endpoint: the auxiliary values just outside the grid satisfy $a_0=a_2$ and $a_{m+1}=a_{m-1}$.
Denoting the $m\times m$ identity matrix by $I_m$,
replacing $-\partial_z^2$ in Eq.~\eqref{eq:latent-operator} by
$L_{\mathrm{grid}}$ yields
\begin{equation}
\label{eq:latent-grid-drift}
    M_z
    = Q \otimes I_m + K \otimes L_{\mathrm{grid}}.
\end{equation}
The first term couples countries at the same latent node,
while the second couples neighboring maturities within
each country. Although this matrix has the same Kronecker structure as Eq.~\eqref{eq:naive-grid-drift}, the observed fields are now evaluated through country-specific clocks. Consequently, two equal physical tenors need not select the same latent node, which removes the common-clock restriction derived in Sec.~\ref{subsubsec:naive-extension}.

In Model B, the primitive shocks are instead white per unit of physical tenor, and each country is represented on a regular physical-tenor grid. To express the drift
on this grid, I first write the velocity field as a function of physical tenor,
\begin{equation}
    \breve A_c(t,\theta)
    \equiv A_c\bigl(t,z_c(\theta)\bigr).
\end{equation}
This composition, called the pullback of $A_c$ by that country's own clock, assigns to each physical tenor the value of the latent field at the corresponding latent maturity.

The drift must be expressed in the same coordinate. Writing $(UA)_c(\theta)=A_c(z_c(\theta))$, its physical representation is
\begin{equation}
\label{eq:variable_change_equiv}
    \mathcal L_\theta
    = U\mathcal L_z U^{-1}.
\end{equation}
The change of coordinate transforms the maturity derivatives and preserves coupling at equal latent maturity. At quoted tenor $\theta$ of country $c$, country $d$ is read at tenor $z_d^{-1}(z_c(\theta))$. Consequently, discretizing this coupling requires a transfer between the two country grids. Together with the discretized maturity derivatives, this
transfer gives the country blocks of the drift matrix,
\begin{align}
    (M_\theta)_{cd}
    &= Q_{cd}P_{d\to c}
    + \delta_{cd}\,k_c\,G_c^{-1}L_c,\\
    P_{d\to c} &=G_c^{-1}G_{cd},
\end{align}
where $G_c$ and $L_c$ are the mass and stiffness matrices on the latent nodes $z_c(\theta_n)$ induced by country $c$'s physical grid, and $G_{cd}$ is the cross-mass matrix between the two country bases. The transfer $P_{d\to c}$ represents the coupling at equal latent maturity, while $G_c^{-1}L_c$ discretizes $-\partial_z^2$ with Neumann boundary conditions\footnote{
In physical coordinates,
$(\partial_z^2A_c)\circ z_c
=h_c^2\partial_\theta^2\Aph_c
+h_ch_c'\partial_\theta\Aph_c$,
where $h_c=(\dd z_c/\dd\theta)^{-1}=1+\theta/\psi_c$.
A first-derivative term emerges from the change of variable, as in
\citet{lecoz_bouchaud_2024}. The Galerkin construction
on the induced latent nodes incorporates both terms
(Appendices~\ref{app:variational}
and~\ref{app:galerkin}).
}. Appendix~\ref{app:variational} derives the physical operator, and Appendix~\ref{app:galerkin} defines these matrices and establishes the convergence of the Galerkin discretization used to properly construct $P_{d \to c}$. This rather involved construction is needed to express interactions at equal latent maturity, as this same mechanism underlies both Model A and B.

As the grid spacings vanish, the two discretizations recover the same deterministic evolution after the coordinate change of Eq.~\eqref{eq:variable_change_equiv}, as established in Appendix~\ref{app:galerkin}. Their native-white forcing conventions nevertheless still lead to distinct stochastic limits, because whiteness is defined relative to different maturity measures. Expressed in physical tenor, latent-white noise has covariance proportional to
\begin{equation}
  \Bigl(1+\frac{\theta}{\psi_c}\Bigr)\,\delta(\theta-\theta'),
  \label{eq:noise-jacobian-preview}
\end{equation}
where the Jacobian factor $1+\theta/\psi_c=(\dd z_c/\dd\theta)^{-1}$ measures the stretch of physical tenor per unit of latent maturity. The two stochastic laws coincide only if their noise covariances match when written in the same coordinate. Under a single common amplitude $D$, this matching requires every clock to be the same linear map, so under the logarithmic country clocks of Eq.~\eqref{eq:log-clock} the two realizations define distinct stochastic laws (Appendix~\ref{app:native-white}). Table~\ref{tab:models} summarizes the main differences between Model A and Model B.

\begin{table*}[t]
\caption{Two native-white realizations of the same elastic geometry.}
\label{tab:models}
\begin{ruledtabular}
\begin{tabular}{lll}
 & \textbf{Model A: latent-native} & \textbf{Model B: physical-native}\\
\hline
Native grid & \tcell{Common regular $z$-grid. The clocks enter when the velocity field is read at quoted tenors, $A_c(t,z_c(\theta))$.} & \tcell{Regular physical-tenor grid per country. At tenor $\theta$ of country $c$, country $d$ is read at equal latent maturity, $\Aph_d(t,z_d^{-1}(z_c(\theta)))$.}\\[0.3em]
Primitive shocks & White per unit latent maturity. & White per unit physical tenor.\\[0.3em]
Dynamics & $\tau\,\dd a_t=-M_za_t\,\dd t+\sqrt{2D}\,J_z\,\dd W_t$ & $\tau\,\dd a_t=-M_\theta a_t\,\dd t+\sqrt{2D}\,J_\theta\,\dd W_t$\\[0.3em]
Observation & $O_A(\psi)$: interpolation at the latent points $z_c(\theta_b)$. & $O_B$: selection of the quoted bonds.\\
\end{tabular}
\end{ruledtabular}
\end{table*}

Both realizations reproduce the daily correlation surface and the intraday build-up of correlations. In Appendix~\ref{app:models} I analyze whether the data can discriminate between the two specifications and find that neither can be rejected in the sense of being misspecified. Nevertheless, the state vector $A(t,z)$ of Model A lives on a single regular latent grid whose dimension does not depend on the clocks, which enter solely through the observation operator $O_A(\psi)$. Latent-white forcing also remains isotropic across countries when averaged over maturity. This simple structure yields a closed-form relation between the coupling matrix $Q$ and the neutral factors that I use to initialize the calibration (Sec.~\ref{subsec:calibration}). Model~A is also found to reconstruct the daily correlation surface more accurately and with little to no bias compared to Model B. Throughout the remainder of the main text, I therefore focus on Model~A, which I simply refer to as the model going forward. The fit accuracy metrics, comparing Model A and Model B, are nonetheless reported in Table~\ref{tab:AB} (Appendix).

\subsubsection{Covariance over a finite window}\label{subsec:finite-window}

For a fixed spatial discretization ($M=M_z$ and $J=J_z$, with $\Bsrc=JJ^{\T}$; Table~\ref{tab:models}), the nodal values across all countries on the latent grid are stacked into the discretized velocity field $a_t\in\R^N$, which follows the Ornstein-Uhlenbeck dynamics~\eqref{eq:general_dynamics}. Assuming that every eigenvalue of $M$ has a strictly positive real part, the equal-time stationary covariance of the discretized velocity field is $\Gamma \triangleq (2D/\tau)X$, where $X$ is the unique solution of the Lyapunov equation \citep{gardiner_2009}
\begin{equation}
  MX+XM^{\T}=\Bsrc .
  \label{eq:lyapunov-main}
\end{equation}
Calendar time enters the covariance of yield changes over an aggregation window $\Delta t$ through the ratio $\Delta t/\tau$ in the window function
\begin{equation}
  \Phi_{\Delta t,\tau}(M)
  \triangleq\Delta t\,M^{-1}+\tau M^{-2}
  \bigl[e^{-\Delta t\,M/\tau}-I\bigr].
  \label{eq:phi}
\end{equation}
Because the yield change over $\Delta t$ is, in model units, the time integral of the velocity field (Eq.~\eqref{eq:centered-increment}), the covariance of interest is that of the integrated discretized velocity field $\int_t^{t+\Delta t}a_u\,\dd u$, integrated over calendar dates at fixed latent nodes:
\begin{equation}
  C_{\Delta t}
  =2D\bigl[\Phi_{\Delta t,\tau}(M)X+X\Phi_{\Delta t,\tau}(M)^{\T}\bigr]
  \label{eq:exact-covariance}
\end{equation}
Expanding Eq.~\eqref{eq:exact-covariance} in the long-horizon ($\Delta t/\tau\to\infty$) and short-horizon ($\Delta t/\tau\to0^+$) limits separates spatial geometry from temporal relaxation (derivations in Appendix~\ref{app:finite-tau}):
\begin{align}
  C_{\Delta t}
  &=2D\,\Delta t\,M^{-1}\Bsrc M^{-\T}+O(D\tau),
  \label{eq:long-limit}\\
  C_{\Delta t}
  &=\frac{2D(\Delta t)^2}{\tau}X
  -\frac{D(\Delta t)^3}{3\tau^2}\Bsrc
  +O\!\left(\frac{D(\Delta t)^4}{\tau^3}\right).
  \label{eq:short-limit}
\end{align}
At long horizons (Eq.~\eqref{eq:long-limit}), the leading kernel $M^{-1}\Bsrc M^{-\T}$ does not depend on $\tau$, so correlations are governed by the spatial geometry alone. At short horizons (Eq.~\eqref{eq:short-limit}), yield changes inherit the equal-time covariance $X$ of the velocity field. Observed bonds are then read from the latent grid through the observation matrix $O \equiv O_A(\psi)$, which interpolates the latent velocity field at the points $z_c(\theta_b)$ and therefore carries the dependence on the clocks. The covariance of the integrated raw velocity field read at the quoted bonds corresponds to the covariance of the observed yield changes and reads
\begin{equation}
  C^{\mathrm{obs}}_{\Delta t} \triangleq O\,C_{\Delta t}\,O^{\T}.
  \label{eq:observed-covariance}
\end{equation}

The drift matrix defines a local interaction graph, with one node per country and latent grid point. Each node couples to neighboring maturities within its country and to the same latent maturity in country $d$ whenever $Q_{cd}\neq0$. The stationary field is Gaussian, but its conditional-independence graph is determined by $\Gamma^{-1}=(\tau/2D)X^{-1}$, with $X$ given by Eq.~\eqref{eq:lyapunov-main}. Under the source convention of Appendix~\ref{app:galerkin}, this precision matrix need not share the sparsity pattern of the drift. The observed yield changes remain jointly Gaussian, because time integration and the observation map $O$ are linear, but their spatial precision matrices are generally dense and the observed process need not remain Markov in calendar time (Appendix~\ref{app:models}).

The covariance in yield units is obtained by multiplying entry $(i,j)$ by $\sigma_i\sigma_j/(\sigma_{A,i}\sigma_{A,j})$, where $\sigma_i$ and $\sigma_{A,i}$ are the two amplitudes of Eq.~\eqref{eq:yield-dynamics} for bond $i$. These positive bondwise factors cancel from Pearson correlations. The reference normalization is evaluated after observation, $\sigma_{A,i}^2=(C^{\mathrm{obs}}_{\Delta t_{\mathrm{ref}}})_{ii}/\Delta t_{\mathrm{ref}}$. The common amplitude $D$ multiplies every covariance entry and cancels out of all correlations, and as such has no physical meaning and can be taken as $1$ without loss of generality.

\subsubsection{Scale invariance of model-implied correlations}\label{subsubsec:scale-invar}

The postulated model implies that correlations, at any horizon $\Delta t$, are invariant under the following scaling for any real $c>0$: 
\begin{equation}
\label{eq:corr-scale-invar}
   (Q,K,\tau,\epsilon)\to(cQ,cK,c\tau,\epsilon/c^2), 
\end{equation}
where the noise intensities $\epsilon$, introduced in Sec.~\ref{subsec:microstructure}, is a micro-structure noise intensity responsible for the Epps effect.
The characteristic relaxation time $\tau$ is physically defined only once the scale of the operator $M$ is fixed. The normalization according to which $\tr Q=\Nc$ recovers the normalization of \citet{lecoz_bouchaud_2024} for $N_c = 1$, allowing at the same time to identify $\tau$ here and compare it to the one found on single-curve asset classes. In the absence of forcing, the amplitude $x(t)$ of an eigenmode $Mu=\lambda u$ satisfies $\tau\dot x=-\lambda x$ and therefore decays over the time scale $\tau/\lambda$. A smaller positive eigenvalue means a weaker restoring drift and slower relaxation. I call such a mode \emph{soft}, relative to modes with larger eigenvalues, which are \emph{stiffer}. When $\Delta t/\tau\gg1$, daily correlations are governed by the long-horizon kernel $M^{-1}\Bsrc M^{-\T}$, so they identify the spatial geometry of the model (clocks, tensions, and coupling) independently of $D$ and $\tau$. An eigenmode of eigenvalue $\lambda$ reaches the long-horizon limit of Eq.~\eqref{eq:long-limit} only when $\Delta t\gg\tau/\lambda$, and the softest mode corresponding to the common level of all curves relaxes over more than one day (Appendix~\ref{app:neutral}). This slow mode leaves daily correlations unaffected because empirical correlations reach their long-horizon plateau well within a single trading day (see Fig.~\ref{fig:epps-empirical}), so the daily correlation surface shows no measurable dependence on $\tau$. Using the joint fit on the intra-day sample, I also verified that artificially pushing to $\Delta t = 1$ month yields a numerically negligible difference with the model-implied correlations at $\Delta t = 1$ day.

\section{Daily calibration}\label{sec:daily}

\subsection{Neutral modes and the fitting procedure}\label{subsec:calibration}

In the daily calibration, I estimate the clocks, the tensions, and the coupling matrix $Q$ jointly against the full correlation surface. To initialize $Q$, I average each country's velocity field over latent maturity, which eliminates the second-derivative tension term of the latent operator~\eqref{eq:latent-operator} under the Neumann boundary conditions and isolates $Q$. On the discrete grid, the exact latent average is a weighted sum of the nodal values,
\begin{equation}
  r_c(t)\triangleq\frac{\bm 1_c^{\T}G_c\,a_c(t)}{\bm 1_c^{\T}G_c\bm 1_c},
  \label{eq:neutral-factor}
\end{equation}
where $a_c$ is the subvector of $a_t$ corresponding to country $c$, $\bm 1_c$ is the vector of ones, and $G_c$ is the mass matrix of the common latent grid (Appendix~\ref{app:galerkin}). I call $r_c$ the neutral factor of country $c$. Because the discrete tension operator annihilates constants and the coupling preserves them, the neutral factors satisfy a closed $\Nc$-dimensional Ornstein-Uhlenbeck system under $Q$ (Appendix~\ref{app:neutral}),
\begin{equation}
  \tau\,\dd r_t=-Qr_t\,\dd t+\sqrt{2D}\,J_r\,\dd W_t,
  \qquad V\triangleq J_rJ_r^{\T},
  \label{eq:neutral-ou-main}
\end{equation}
where $V$ is the covariance of the primitive shocks projected onto the neutral factors. In Model~A, this projected source covariance is isotropic across countries, $V=v\,I_{\Nc}$ with $v>0$, so the long-horizon covariance $C_r$ of the time-integrated neutral factors satisfies $C_r\propto Q^{-2}$ (the equal-time covariance of $r_t$ scales as $Q^{-1}$), and hence
\begin{equation}
  Q\propto C_r^{-1/2}.
  \label{eq:Q-update-A}
\end{equation}
At the model level, this relation identifies all entries of $Q$ simultaneously from the covariance matrix of the $\Nc$ raw neutral factors and yields a positive-definite estimate by construction. Empirically, a proxy for the neutral factor of each country is computed by interpolating its volatility-standardized yield changes in latent maturity and averaging over the latent interval covered by all countries, producing a weighted sum of the country's standardized bond yield changes with clock-dependent weights (Appendix~\ref{app:neutral}). Applied to these empirical proxies, the identity~\eqref{eq:Q-update-A} provides the initial estimate of $Q$ before the joint optimization over the clocks, the tensions, and $Q$ (with $Q$ parameterized by its Cholesky factor under $\tr Q=\Nc$ and the observation matrix rebuilt at every clock update; Appendices~\ref{app:neutral} and~\ref{app:numerics}).

Following \citet{lecoz_bouchaud_2024}, I measure the quality of a fit by the typical error $\Sigma$, the standard deviation of the residuals over the set $\mathcal I$ of fitted entries of the correlation matrix,
\begin{equation}
  \begin{gathered}
  \mathcal E_{ij}\triangleq\rho^{\mathrm{model}}_{ij}-\rho^{\mathrm{emp}}_{ij},
  \qquad
  \bar{\mathcal E}\triangleq\frac{1}{|\mathcal I|}\sum_{(i,j)\in\mathcal I}\mathcal E_{ij},\\
  \Sigma^2\triangleq\frac{1}{|\mathcal I|}
  \sum_{(i,j)\in\mathcal I}\bigl(\mathcal E_{ij}-\bar{\mathcal E}\bigr)^2 .
  \end{gathered}
  \label{eq:fit-error}
\end{equation}

\subsection{Reconstruction of the daily surface}\label{subsec:fit}

\begin{figure*}[t]
  \centering
  \includegraphics[width=\textwidth]{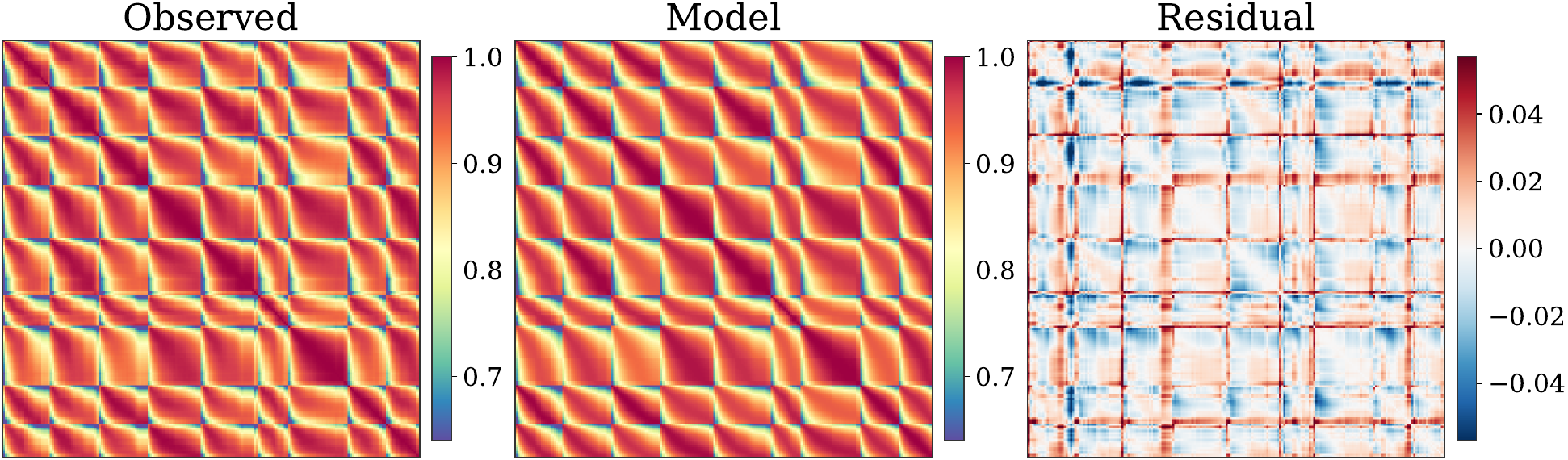}
  \caption{Correlation matrix of daily yield changes for the nine issuers (daily sample): observed (left), reconstructed by the model (center), and residual $\mathcal E_{ij}=\rho^{\mathrm{model}}_{ij}-\rho^{\mathrm{emp}}_{ij}$ (right). Countries are in alphabetical order, AT, BE, DE, ES, FR, IE, IT, NL, PT, and, within each country, bonds are ordered by increasing tenor, totaling 196 bonds with tenors from 1 to 95 years (AT 22 bonds, 1--95 years; BE 23, 2--46 years; DE 23, 1--31 years; ES 25, 1--46 years; FR 27, 1--46 years; IE 14, 1--30 years; IT 28, 1--46 years; NL 18, 1--28 years; PT 16, 1--28 years). The typical error defined in Eq.~\eqref{eq:fit-error} is $\Sigma=1.7\%$.}
  \label{fig:daily-surface}
\end{figure*}

Across the 196-bond, nine-country daily surface, the calibrated model reaches a typical error of
\begin{equation}
  \Sigma=1.7\%,
  \label{eq:sigma-value}
\end{equation}
comparable to the $\sim 1\%$ error reported for a single curve of interest-rate futures \citep{lecoz_bouchaud_2024}. The reconstruction captures the block structure of the surface, the broadening of the ridges with tenor, and the relative level of the cross-country blocks (Fig.~\ref{fig:daily-surface}; three-dimensional views in Fig.~\ref{fig:frit-3d} for France and Italy and Fig.~\ref{fig:all9-3d} of Appendix~\ref{app:numerics} for all nine issuers). The few isolated stripes in the residual map come from idiosyncratic fluctuations of individual bonds that a smooth field does not aim to fit. The reported calibration uses a uniform latent grid of 96 nodes. Refining the grid to 128 nodes leaves $\Sigma$ unchanged at the reported precision, although separating two bonds quoted at closer maturities operationally would require a finer mesh, see Appendix~\ref{app:numerics}).

\begin{figure}[t]
  \centering
  \includegraphics[width=\columnwidth]{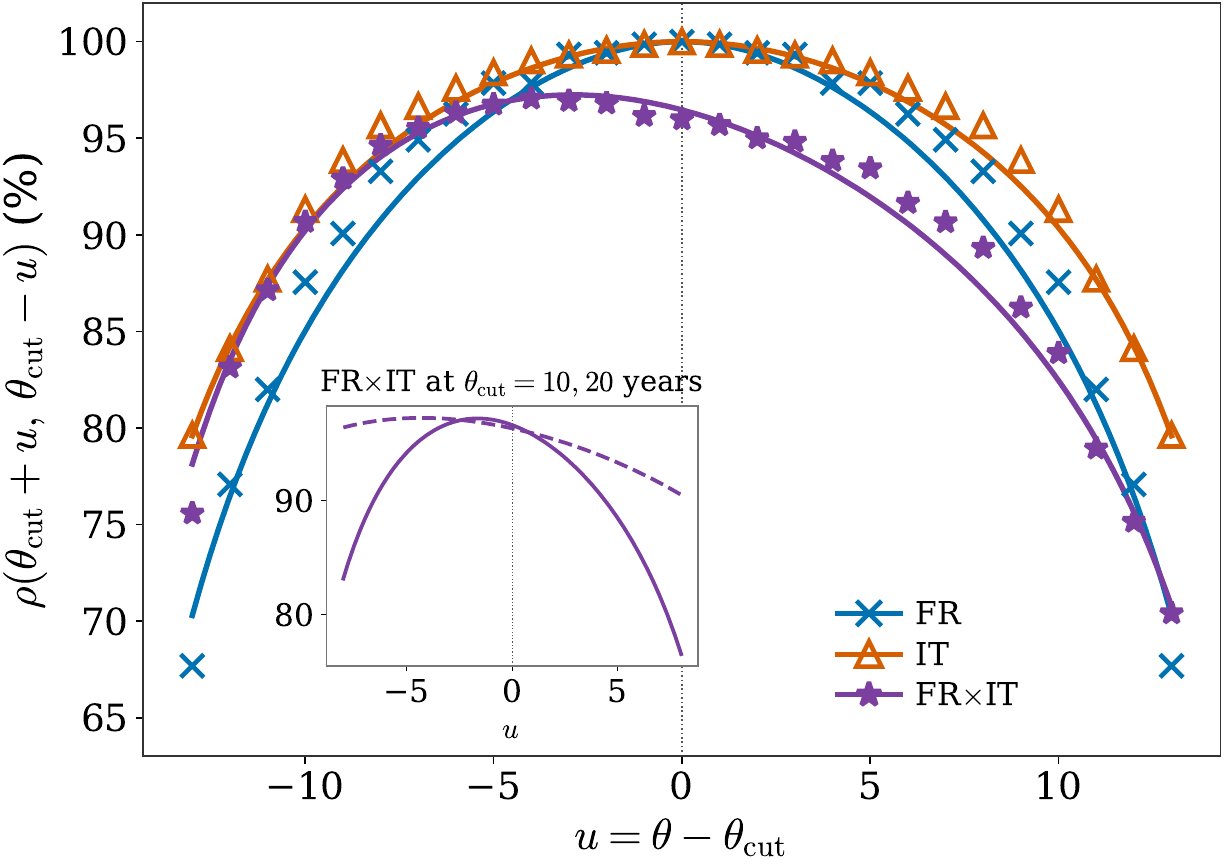}
  \caption{Correlation along the cut perpendicular to the diagonal through the center tenor $\theta_{\mathrm{cut}}=15$ years: $\rho(\theta_{\mathrm{cut}}+u,\theta_{\mathrm{cut}}-u)$ as a function of the signed shift $u$, within France (FR), within Italy (IT), and in the ordered cross block (FR$\times$IT, France at $\theta_{\mathrm{cut}}+u$ and Italy at $\theta_{\mathrm{cut}}-u$). Markers show the daily sample and lines show the model. The inset plots the model FR$\times$IT cuts at $\theta_{\mathrm{cut}}=10$ (solid) and $20$ years (dashed), whose maxima sit at negative $u$.}
  \label{fig:antidiag}
\end{figure}

The model predicts that each cross-country correlation ridge follows the equal-latent locus $z_c(\theta)=z_d(\theta')$ of Sec.~\ref{subsec:clocks}. Once the nine curves are mapped onto the shared coordinate $z$ where the coupling $Q$ acts, all cross-country ridges straighten and align along the main diagonal of every block (Fig.~\ref{fig:latent}). In physical tenors, cuts perpendicular to the diagonal test this ridge displacement (Fig.~\ref{fig:antidiag}). Around a center tenor $\theta_{\mathrm{cut}}$, let $\theta=\theta_{\mathrm{cut}}+u$ and $\theta'=\theta_{\mathrm{cut}}-u$. In the France-Italy cross block, where $\psi_{\cIT}<\psi_{\cFR}$, an Italian bond co-moves most strongly with a shorter-maturity French bond: the profile peaks below one and is flatter, reaching a model maximum of $97.2\%$ at $u\simeq-2.8$~years for $\theta_{\mathrm{cut}}=15$~years, close to the equal-latent shift $z_{\cFR}(\theta_{\mathrm{cut}}+u^\ast)=z_{\cIT}(\theta_{\mathrm{cut}}-u^\ast)$ . The model implied correlations depart at most of $3 \%$ on the extremes of the anti-diagonal. Across all 36 country pairs (Fig.~\ref{fig:allpairs-cuts}, Appendix~\ref{app:numerics}), the model tracks the empirical profiles and the shift of the cross-block peak follows the clock differences of Table~\ref{tab:daily-params}, staying within $0.4$~years of the diagonal for issuer pairs with nearly identical clocks.

The clocks also predict the cell-by-cell sign of the ordered-block asymmetry tested in Fig.~\ref{fig:asymmetry}. Because $\psi_{\cIT}<\psi_{\cFR}$, the model implies $\rho(\Delta Y_{\cFR,i},\Delta Y_{\cIT,j})>\rho(\Delta Y_{\cFR,j},\Delta Y_{\cIT,i})$ for every pair of tenors $\theta_i<\theta_j$. Empirically, this sign holds in 218 of the 231 cells, and in 169 of the 170 Romano-Wolf rejections (Fig.~\ref{fig:asymmetry}). Refitting the model on two-thirds of the bonds shifts each $\psi_c$ by less than $1\%$, which is under one-tenth of $\psi_{\cFR}-\psi_{\cIT}$ (Appendix~\ref{app:models}).

Because the clocks reparameterize maturity, the cross-country correlation equals the latent correlation evaluated at the two clock depths, $\rho_{cd}(\theta,\theta') = \varrho_{cd}\big(z_c(\theta),z_d(\theta')\big)$. In panel (a), where France and Italy are interchangeable in tensions and coupling, the calibrated clocks alone bend the ridge of maximal correlation onto the equal-latent locus.

Because the logarithmic clock is concave, the displaced ridge also exhibits flank asymmetry, with the flank facing the main diagonal flatter than the outer flank in a ratio determined by the clocks alone (Appendix~\ref{app:naive}), matching the empirical France-Italy cut in Fig.~\ref{fig:antidiag}. A second, complementary source of maturity-index asymmetry enters through the latent correlations themselves. 
For an exchange-symmetric latent covariance block, the variance ratio $\Lambda_{cd}(z)=
C^{\mathrm{lat}}_{cc}(z,z)/
C^{\mathrm{lat}}_{dd}(z,z)$ gives
$\varrho_{cd}(z,z')/\varrho_{cd}(z',z)
=\sqrt{\Lambda_{cd}(z')/\Lambda_{cd}(z)}$. This relation isolates the variance-ratio channel from the boundary generated covariance asymmetry described in Appendix~\ref{app:naive}. Figure~\ref{fig:channels} separates the clock mechanism from the complementary variance-ratio channel.

At the calibrated tensions and coupling, $\Lambda_{cd}(z)$ is nearly constant across maturity, so imposing a common France-Italy clock leaves the cross block almost symmetric under exchange of its maturity indices (Fig.~\ref{fig:channels}(b)), whereas the full calibrated block (Fig.~\ref{fig:channels}(c)) inherits the bent, lopsided ridge generated by the clocks in panel~(a). Separately from this variance-ratio channel, the tensions and coupling control the width of the cross-country ridge, which determines how strongly the clock displacement $u^\ast$ translates into correlation asymmetry. Two counterfactuals confirm these roles: coupling curves at equal physical tenor without logarithmic clocks raises the global typical error from $1.7\%$ to $6.0\%$, whereas a shared logarithmic clock can approximate the ordered-block asymmetry only by distorting maturity-dependent variance ratios (more details in Appendix~\ref{app:naive}).

\begin{figure*}[t]
  \centering
  \includegraphics[width=1.05\textwidth]{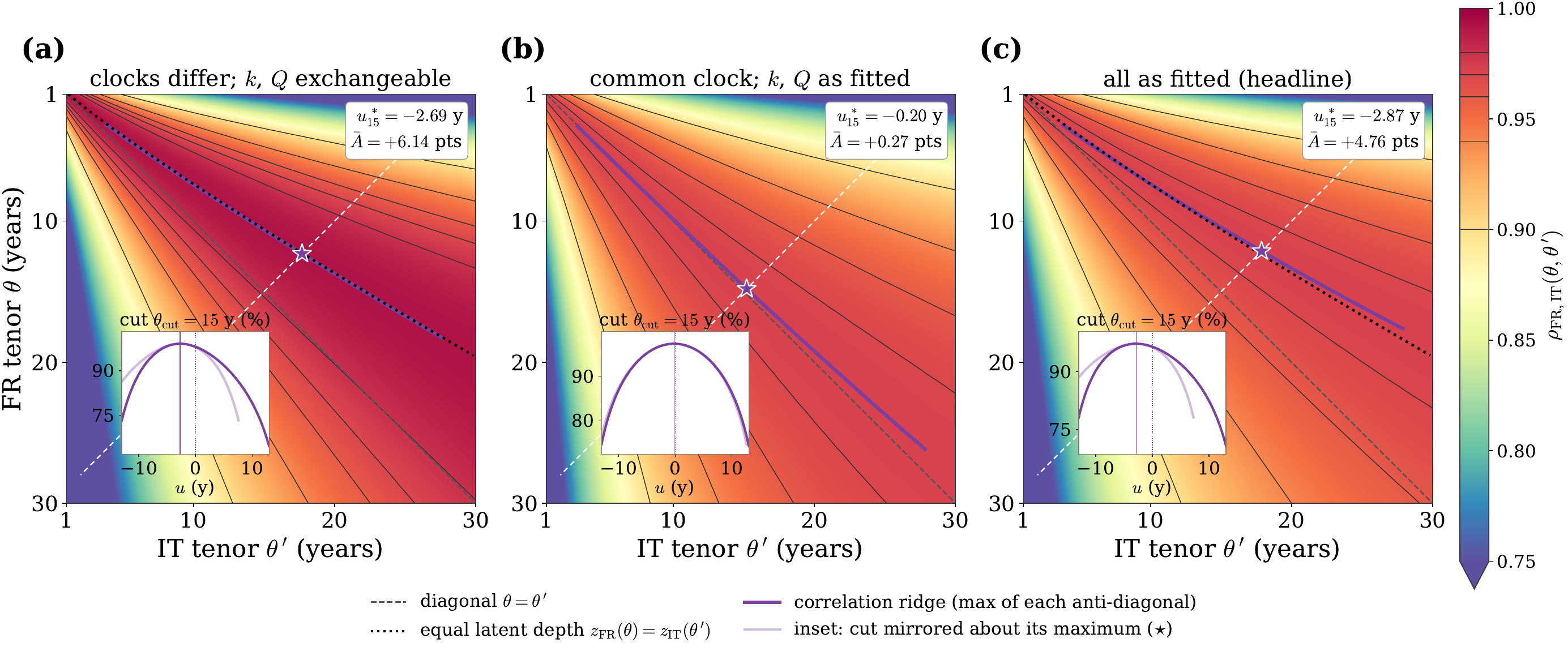}
  \caption{Clocks and variance ratios in the ordered-block asymmetry. France-Italy correlation blocks $\rho_{\cFR,\cIT}(\theta,\theta')$ of the calibrated Model~A with (a) the calibrated clocks, France and Italy being interchangeable in tensions and coupling, (b) a common France-Italy clock ($\psi=0.335$) with the calibrated tensions and coupling, and (c) all parameters as calibrated. The purple curve marks the ridge of maximal correlation along each anti-diagonal, the dotted curve the equal-latent locus $z_{\cFR}(\theta)=z_{\cIT}(\theta')$, the dashed line the diagonal, and the thin lines the iso-correlation contours. Insets show the cut $\rho(15+u,15-u)$ (\%) and its mirror image about the maximum ($\star$), whose gap is the lopsidedness of the ridge. Each panel reports the ridge position $u^\ast_{15}$ at $\theta_{\mathrm{cut}}=15$~years and the mean ordered-block asymmetry $\bar A$ over the 231 common-tenor cells of Sec.~\ref{subsec:motivation}.}
  \label{fig:channels}
\end{figure*}

\begin{figure}[t]
  \centering
  \includegraphics[width=1.02\columnwidth]{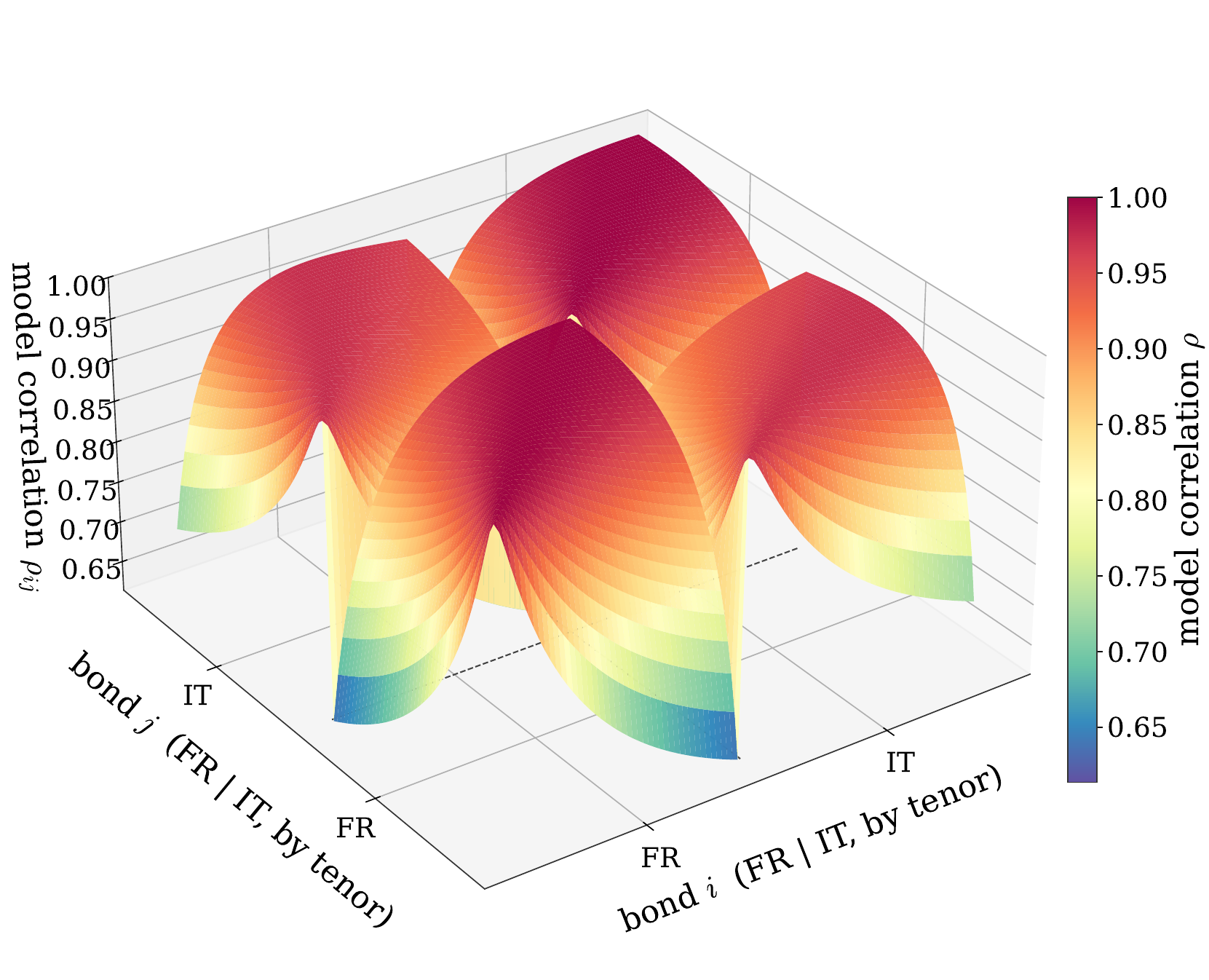}
  \caption{Model correlation surface for French and Italian bonds (daily calibration). Bonds are ordered by country and, within each country, by tenor. The surface comprises the two diagonal blocks, which peak at one along their ridges, and the two cross blocks, which form slightly lower, smoother ridges.}
  \label{fig:frit-3d}
\end{figure}

To test whether the calibrated geometry generalizes to maturities outside the fitting set, I withhold 21 of the 196 bonds by removing every tenth bond in each country starting from the third shortest, so that every issuer retains its shortest and longest maturities. I then refit the model on the remaining 175 bonds. The model-implied correlations on those held-out bonds achieve a typical error of $1.59\%$ (Table~\ref{tab:AB}), with a $95\%$ confidence interval of $[-0.58,+0.01]$\% for the withheld-minus-in-sample error difference under a ten-day block bootstrap across 500 replicates \citep{kunsch_1989,politis_romano_1994,lahiri_2003}.  

\subsection{Clocks and tensions}\label{subsec:clock-results}

\begin{table}[b]
\caption{Daily calibration: clock scales $\psi_c$, tensions $k_c$, and diagonal couplings $Q_{cc}$, with the normalization $\tr Q=\Nc$ (full coupling matrix in Table~\ref{tab:Q}). Typical error $\Sigma=1.7\%$ on a regular latent grid of 96 nodes.}
\label{tab:daily-params}
\begin{ruledtabular}
\begin{tabular}{lccc}
Country & $\psi_c$ (yr) & $10^3\,k_c$ (yr$^2$) & $Q_{cc}$\\
\hline
AT & 0.355 & 0.233 & 0.057\\
BE & 0.352 & 8.59 & 1.302\\
DE & 0.357 & 2.13 & 0.448\\
ES & 0.341 & 142 & 3.238\\
FR & 0.357 & 5.91 & 0.759\\
IE & 0.355 & 0.412 & 0.208\\
IT & 0.315 & 28.3 & 0.539\\
NL & 0.357 & 1.34 & 0.810\\
PT & 0.339 & 16.7 & 1.638\\
\end{tabular}
\end{ruledtabular}
\end{table}

All nine calibrated clocks share the same shape: every country's clock $z_c(\theta)$ is nearly linear below four months and exhibits logarithmic behavior at longer maturities, with $\psi_c$ clustering between $0.31$~years (IT) and $0.36$~years (DE, FR, NL; Table~\ref{tab:daily-params}). At a ten-year tenor, one additional year of physical tenor corresponds to $1.6$ to $1.8$~weeks of latent maturity, close to the slightly stronger compression found by \citet{lecoz_bouchaud_2024} for interest-rate futures, where the linear-to-logarithmic crossover occurs near two months and one year at the ten-year point spans roughly one week of latent time. 

Within this narrow range, the small cross-country differences in $\psi_c$ that generate the ridge shifts of Sec.~\ref{subsec:fit} order with sovereign credit spreads, decreasing from $0.352$--$0.357$~years in the six lower-spread issuers to $0.339$--$0.341$~years in Portugal and Spain and $0.315$~years in Italy (Table~\ref{tab:daily-params}). A given tenor on a peripheral, higher-spread curve therefore co-moves most strongly with a shorter tenor on a core curve. I think this ordering reflects the shorter effective duration of higher-yield coupon bonds and a compression of the horizon over which market participants discount future rate uncertainty when credit risk is higher. 

\begin{figure}[t]
  \centering
  \includegraphics[width=0.90\columnwidth]{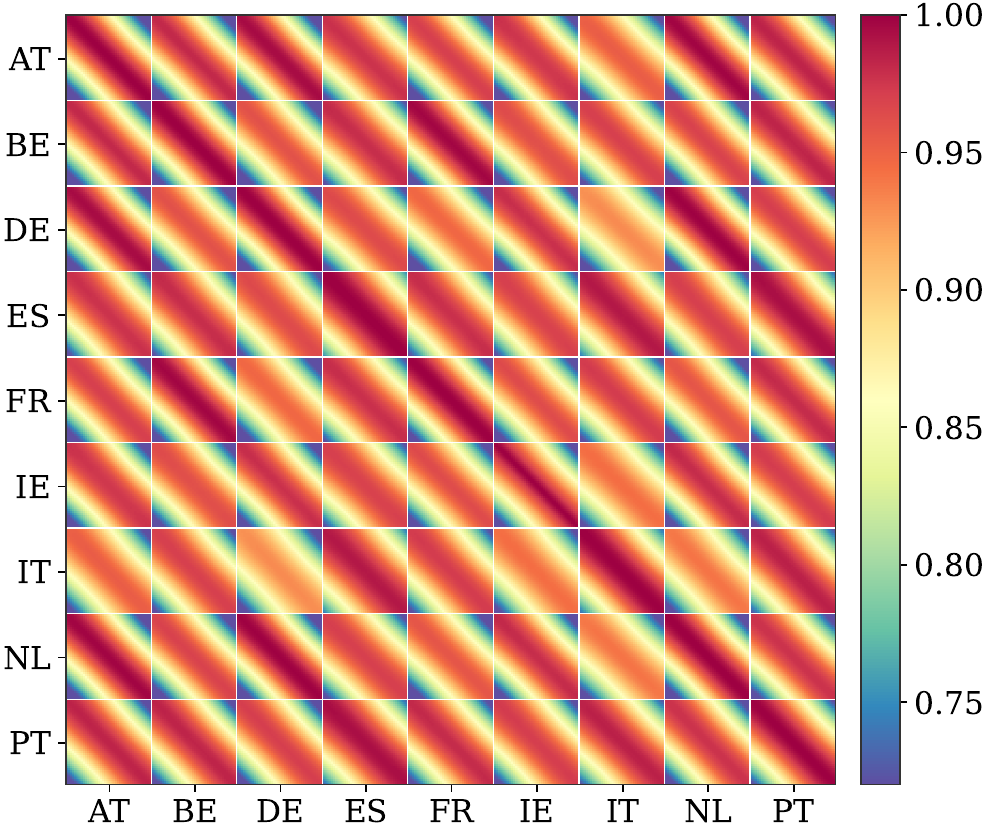}
  \caption{Model correlation surface of the latent velocity field on the common latent grid, countries in alphabetical order (daily calibration). Every block, diagonal or not, has its ridge along its own diagonal. The clocks map the unequal physical alignments onto a common latent maturity.}
  \label{fig:latent}
\end{figure}

\subsection{Country coupling}\label{subsec:coupling}

As noted in Sec.~\ref{subsec:clocks}, the entries of $Q$ are precision-matrix couplings conditional on all other issuers (Table~\ref{tab:Q} in Appendix~\ref{app:numerics} and Fig.~\ref{fig:qgraph}). Unconditionally, the neutral factors $r_c$ of Eq.~\eqref{eq:neutral-factor} are almost collinear: their long-horizon covariance matrix $C_r\propto Q^{-2}$ implies pairwise correlations between $0.948$ and $0.9995$ across all country pairs. This shared comovement is the common level of the nine curves and is carried by the softest eigenmode of $Q$, whose eigenvalue is $1.0\times10^{-3}$ relative to $\tr Q/\Nc=1$ and whose eigenvector components all share the same sign, led by AT ($+0.93$), DE ($+0.26$), and NL ($+0.19$). A zero eigenvalue would leave the common level unanchored, and the calibrated coupling lies close to this limit. A negative entry $Q_{cd}<0$ indicates a pair that co-moves more closely than implied by the rest of the network, whereas a positive entry $Q_{cd}>0$ indicates a pair whose spreads tend to move in opposite directions around the common level.

Six country pairs maintain the same coupling sign across all tested subperiods of the daily sample: BE-FR, DE-IT, DE-IE, ES-PT, AT-DE, and AT-IT. The remaining off-diagonal entries change sign across subperiods. Because the baseline specification treats $Q$ as constant over the calibration window, Table~\ref{tab:Q} captures the average coupling over the sample. Promoting $Q$ to a dynamical state variable is left to future work.

\begin{figure}[!tbp]
  \centering
  \includegraphics[width=0.80\columnwidth]{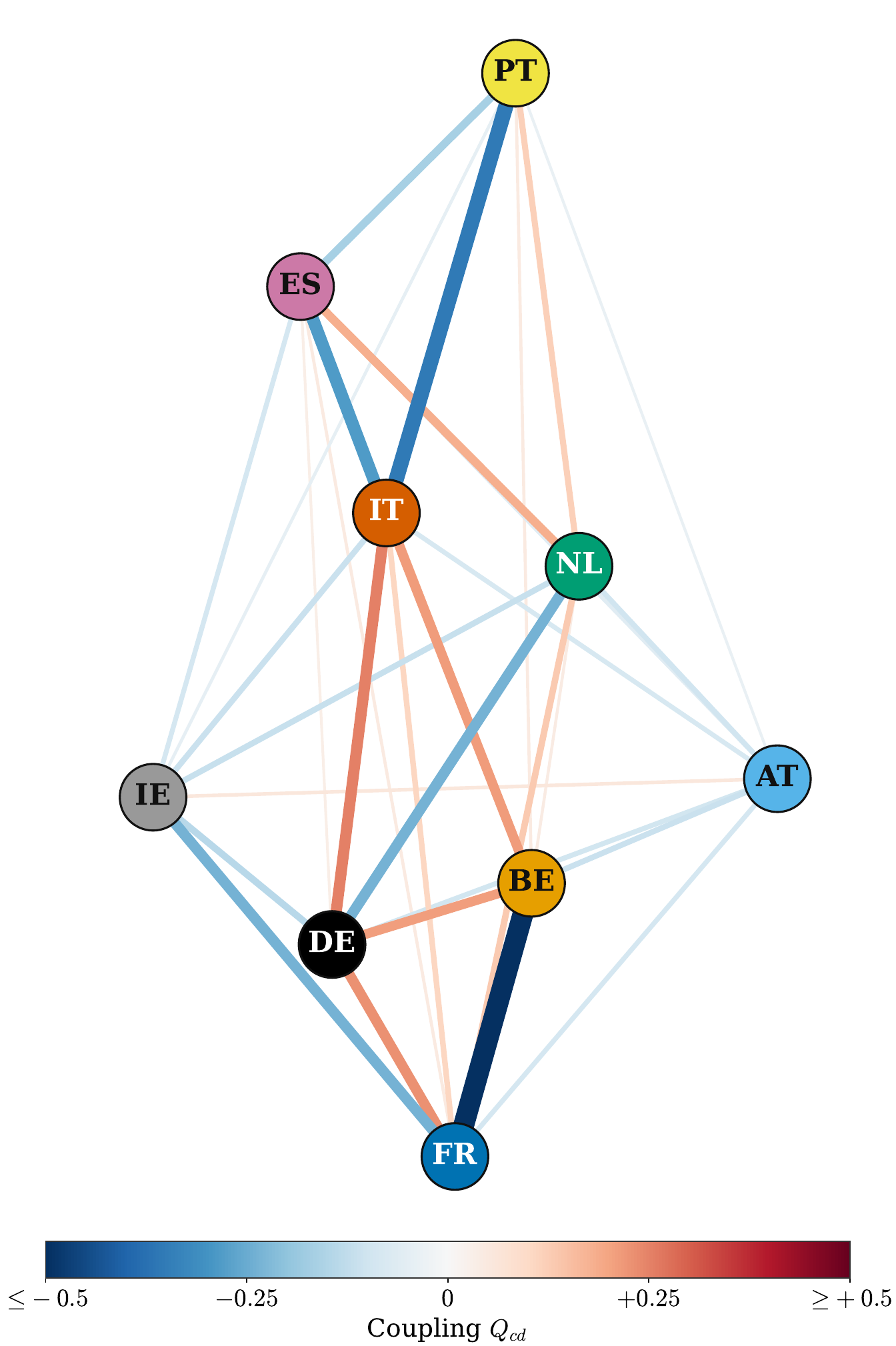}
  \caption{Spring-layout network of the off-diagonal country couplings $Q_{cd}$ from the daily calibration (Table~\ref{tab:Q}, normalized by $\tr Q=\Nc$). Edge width and color saturation increase with $|Q_{cd}|$: blue edges denote negative couplings ($Q_{cd}<0$, favoring co-movement) and red edges denote positive couplings ($Q_{cd}>0$, penalizing co-movement relative to the common level).}
  \label{fig:qgraph}
\end{figure}

\subsection{Spread factors}\label{subsec:spread-factors}

The spectral decomposition of $Q$ organizes the cross-country coupling into orthogonal spread factors. Let $\varpi_1\leq\cdots\leq\varpi_{\Nc}$ denote the eigenvalues of $Q$ and $e_\alpha$ its orthonormal eigenvectors. Because the projected source covariance of the neutral factors is isotropic ($V=v\,I_{\Nc}$; Sec.~\ref{subsec:calibration}), it remains isotropic under an orthogonal change of basis. The modal projections $\tilde r_\alpha=e_\alpha^{\T}r$ of Eq.~\eqref{eq:neutral-ou-main} therefore evolve as independent scalar Ornstein-Uhlenbeck processes,
\begin{equation}
  \tau\,\dd\tilde r_{\alpha,t}=-\varpi_\alpha\tilde r_{\alpha,t}\,\dd t+\sqrt{2Dv}\,\dd W_{\alpha,t},
  \label{eq:spread-modes}
\end{equation}
driven by independent Brownian motions $W_{\alpha}$ (Appendix~\ref{app:neutral}). These modes describe the velocity of yield-curve deformations. To be clear, the corresponding yield changes are obtained by integrating the modes over calendar time, so Eq.~\eqref{eq:spread-modes} does not by itself imply mean reversion of yield-spread levels.
Mode $\alpha$ relaxes over the time scale $\tau/\varpi_\alpha$, with equal-time variance $Dv/(\tau\varpi_\alpha)$ and long-window variance $2Dv\,\Delta t/\varpi_\alpha^2$. Soft modes have large amplitude and slow mean reversion, whereas stiff modes ($\varpi_\alpha$ large) have small amplitude and fast mean reversion. The softest mode ($\alpha=1$) is the common level described in Sec.~\ref{subsec:coupling}. The remaining eight orthogonal modes ($\alpha=2,\ldots,9$) can be read as spread factors.

\begin{table}[t]
\caption{Spectrum of the daily coupling $Q$ (Table~\ref{tab:Q}): eigenvalues $\varpi_\alpha$, relaxation rate relative to the common level $\varpi_\alpha/\varpi_1$, and components of the eigenvector $e_\alpha$ larger than $0.25$ in absolute value, the overall sign being fixed by the largest component.}
\label{tab:spread-modes}
\footnotesize
\begin{ruledtabular}
\begin{tabular}{rrrl}
$\alpha$ & $\varpi_\alpha$ & $\varpi_\alpha/\varpi_1$ & Leading components (loadings) of $e_\alpha$\\
\hline
1 & $1.02\times10^{-3}$ & 1 & \lcell{AT $+0.93$, DE $+0.26$ (all positive)}\\
2 & 0.03629 & 35 & \lcell{DE $+0.68$, FR $-0.51$, BE $-0.32$, NL $+0.30$}\\
3 & 0.08617 & 84 & \lcell{IE $+0.92$}\\
4 & 0.1955 & 191 & \lcell{IT $+0.81$, FR $-0.37$, BE $-0.33$}\\
5 & 0.8933 & 874 & \lcell{NL $+0.89$, DE $-0.40$}\\
6 & 1.056 & 1033 & \lcell{FR $+0.54$, DE $+0.50$, IT $+0.44$, IE $-0.34$}\\
7 & 1.663 & 1626 & \lcell{BE $+0.84$, FR $-0.49$}\\
8 & 1.766 & 1727 & \lcell{PT $+0.92$, IT $-0.31$}\\
9 & 3.302 & 3229 & \lcell{ES $+0.99$}\\
\end{tabular}
\end{ruledtabular}
\end{table}

In the calibrated spectrum (Table~\ref{tab:spread-modes}), the common level ($\alpha=1$) is nearly free and relaxes 35 times more slowly than the softest spread mode ($\varpi_2/\varpi_1\simeq35$). In such a single-currency setting, the overall level of euro-area yields follows the single monetary policy, so no restoring force internal to the sovereign bond network pulls the common level back, whereas relative spreads across issuers remain anchored. Two of the three softest spread modes correspond to the main benchmark spreads of the euro-area market \citep{dunne_et_al_2007,monfort_renne_2014}: Mode~2 opposes Germany and the Netherlands to France and Belgium, capturing the semi-core against core spread led by the OAT-Bund differential. Mode~4 opposes Italy to France and Belgium, capturing the Italian spread against the semi-core. The French semi-core spread mode is found to be softer than the Italian spread mode, giving it a larger variance and slower mean reversion. This ordering could reflect the political and fiscal uncertainty in France beginning in mid-2024.

In Table~\ref{tab:spread-modes}, the eigenvectors act on the raw neutral factors under isotropic latent forcing. Because the calibration targets Pearson correlations, these loadings are not cash hedge ratios, and the uneven weights of the common mode come from the diagonal scaling of $Q$. Consecutive spread eigenvalues also differ by factors of only $1.06$ to $4.6$, so neighboring eigenvectors such as Modes~7 and~8 can mix under finite-sample perturbations, while daily correlations constrain the softest, most persistent directions of $Q$ most tightly (Appendix~\ref{app:neutral}).

\section{Intraday horizons: the Epps effect}\label{sec:epps}

\subsection{Finite relaxation}\label{subsec:relaxation}

Because an observed yield change in model units is the time integral of a velocity field that relaxes on the finite time scale $\tau$ of Sec.~\ref{subsec:baseline}, Eqs.~\eqref{eq:long-limit}--\eqref{eq:short-limit} imply two asymptotic regimes for the covariance of the integrated raw velocity field:
\begin{equation}
  C_{\Delta t}\simeq
  \begin{cases}
  \dfrac{2D(\Delta t)^2}{\tau}X, & \Delta t\ll\tau,\\[0.9em]
  2D\,\Delta t\,M^{-1}\Bsrc M^{-\T}, & \Delta t\gg\tau,
  \end{cases}
  \label{eq:epps-field-limits}
\end{equation}
where $X$ solves the Lyapunov equation~\eqref{eq:lyapunov-main}, and observed bonds inherit both regimes through the observation map of Eq.~\eqref{eq:observed-covariance}. At short horizons ($\Delta t\ll\tau$), the velocity field changes little within the observation window, so the covariance scales quadratically with $\Delta t$. Because variances and covariances carry the same $(\Delta t)^2$ factor, correlations at $\Delta t\ll\tau$ approach those of the instantaneous velocity field, $\Corr(X)$. Finite relaxation therefore interpolates smoothly between $\Corr(X)$ at short horizons and $\Corr(M^{-1}\Bsrc M^{-\T})$ at long horizons. When $M$ is symmetric and the source is white, $X\propto M^{-1}$ whereas $M^{-1}\Bsrc M^{-\T}\propto M^{-2}$: the long-horizon kernel places greater weight on the slowest, smoothest eigenmodes, raising the correlations they induce. Once the exponential transients have decayed after a few relaxation times, convergence to the long-horizon plateau is algebraic. By Eq.~\eqref{eq:exact-covariance},
\begin{equation}
  C_{\Delta t}=2D\bigl[\Delta t\,M^{-1}\Bsrc M^{-\T}
  -\tau\bigl(M^{-2}X+XM^{-2\T}\bigr)\bigr]
  \label{eq:plateau-approach}
\end{equation}
up to exponentially small terms, so correlations approach their plateau with corrections of relative order $\tau/\Delta t$, much more slowly than a single exponential decay.

\subsection{A microstructure intensity for each bond}\label{subsec:microstructure}

To account for the near-zero correlations observed at the shortest intraday horizons, I add an independent microstructure noise as in \citet{lecoz_bouchaud_2024}, generalized to allow a bond-specific noise intensity $\epsilon_i\geq0$ for each bond $i$ in the intraday panel:
\begin{align}
  \Delta Y^{\mathrm{obs}}_t
  ={}&O\!\int_t^{t+\Delta t}\!a_u\,\dd u\notag\\
  &+\sqrt{2D}\,\diag(\epsilon)^{1/2}\bigl[\widetilde W_{t+\Delta t}-\widetilde W_t\bigr],
  \label{eq:micro-observation}\\
  \Cov\bigl(\Delta Y^{\mathrm{obs}}_t\bigr)
  ={}&C^{\mathrm{obs}}_{\Delta t}+2D\,\Delta t\,\diag(\epsilon),
  \label{eq:micro-covariance}
\end{align}
where $O$ is the observation matrix of Eq.~\eqref{eq:observed-covariance}, $\epsilon=(\epsilon_i)$ collects the bond-specific intensities, and $\widetilde W$ is a standard Brownian motion independent of the discretized velocity field $a_t$ (the scalar intensity of \citet{lecoz_bouchaud_2024} corresponds to the homogeneous case $\epsilon_i\equiv\epsilon$). Because this additive noise term is linear in $\Delta t$ and diagonal across bonds, it leaves the spatial geometry of the velocity field untouched. In the factored model units of Eq.~\eqref{eq:observed-covariance}, the omitted positive amplitude $\sigma_i/\sigma_{A,i}$ multiplies the entire observation of bond $i$, including its independent noise, and therefore cancels from the correlations implied by Eq.~\eqref{eq:micro-covariance}. With $\sigma_i$ the empirical volatility of this complete observation, the reference normalization extends to $\sigma_{A,i}^2=[C^{\mathrm{obs}}_{\Delta t_{\mathrm{ref}}}+2D\Delta t_{\mathrm{ref}}\,\diag(\epsilon)]_{ii}/\Delta t_{\mathrm{ref}}$, which reduces to the field-only definition when $\epsilon=0$.

Let $X^{\mathrm{obs}}=OXO^{\T}$. At short horizons ($\Delta t\ll\tau$), the velocity-field variance of bond $i$ scales as $2D(\Delta t)^2X^{\mathrm{obs}}_{ii}/\tau$, whereas its microstructure variance scales linearly as $2D\epsilon_i\Delta t$. Microstructure noise therefore dominates the variance of bond $i$ below the crossover time
\begin{equation}
  \tau_i^{\star}\triangleq\frac{\tau\,\epsilon_i}{X^{\mathrm{obs}}_{ii}} .
  \label{eq:crossover}
\end{equation}
The crossover $\tau_i^{\star}$ is invariant under the rescaling \eqref{eq:corr-scale-invar} detailed in Sec.~\ref{subsubsec:scale-invar}. This also means that $\tau_i^{\star}$ is identified physically, while $\tau$ becomes identifiable only once a normalization is chosen of $Q$. Below it, the yield changes of bond $i$ are dominated by idiosyncratic noise, and above it, by the elastic velocity field. Below the crossover times $\tau_i^{\star}$ and $\tau_j^{\star}$ of both bonds $i\neq j$, their correlation vanishes linearly with $\Delta t$:
\begin{equation}
  \rho_{ij}(\Delta t)\simeq\frac{\Delta t}{\tau}\,
  \frac{X^{\mathrm{obs}}_{ij}}{\sqrt{\epsilon_i\epsilon_j}}
  \;\longrightarrow\;0,
  \qquad \Delta t\to0^+.
  \label{eq:epps-zero}
\end{equation}
If only one of the two bonds carries a positive intensity, $\rho_{ij}(\Delta t)$ still vanishes as $(\Delta t/\tau)^{1/2}$ (Appendix~\ref{app:epps-proof}). If neither bond carries a positive intensity, $\rho_{ij}(\Delta t)$ converges to the instantaneous velocity-field correlation $\Corr(X^{\mathrm{obs}})_{ij}$. At long horizons ($\Delta t\gg\tau$), the microstructure term lowers the correlation plateau only by a relative fraction of order $\epsilon_i/(OM^{-1}\Bsrc M^{-\T}O^{\T})_{ii}$, which is negligible for small intensities.

Identification of $\epsilon_i$ is therefore concentrated at short horizons: a positive intensity is needed on bond $i$, and can be statistically resolved, only when its short-horizon correlations fall below the level implied by the instantaneous velocity field alone. Consequently, the spatial geometry pins down the long-horizon plateau, while the short-horizon rise of the Epps curves identifies $\epsilon_i$.

\subsection{Calibration on the intraday sample}\label{subsec:intraday-fit}

Estimating the clocks, the coupling matrix $Q$, the tensions, the microstructure intensities, and the relaxation parameter jointly against the exact finite-window covariance of Eq.~\eqref{eq:exact-covariance} yields a global relaxation parameter of
\begin{equation}
  \tau\approx222~\mathrm{s}\approx3.7~\mathrm{min}
  \label{eq:tau-value}
\end{equation}
under $\tr Q=\Nc$, which is reproducible across optimizer starts and lies well inside the search domain. Through Eq.~\eqref{eq:crossover}, $\tau$ and the bond-specific intensities $\epsilon_i$ set the crossover times $\tau_i^{\star}$: sub-second for French and Belgian bonds ($0.18$~s and $0.46$~s), $4$--$7$~s for German and Spanish bonds, and $10$--$13$~s for the other five issuers (country medians). In this baseline joint calibration, the clocks are initialized at the daily geometry estimated over the calendar window of the intraday dataset, and because the long-horizon neutral-factor relation~\eqref{eq:Q-update-A} is only approximate when $\Delta t\sim\tau$, I use the empirical neutral factors of the intraday sample to initialize $Q$ and then optimize all parameters jointly via the Nelder-Mead simplex algorithm from multiple starting points (Appendix~\ref{app:numerics}). Across sampling horizons $\Delta t$ from one second to one hour, the correlation build-up proceeds in two regimes. Below the crossover times $\tau_i^{\star}$ (country medians $0.2$--$13$~s), correlations rise steeply from $21$--$36\%$, governed by the competition between the instantaneous velocity-field covariance and microstructure noise in Eqs.~\eqref{eq:crossover}--\eqref{eq:epps-zero}. At longer horizons, correlations approach their plateau over a few minutes and, for certain pairs, continue to drift upward gradually up to one hour according to the algebraic $O(\tau/\Delta t)$ convergence of Eq.~\eqref{eq:plateau-approach}. To examine variation across individual pairs and country blocks, Fig.~\ref{fig:epps} (three representative pairs), Fig.~\ref{fig:pairwise} in Appendix~\ref{app:neutral} (France-Germany pairs), and Fig.~\ref{fig:epps-blocks} and Table~\ref{tab:block-fits} in Appendix~\ref{app:numerics} (all $22{,}155$ empirical Epps curves across the $45$ within- and cross-country blocks) report conditional pairwise and block-wise recalibrations of the time and noise parameters with a common adjusted spatial geometry held fixed (construction, conventions, and comparison with the baseline joint fit in Appendix~\ref{app:block-fits}).

\begin{figure}[t]
  \centering
  \includegraphics[width=\columnwidth]{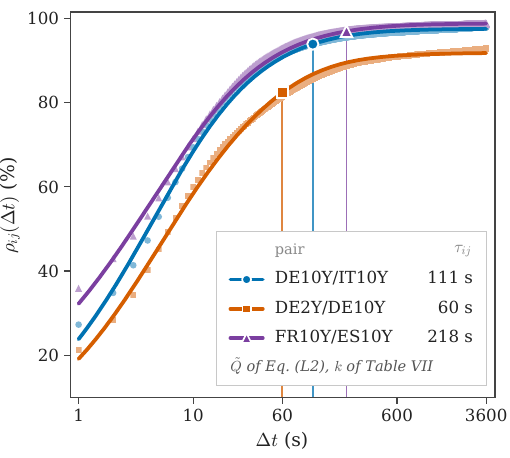}
  \caption{Epps effect on the intraday sample: Pearson correlation of yield changes as a function of the sampling horizon $\Delta t$, for two ten-year bonds of different issuers (DE10Y/IT10Y and FR10Y/ES10Y) and two maturities of the German curve (DE2Y/DE10Y). Markers show empirical correlations. Curves show conditional pairwise recalibrations of $(\tau_{ij},\epsilon_i,\epsilon_j)$ on the fixed spatial geometry of Appendix~\ref{app:block-fits}. Here $\tau_{ij}$ denotes the pair-level relaxation parameter.}
  \label{fig:epps}
\end{figure}

\begin{figure*}[t]
  \centering
  \includegraphics[width=0.50\textwidth]{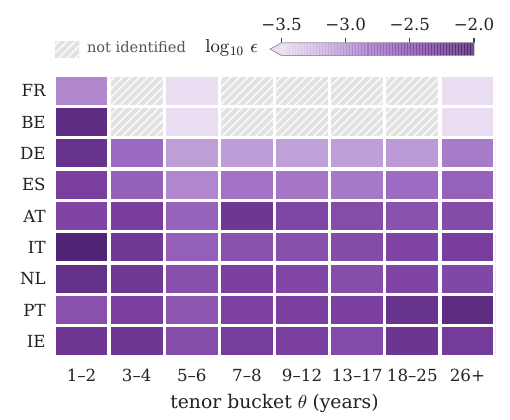}\hfill
  \includegraphics[width=0.50\textwidth]{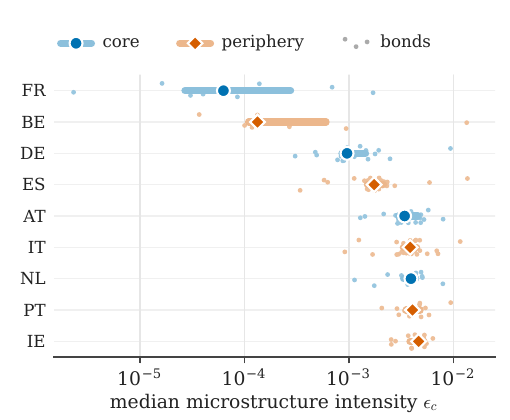}\\[-0.2em]
  \makebox[0.53\textwidth]{\small (a)}\hfill\makebox[0.44\textwidth]{\small (b)}
  \caption{Microstructure intensities of the intraday calibration. (a) $\log_{10}\epsilon$ (normalization $\tr Q=\Nc$) by country and tenor bucket. Each bucket aggregates the bonds whose intensity is identified, and hatched grey cells mark buckets in which no intensity is identified (unstable across starts of the optimizer). Countries are ordered by median intensity. (b) Median intensity $\epsilon_c$ of each country (large markers), uncertainty intervals (horizontal bars), and individual bond estimates (small dots) over the bonds with an identified intensity. Blue circles correspond to core issuers (AT, DE, FR, and NL) and orange diamonds to periphery issuers (BE, ES, IE, IT, and PT) under a two-way split, whereas Sec.~\ref{subsec:spread-factors} groups France and Belgium as semi-core.}
  \label{fig:eps}
\end{figure*}

The calibrated microstructure intensities $\epsilon_i$ are small across the universe and are identified only where short-horizon decorrelation exceeds the prediction of the instantaneous velocity field (Fig.~\ref{fig:eps}). Across several French and Belgian maturity buckets, the estimated intensity is too small to be resolved and varies across optimizer starts, so those buckets are left unreported (hatched grey cells). Where intensities are identified, the largest values occur in the 1--2-year maturity bucket, where sub-minute decorrelation is strongest, and the country medians span nearly two orders of magnitude [Fig.~\ref{fig:eps}(b)].

Following \citet{lecoz_et_al_2025, bacry_et_al_2013}, 
the variance amplitude $\epsilon_i$ can be read as an empirical proxy for microstructure noise and illiquidity, driven by idiosyncratic order flow and bid-ask bounce.
Mapped across issuers and tenors, $\epsilon_i$ provides a high-frequency liquidity map of the euro-area sovereign market whose country ranking in Fig.~\ref{fig:eps}(b) departs from a simple core-periphery split. In particular, Austria and the Netherlands have comparatively high estimated intensities despite belonging to the core group. In the unconstrained bond-by-bond fit, the medians for France and Belgium rely on a smaller number of bonds and carry wider uncertainty intervals.  Parameterizing the short-end concentration via a power-law tenor profile $\epsilon_i\propto\theta_i^{-\beta_{cd}}$ resolves the identification of the FR, BE, and BE-FR blocks with tight bootstrap intervals at the bottom of the intensity scale, see the block-wise calibration of Appendix~\ref{app:numerics} and reported values in Table~\ref{tab:block-fits}.

\subsection{Relaxation times across markets and pairs}\label{subsec:relaxation-times}

\citet{lecoz_bouchaud_2024} find a global relaxation time of $\tau\approx36$~minutes for SOFR futures. I however find a global relaxation time of  $\tau\approx3.7$~minutes for the studied euro-zone sovereign bond market, an order of magnitude smaller.  The trace normalization of the precision matrix $Q$ ensures that the two models are comparable as they coincide in the single curve $N_c = 1$ case. The difference in relaxation times is therefore primarily due to the difference in the nature of the two markets. The comparison of SOFR Epps curves with that of bonds also draws the same conclusion independently of the model, as seen on Fig.~\ref{fig:epps-empirical} and Fig.~1 of \cite{lecoz_bouchaud_2024}. 

I propose to interpret the Epps effect as being the consequence of different market activities that occur at separate time-scales. This reading would qualify why the characteristic relaxation time $\tau$ is, somewhat paradoxically, higher for highly liquid assets such as interest-rate futures. At high frequencies, algorithmic flows commonly trade individual instruments independently of the rest of the market. At lower frequencies, client and dealer driven flow trades the yield curves as a whole. I hypothesize that the relative isotropy of high-frequency strategies enforces the micro-structure amplitudes $\epsilon$ while longer-frequency practices enforce the interest-rate market elastic geometry. The crossover times $\tau_i^{\star}$ of Eq.~\eqref{eq:crossover} would thus mark the horizon between these two regimes: below $\tau_i^{\star}$, instrument-level noise dominates bond $i$, and above it the curve-level velocity field takes over. In interest-rate futures, where high-frequency algorithmic volume dominates since those assets are routinely used for hedging, cross-maturity correlations take several minutes to emerge. In cash government bonds, where algorithmic trading represents a smaller share of the total activity, the curve-level flow takes over within $\tau_i^{\star}\lesssim15$~s and observed correlations approach most of their long-horizon plateau within the first few minutes.

 Under the trading-flow interpretation above, one would expect an interest-rate futures contract such as the Bund future to exhibit a longer relaxation time than its underlying deliverable cash bonds over a common sample.

\section{Conclusion}\label{sec:conclusion}

Sovereign yield curves are modeled as elastic strings coupled on a graph at equal latent maturity, where each country reads a shared latent maturity coordinate through its own logarithmic clock. The coupling is encoded in a precision matrix $Q$, and country tensions $k_c$ encode the elastic nature of each string. This unified picture accounts for all three \textit{stylized facts} of euro-zone government bonds: smooth correlation surfaces that decay monotonically with maturity separation, ordered cross-country blocks that are generally asymmetric under exchange of tenors, and correlations that build up over time. Calibrated on daily yield changes across nine euro-zone sovereign issuers, the proposed model reconstructs the full cross-curve correlation surface with a typical error of $1.7\%$, an accuracy maintained for instruments held-out of the calibration. Small differences in clock scales across issuers bend and shift the cross-country correlation ridges away from the, perhaps expected, equal-tenor diagonal. The logarithmic clocks make the shifted correlation ridges lopsided, while the tensions and coupling control their width. Finally, the spectrum of $Q$ leaves the common euro-zone yield level nearly unanchored, consistent with a single monetary policy.

At intraday frequencies, the baseline joint calibration fits a global relaxation parameter $\tau\approx3.7$~minutes. Bond-specific microstructure intensities $\epsilon_i$ faithfully account for the Epps effect when the crossover times $\tau_i^{\star}$ are allowed to vary between bond-pairs. Differences in trading activity may explain the variation in relaxation times across markets and bond pairs. A potential outlook would be to formalize a unified model with a matrix-valued relaxation time kernel, which I leave for future work.

Because several off-diagonal couplings in $Q$ change sign across subperiods (Sec.~\ref{subsec:coupling}), extending the model to non-stationary market regimes calls for treating the clocks, tensions, and coupling matrix as time-varying state variables. Such an analysis could account for shifting economic regimes, which I leave for future work.

\section*{Acknowledgments}

I thank Victor Le Coz and Damien Challet for many insightful discussions. I am also grateful to Jean-Philippe Bouchaud and Christian Bongiorno for valuable exchanges.
I acknowledge funding from ANRT, under the CIFRE contract nr.~2025/0279.

This work used HPC resources from the ``Mésocentre'' computing center of CentraleSupélec and École Normale Supérieure Paris-Saclay supported by CNRS and Région Île-de-France.

\bibliographystyle{apsrev4-2}
\bibliography{references}

\appendix

\section{Notation}\label{app:notation}

Table~\ref{tab:notation} summarizes the notation used throughout this study. The symbol $\triangleq$ marks definitions, and $\avg{\cdot}$ denotes expectations.

\begin{table*}[t]
\caption{Notation.}
\label{tab:notation}
\footnotesize
\begin{ruledtabular}
\begin{tabular}{p{0.14\textwidth}p{0.33\textwidth}p{0.14\textwidth}p{0.33\textwidth}}
Symbol & Definition & Symbol & Definition\\
\hline
$c,d$; $\Nc$ & Country indices; number of countries ($\Nc=9$) & $\eta$, $D$ & Primitive Gaussian shock; its amplitude\\
$i,j$ & Indices of observed bonds & $\tau$ & Global relaxation parameter (baseline joint fit, $\tr Q=\Nc$)\\
$t$; $s$ & Date; time lag & $a_t\in\R^N$ & Discretized raw velocity field (internal grid)\\
$\theta$ & Physical tenor, in years & $M$ ($M_z$, $M_\theta$) & Drift matrix (Models A, B)\\
$\Delta t$ & Aggregation horizon & $J$ ($J_z$, $J_\theta$), $\Bsrc$ & Source loading; source covariance $\Bsrc=JJ^{\T}$\\
$Y_c(t,\theta)$, $\Delta Y$ & Yield of country $c$; yield change (model units from Eq.~\eqref{eq:centered-increment}) & $W_t$, $\widetilde W_t$ & Brownian motions (velocity field; microstructure)\\
$\gamma$, $\sigma$; $R_\gamma$ & Yield drift and univariate volatility (estimated at $\Delta t_{\mathrm{ref}}$), Eq.~\eqref{eq:yield-dynamics}; drift-to-fluctuation ratio & $X$ & Solution of $MX+XM^{\T}=\Bsrc$\\
$f$, $\gamma_f$, $\sigma$, $\Kf$ & Forward rate; its drift, diffusion volatility, and correlation kernel (App.~\ref{app:drift}) & $\Gamma$, $\Gamma(s)$ & Velocity-field covariance $(2D/\tau)X$; lag covariance\\
$\rho_{ij}$; $\rho_{\theta\theta'}$ & Pearson correlation of bonds $i,j$; of tenors $\theta,\theta'$ & $\Phi_{\Delta t,\tau}$ & Window function, Eq.~\eqref{eq:phi}\\
$\mathcal E_{ij}$, $\Sigma$ & Fit residual; typical error & $C_{\Delta t}$, $C_{\Delta t}(s)$ & Covariance of integrated raw velocity field; lag covariance\\
$\avg{\cdot}$; $\triangleq$ & Expectation; definition & $O$ ($O_A(\psi)$, $O_B$); $C^{\mathrm{obs}}_{\Delta t}$ & Observation matrix (Models A, B); readout covariance $OC_{\Delta t}O^{\T}$\\
$z$, $I=[0,\zmax]$ & Latent maturity; latent interval & $\epsilon_i$; $\epsilon_c$, $\epsilon_{cd}$ & Microstructure intensity of bond $i$; country median, block mean\\
$z_c(\theta)$, $z_c^{-1}(z)$ & Clock of country $c$ ($z(\theta)$ for one curve); its inverse & $\tau_i^{\star}$ & Noise crossover time $\tau\epsilon_i/X^{\mathrm{obs}}_{ii}$, Eq.~\eqref{eq:crossover}\\
$h_c$ & Clock Jacobian $(\dd z_c/\dd\theta)^{-1}$, Eq.~\eqref{eq:clock-jacobian} & $r_c(t)$, $C_r$ & Neutral factors (raw velocity field); long-horizon covariance of their integrals\\
$\psi_c$; $\psi$ & Clock scale; vector of clock scales & $V$ ($V_A$, $V_B(\psi)$) & Projected source covariance of the neutral factors\\
$Q$; $\widetilde Q$ & Country coupling ($\tr Q=\Nc$); adjusted coupling, Eq.~\eqref{eq:Q-common-mode} & $G_c$, $G_{cd}$, $L_c$ & Mass, cross-mass, and stiffness matrices\\
$K=\diag(k_c)$ & Country tensions & $P_{d\to c}$, $H$ & Galerkin transfer; discrete Hessian of the action\\
$\Sact$ & Action & $\tau_{ij}$; $\tau_{cd}$ & Conditional pair-fit (Fig.~\ref{fig:epps}) and block-fit (Table~\ref{tab:block-fits}) relaxation parameters (App.~\ref{app:block-fits})\\
$\Lop$, $\Lop_z$, $\Lop_\theta$ & Single-curve, latent, physical operators & $z_\pm$; $w_b$, $\hat r_c$ & Common latent support; weights and empirical neutral factor (App.~\ref{app:neutral})\\
$A$, $\Aph$ & Raw velocity field in latent; in physical coordinates & $\varpi_\alpha$, $e_\alpha$; $\tilde r_\alpha$ & Eigenvalues and eigenvectors of $Q$; spread factor $e_\alpha^{\T}r$\\
$\mu$, $\nu$ & Line tension and stiffness parameters of the single-curve action & $\Delta_{ij}$, $Z_{ij}$ & Fisher correlation asymmetry; studentized statistic (App.~\ref{app:statistics})\\
$q$ & Normalized pairwise coupling $Q_{cd}/\sqrt{Q_{cc}Q_{dd}}$ & $L$, $B$; $T_2$, $T_{\max}$ & Mean block length, bootstrap replications; global symmetry statistics (App.~\ref{app:statistics})\\
$\sigma_A$; $\Delta t_{\mathrm{ref}}$ & Integrated velocity-field amplitude, Eq.~\eqref{eq:sigma-A}; reference horizon (one trading day) & & \\
\end{tabular}
\end{ruledtabular}
\end{table*}

\section{Testing the ordered-block asymmetry}\label{app:statistics}

\subsection{Statistic and block-bootstrap test}\label{app:asym-protocol}

The maturity-index symmetry test of Sec.~\ref{subsec:motivation} uses the bonds of Fig.~\ref{fig:fr-it}: daily yield changes over 2024-2026, with one bond per country and rounded-year tenor label (the bond closest to its label at the mid-sample date). Retaining the 22 labels present in both France and Italy ($1$, $2$, $3$, $4$, $5$, $6$, $7$, $8$, $9$, $10$, $12$, $13$, $14$, $15$, $19$, $22$, $24$, $26$, $27$, $28$, $29$, and $46$~years) gives $231$ distinct cells $i<j$. The diagonal $i=j$ compares a correlation with itself and is excluded.

For each pair of tenors $(i,j)$, define the Fisher-transformed correlation difference
\begin{equation}
  \begin{aligned}
  \Delta_{ij}\triangleq{}&\operatorname{atanh}\rho(\Delta Y_{\cFR,i},\Delta Y_{\cIT,j})\\
  &-\operatorname{atanh}\rho(\Delta Y_{\cFR,j},\Delta Y_{\cIT,i}),
  \end{aligned}
  \label{eq:asym-delta}
\end{equation}
which is antisymmetric ($\Delta_{ji}=-\Delta_{ij}$). The null hypothesis of maturity-index symmetry of the ordered cross block is $H_0:\Delta_{ij}=0$ for all $i<j$.

Closed-form asymptotic variances for dependent correlations assume independent Gaussian dates. In the daily sample, yield changes have a median excess kurtosis of $3.5$ (ranging up to $9.5$) and display volatility clustering, with a lag-$1$ autocorrelation of about $0.27$ for squared yield changes and for $\cFR\times\cIT$ cross-products. Resampling whole trading dates jointly across all bonds preserves the dependence between the two correlations of each cell and across cells without parametric assumptions.

I resample trading dates using the stationary bootstrap of \citet{politis_romano_1994}, drawing circular blocks of geometric length with mean $L$. Applying the automatic block-length selection rule of \citet{politis_white_2004}, with the correction of \citet{patton_politis_white_2009}, across the $484$ $\cFR\times\cIT$ cross-product series gives a median block length of $4.13$~days. I therefore set $L=4$ and draw $B=10{,}000$ bootstrap replications. With $\sigma^\ast_{ij}$ denoting the bootstrap standard deviation of the bootstrap replicates $\Delta^\ast_{ij}$, the studentized cell statistic and its centered replicates (which mimic the null distribution) are
\begin{equation}
  Z_{ij}\triangleq\frac{\Delta_{ij}}{\sigma^\ast_{ij}},
  \qquad
  Z^\ast_{ij}\triangleq\frac{\Delta^\ast_{ij}-\Delta_{ij}}{\sigma^\ast_{ij}},
  \label{eq:asym-Z}
\end{equation}
and the per-cell two-sided bootstrap $p$-value is
\begin{equation}
  p_{ij}\triangleq\frac{1+\#\bigl\{\lvert Z^\ast_{ij}\rvert\geq\lvert Z_{ij}\rvert\bigr\}}{B+1}.
  \label{eq:asym-pval}
\end{equation}
To test $H_0$ jointly across the whole block with a single $p$-value, I form the global sum-of-squares and extreme-value statistics
\begin{equation}
  T_2\triangleq\sum_{i<j}Z_{ij}^2,
  \qquad
  T_{\max}\triangleq\max_{i<j}\lvert Z_{ij}\rvert,
  \label{eq:asym-global}
\end{equation}
and compare them with the same statistics evaluated on the centered replicates $Z^\ast_{ij}$. With $B=10{,}000$, the smallest attainable $p$-value is $1/(B+1)\approx1\times10^{-4}$.

\subsection{Familywise error control}\label{app:asym-rw}

To control the familywise error rate at $5\%$ across the $231$ cells under arbitrary cross-cell dependence, I apply the step-down procedure of \citet{romano_wolf_2005} to $\max\lvert Z\rvert$. At each step, the critical value is the $95\%$ quantile of the maximum of $\lvert Z^\ast_{ij}\rvert$ over the remaining unrejected cells across the $B$ joint bootstrap replications. Cells exceeding this threshold are rejected and removed, and the step is repeated until no further cell is rejected. Outlined cells in Figs.~\ref{fig:asymmetry} and~\ref{fig:asymmetry-aligned} mark the Romano-Wolf rejections.

\subsection{Results and block-length robustness}\label{app:asym-robustness}

The rejection of maturity-index symmetry does not depend on the mean block length $L$. Across block lengths from $L=1$ to $L=20$ and for non-overlapping two-day yield changes, both global tests reject $H_0$ with $p<1\times10^{-4}$ and between $148$ and $180$ of the $231$ cells are rejected at the $5\%$ familywise error rate. On the bonds of Fig.~\ref{fig:fr-it} with $L=4$ (Table~\ref{tab:asym-tests}), $T_2=3{,}751$ exceeds its $99.9\%$ null quantile of $1{,}625$ by a factor of $2.3$, $T_{\max}=6.26$ exceeds its $99.9\%$ null quantile of $4.34$ by a factor of $1.4$, and $170$ of the $231$ pairs ($73.6\%$) are rejected. The violations are directional: in $94.4\%$ of all cells $i<j$, $\rho(\Delta Y_{\cFR,i},\Delta Y_{\cIT,j})>\rho(\Delta Y_{\cFR,j},\Delta Y_{\cIT,i})$, so shorter French maturities co-move more strongly with longer Italian maturities than the mirror pairs. Across alternative mean block lengths, the number of Romano-Wolf rejections and $T_{\max}$ (with its $99.9\%$ null quantile in parentheses) are $172/231$ and $6.40$ ($4.38$) for $L=1$ (the independent and identically distributed bootstrap), $169/231$ and $6.24$ ($4.31$) for $L=5$, $156/231$ and $6.26$ ($4.76$) for $L=10$, $148/231$ and $6.33$ ($4.86$) for $L=20$, and $180/231$ and $T_{\max}=6.76$ ($4.65$) for non-overlapping two-day yield changes ($L=2$), which guards against non-synchronous closing quotes.

\begin{table}[t]
\caption{Block-bootstrap tests of maturity-index symmetry in the France-Italy cross block ($B=10{,}000$ replications): Romano-Wolf rejections at $5\%$ familywise error, global statistics $T_2$ and $T_{\max}$ of Eq.~\eqref{eq:asym-global} (with their $99.9\%$ null quantiles in parentheses), and global $p$-value.}
\label{tab:asym-tests}
\footnotesize
\begin{ruledtabular}
\begin{tabular}{lcccc}
Test & Rejections & $T_2$ & $T_{\max}$ & $p$\\
\hline
Fig.~\ref{fig:fr-it} ($L=4$) & 170/231 (73.6\%) & 3751 (1625) & 6.26 (4.34) & $<10^{-4}$\\
Aligned (22) & 176/231 (76.2\%) & 3931 (1722) & 7.06 (4.41) & $<10^{-4}$\\
Aligned (18) & 111/153 (72.5\%) & 2366 (1153) & 5.84 (4.28) & $<10^{-4}$\\
\end{tabular}
\end{ruledtabular}
\end{table}

\subsection{Exact-maturity alignment}\label{app:asym-aligned}

Because benchmark labels are rounded to integer years, two bonds carrying the same label in France and Italy do not share the exact same maturity date, and a systematic maturity offset between the two issuers could in principle bend the cross block. To verify that maturity mismatch is not the source of the ordered-block asymmetry, I construct daily constant-maturity yield changes at the exact maturities of the 22 labels for each country. Specifically, on each trading day I linearly interpolate in actual time to maturity between the two bracketing bonds available on that day across all fixed-rate nominal bonds in the sample (54 French and 93 Italian bonds, with linear extrapolation from the two outermost bonds up to $0.5$~years). The maturity difference between the two countries is then identically zero on every day, and the ageing of individual bonds over the sample is eliminated. Interpolating yield changes with that day's weights omits only a roll-down term of order $(\text{curve slope})/365$, which is negligible (Appendix~\ref{app:drift}).

\begin{figure}[t]
  \centering
  \includegraphics[width=\columnwidth]{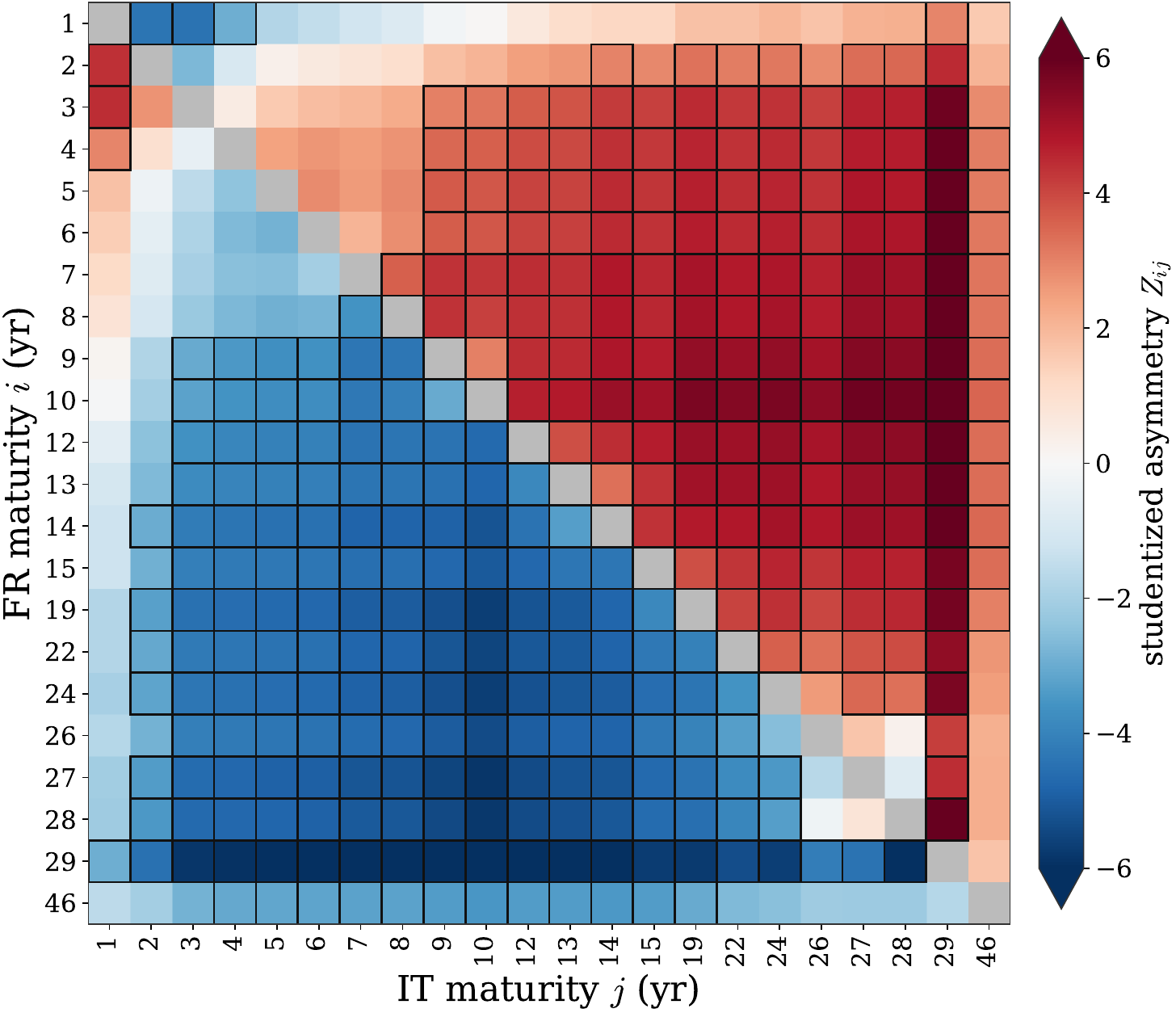}
  \caption{Ordered-block asymmetry of the France-Italy cross block on constant-maturity yield changes at exact maturities (daily sample). Each cell $(i,j)$ shows the antisymmetric studentized block-bootstrap statistic $Z_{ij}$ of $\operatorname{atanh}\rho(\Delta Y_{\cFR,i},\Delta Y_{\cIT,j})-\operatorname{atanh}\rho(\Delta Y_{\cFR,j},\Delta Y_{\cIT,i})$. Outlined cells are significant at $5\%$ familywise error (Romano-Wolf), and the global maturity-index symmetry test gives $p<1\times10^{-4}$.}
  \label{fig:asymmetry-aligned}
\end{figure}

Applying the same block-bootstrap test unchanged yields Fig.~\ref{fig:asymmetry-aligned} and the second and third rows of Table~\ref{tab:asym-tests}. Across all 22 maturities, 176 of the 231 cells ($76.2\%$) are rejected at $5\%$ familywise error, with $T_2=3{,}931$ ($99.9\%$ null quantile $1{,}722$), $T_{\max}=7.06$ ($4.41$), and global $p<1\times10^{-4}$. Restricting the test to the 18 maturities whose bracketing bonds are at most 3~years apart (which drops 19, 22, 29, and 46~years) gives 111 rejections out of 153 cells ($72.5\%$), with $T_2=2{,}366$ ($1{,}153$), $T_{\max}=5.84$ ($4.28$), and $p<1\times10^{-4}$. Across the 231 cells, the aligned map of $\Delta_{ij}$ reproduces the label-based map with a correlation of $0.93$ and the same sign in $95\%$ of cells.

\subsection{Bond substitution}\label{app:asym-substitution}

At 11 labels ($1$, $2$, $3$, $4$, $5$, $6$, $7$, $8$, $9$, $14$, and $19$~years), both countries have at least two bonds with a full price history. Building a second $\cFR\times\cIT$ panel from the second-closest bond of each label in each country yields a panel that shares no bond with the panel of Fig.~\ref{fig:fr-it}. On the log-ratio scale $\log\rho(\Delta Y_{\cFR,i},\Delta Y_{\cIT,j})-\log\rho(\Delta Y_{\cFR,j},\Delta Y_{\cIT,i})$, the two asymmetry maps have a correlation of $0.94$ (bootstrap $90\%$ interval $0.82$--$0.96$) and the same sign in $85\%$ of cells, and $94\%$ of the mean squared asymmetry is common to both panels. The ordered-block asymmetry is therefore a property of the two curves.

\section{The drift and its magnitude}\label{app:drift}

This appendix justifies dropping the drift $\gamma$ of Eq.~\eqref{eq:yield-dynamics}. I derive the no-arbitrage drift of forward rates, translate it into the drift of the yield of a given bond, and bound the ratio of drift to fluctuation. These calculations follow standard derivations of \cite{baaquie_2004_book, baaquie_2009_book}.

Let $f(t,x)$ be the instantaneous forward rate at date $t$ for maturity date $x\geq t$, so that the price of a zero-coupon bond maturing at $T$ is $\mathcal P(t,T)=\exp[-\int_t^Tf(t,x)\,\dd x]$ and the short rate is $f(t,t)$. The no-arbitrage argument concerns horizons longer than the relaxation time $\tau$, at which the integrated velocity field behaves as a Brownian field (Eq.~\eqref{eq:long-limit}). At those horizons I write
\begin{equation}
  \dd f(t,x)=\gamma_f(t,x)\,\dd t+\sigma(t,x)\,\dd\Wf(t,x),
  \label{eq:forward-dynamics}
\end{equation}
where $\Wf$ is a Brownian field with $\avg{\dd\Wf(t,x)\,\dd\Wf(t,x')}=\Kf(t;x,x')\,\dd t$, $\Kf(t;x,x)=1$, and $|\Kf|\leq1$. Here $\sigma$ is the diffusion volatility of forward rates, in yield units and distinct from the velocity-field amplitude $\sigma_A$ of Eq.~\eqref{eq:sigma-A}, and $\Kf$ is the correlation kernel of forward-rate changes.

\begin{proposition}[No-arbitrage drift]\label{prop:hjm}
The discounted zero-coupon prices $e^{-\int_0^tf(u,u)\dd u}\,\mathcal P(t,T)$ are martingales for every $T$ if and only if
\begin{equation}
  \gamma_f(t,T)=\sigma(t,T)\int_t^T\Kf(t;T,x')\,\sigma(t,x')\,\dd x'.
  \label{eq:hjm-drift}
\end{equation}
\end{proposition}

\begin{proof}
This is the condition of \citet{hjm_1992}, written for a continuum of correlated factors as in \citet{baaquie_2001,baaquie_2004_book}. Since $\log\mathcal P(t,T)=-\int_t^Tf(t,x)\,\dd x$, differentiating with respect to $t$ gives
\[
  \dd\log\mathcal P=\bigl[f(t,t)-\bar\gamma_T\bigr]\dd t-\int_t^T\sigma(t,x)\,\dd\Wf(t,x)\,\dd x,
\]
with $\bar\gamma_T=\int_t^T\gamma_f(t,x)\,\dd x$ and quadratic variation $(\dd\log\mathcal P)^2=\sigma_T^2\,\dd t$, where
\[
  \sigma_T^2=\int_t^T\!\!\int_t^T\sigma(t,x)\,\Kf(t;x,x')\,\sigma(t,x')\,\dd x\,\dd x'.
\]
By It\^o's formula, the discounted price is a martingale for every $T$ if and only if $\bar\gamma_T=\tfrac12\sigma_T^2$. Since both sides vanish at $T=t$ and $\Kf(t;x,x')=\Kf(t;x',x)$, differentiating with respect to $T$ yields Eq.~\eqref{eq:hjm-drift}.
\end{proof}

For a zero-coupon bond of maturity date $T$ and tenor $\theta=T-t$, the yield is $y(t)=\theta^{-1}\int_t^Tf(t,x)\,\dd x$ (for a coupon bond it is a weighted average of the same kind, leaving the orders of magnitude below unchanged). Differentiating $\theta y(t)=\int_t^Tf(t,x)\,\dd x$ with $\dd\theta=-\dd t$ and inserting Eq.~\eqref{eq:forward-dynamics} gives
\begin{multline}
  \dd y=\Bigl[\frac{y-f(t,t)}{\theta}+\frac1\theta\int_t^T\gamma_f(t,x)\,\dd x\Bigr]\dd t\\
  +\frac1\theta\int_t^T\sigma(t,x)\,\dd\Wf(t,x)\,\dd x.
  \label{eq:yield-drift}
\end{multline}
The drift of the yield of a given bond thus comprises the predictable roll-down $[y-f(t,t)]/\theta$ as the bond ages along the curve, the no-arbitrage drift averaged over the life of the bond, and, under the empirical measure, risk premia.

To bound their magnitude, let $\sigma$ also denote an upper bound of $\sigma(t,x)$. Since $|\Kf|\leq1$, Eq.~\eqref{eq:hjm-drift} gives $|\gamma_f(t,x)|\leq\sigma^2(x-t)$, hence
\begin{equation}
  \Bigl|\frac1\theta\int_t^T\gamma_f(t,x)\,\dd x\Bigr|
  \leq\frac{\sigma^2}{\theta}\int_0^\theta u\,\dd u=\frac{\sigma^2\theta}{2}.
  \label{eq:drift-bound}
\end{equation}
Over a window $\Delta t$ this term moves the yield by at most $\sigma^2\theta\Delta t/2$, whereas the stochastic term has standard deviation at most $\sigma\sqrt{\Delta t}$, and close to it when forward rates are strongly correlated. The ratio of drift to fluctuation is therefore bounded by
\begin{equation}
  R_\gamma\triangleq\tfrac12\,\sigma\,\theta\sqrt{\Delta t}.
  \label{eq:drift-ratio}
\end{equation}
For $\sigma=1\%$ per square-root year, that is, about $6.3$ basis points per day, and $\Delta t=1/252$ year, $R_\gamma\leq0.95\%$ at $\theta=30$ years, $0.32\%$ at ten years, and $3.0\%$ at 95 years, the longest tenor of the daily sample. For one second, within a trading year of $252$ days of $8.5$ hours, $\Delta t\approx1.3\times10^{-7}$ year and $R_\gamma\leq5.4\times10^{-5}$ at thirty years. The roll-down has the same order of magnitude, since a curve slope $y-f(t,t)$ of one percent over ten years gives $10$ basis points per year, $0.04$ basis point per day, or $0.6\%$ of a daily fluctuation of $6.3$ basis points. A term premium of a few tens of basis points per year in yield contributes about $0.1$ basis point per day, on the order of one percent of the daily fluctuation.

Writing the yield change of bond $i$ over a window as $\Delta Y_i=\Delta Y_i^{\gamma}+\Delta Y_i^{A}$ (integrated drift plus integrated velocity-field term, in yield units), any drift component that is constant over the sample is removed by centering. For a time-varying drift,
\begin{align*}
  &\Cov(\Delta Y_i,\Delta Y_j)=\Cov(\Delta Y_i^{A},\Delta Y_j^{A})\\
  &\quad+\Cov(\Delta Y_i^{\gamma},\Delta Y_j^{A})+\Cov(\Delta Y_i^{A},\Delta Y_j^{\gamma})\\
  &\quad+\Cov(\Delta Y_i^{\gamma},\Delta Y_j^{\gamma}),
\end{align*}
and the Cauchy-Schwarz inequality bounds the last three terms by $R_{\gamma,i}+R_{\gamma,j}+R_{\gamma,i}R_{\gamma,j}$ times $(\Var\Delta Y_i^{A}\,\Var\Delta Y_j^{A})^{1/2}$. For small drift-to-fluctuation ratios, the covariance correction normalized by $(\Var\Delta Y_i^{A}\,\Var\Delta Y_j^{A})^{1/2}$ and the absolute change in correlation are therefore of order $R_\gamma$. They are of order $R_\gamma^2$ if every integrated drift component is uncorrelated with every driftless yield increment, so that the two cross terms vanish. Predictability alone does not guarantee this orthogonality for the integrated Ornstein-Uhlenbeck field.

This bound applies at horizons longer than $\tau$, where the integrated velocity field behaves as a Brownian field and prices are semimartingales. At horizons shorter than $\tau$, $\int A\,\dd u$ has finite variation and the absence of arbitrage is enforced by transaction costs. The short-horizon persistence of $A$ is an integral part of the dynamics \citep{lecoz_bouchaud_2024} kept in full by Eq.~\eqref{eq:exact-covariance}.

\section{Definitions and variational equations}\label{app:variational}

Let $I=[0,\zmax]$ be the common latent interval and $\Theta_c=[0,\theta_c^{\max}]$ the physical-tenor interval of country $c$. Assume that the clock $z_c:\Theta_c\to I$ is an increasing $C^2$ diffeomorphism with inverse $z_c^{-1}$, and that the derivatives of both maps are bounded away from zero and infinity. The derivative of the inverse clock, read at the tenor $\theta$,
\begin{equation}
  h_c(\theta)\triangleq\left.\frac{\dd z_c^{-1}}{\dd z}\right|_{z=z_c(\theta)}=\Bigl(\frac{\dd z_c}{\dd\theta}\Bigr)^{-1}>0,
  \label{eq:clock-jacobian}
\end{equation}
is the Jacobian that converts latent maturity into physical tenor, and it appears throughout the appendices. For the logarithmic clock~\eqref{eq:log-clock}, $h_c(\theta)=1+\theta/\psi_c$, the factor of Eq.~\eqref{eq:noise-jacobian-preview}. The state space is $\mathcal H=L^2(I)^{\Nc}$ with form domain $\mathcal V=H^1(I)^{\Nc}$.

The bilinear form associated with Eq.~\eqref{eq:action} is
\begin{equation}
  \mathfrak a(u,v)\triangleq\int_Iu(z)^{\T}Qv(z)\,\dd z
  +\sum_ck_c\int_Iu_c'(z)v_c'(z)\,\dd z,
  \label{eq:bilinear-form}
\end{equation}
so that the action reads $\Sact[A]=-\tfrac12\int\mathfrak a(A,A)\,\dd t$. The form is continuous and coercive on $\mathcal V$ when $Q\succ0$ and $k_c>0$, since $\mathfrak a(u,u)\geq\min(\lambda_{\min}(Q),\min_ck_c)\,\|u\|_{H^1}^2$. Taking a first variation and integrating the gradient term by parts gives
\begin{multline}
  \delta\Sact[A;v]
  =-\int\!\dd t\biggl[
  \int_Iv^{\T}\bigl(QA-K\partial_z^2A\bigr)\dd z\\
  +\sum_ck_c\,v_c\,\partial_zA_c\Big|_{0}^{\zmax}\biggr].
  \label{eq:first-variation}
\end{multline}
The bulk term gives $\delta\Sact/\delta A=-\Lop_zA$, with the operator of Eq.~\eqref{eq:latent-operator}, and free endpoint variations produce the Neumann conditions $\partial_zA_c(0)=\partial_zA_c(\zmax)=0$.

For completeness, the action of \citet{lecoz_bouchaud_2024} contains, in its integrand, the additional curvature term $(\partial_\theta^2A)^2/\nu^4$. With it, a reflecting endpoint requires two conditions, and the even-extension convention is
\begin{equation}
  \partial_\theta A(t,0)=0,
  \qquad
  \partial_\theta^3A(t,0)=0,
  \label{eq:fourth-order-boundary}
\end{equation}
with decay conditions replacing the second endpoint on the half-line. The second-order specialization used in this paper requires only the first condition.

For white forcing and a self-adjoint operator, the stationary covariance $\Gamma=(D/\tau)\Lop^{-1}$ solves the Lyapunov relation $\Lop\Gamma+\Gamma\Lop=2D/\tau$, and integrating $(A,\Lop A)$ by parts under the Neumann condition gives the stationary Gaussian law at a given date,
\begin{equation}
  p_{\mathrm{st}}[A]\propto\exp\biggl\{-\frac{\tau}{2D}\int_0^\infty
  \Bigl[A^2+\frac{(\partial_\theta A)^2}{\mu^2}\Bigr]\dd\theta\biggr\},
  \label{eq:stationary-law}
\end{equation}
as stated in Sec.~\ref{subsec:baseline}, where the exponent is $\tau/D$ times the integrand in $t$ of the action~\eqref{eq:action-single}.

In the physical-native realization of Sec.~\ref{subsec:discretization}, the velocity field is represented in physical coordinates, $\Aph_c(t,\theta)=A_c(t,z_c(\theta))$, and the latent operator~\eqref{eq:latent-operator} becomes the physical operator
\begin{align}
  (\Lop_\theta\Aph)_c(\theta)={}&\sum_dQ_{cd}\,\Aph_d\bigl(z_d^{-1}(z_c(\theta))\bigr)\notag\\
  &-k_c\bigl[h_c^2\,\partial_\theta^2\Aph_c+h_ch_c'\,\partial_\theta\Aph_c\bigr](\theta),
  \label{eq:physical-operator}
\end{align}
where $h_c'=\dd h_c/\dd\theta$. The coupling term reads country $d$ at the tenor $z_d^{-1}(z_c(\theta))$ that shares the latent maturity of country $c$ at tenor $\theta$.

\begin{proposition}[Physical pullback]\label{prop:pullback}
Define $(UA)_c(\theta)=A_c(z_c(\theta))=\Aph_c(\theta)$. Then $\Lop_\theta=U\Lop_zU^{-1}$, with $\Lop_\theta$ given by Eq.~\eqref{eq:physical-operator}. If $0<h_c<\infty$ at an endpoint, the latent and physical Neumann conditions are equivalent there.
\end{proposition}

\begin{proof}
For the country term,
\[
  \begin{aligned}
  (UQA)_c(\theta)
  &=\sum_dQ_{cd}A_d(z_c(\theta))\\
  &=\sum_dQ_{cd}\,\Aph_d\bigl(z_d^{-1}(z_c(\theta))\bigr).
  \end{aligned}
\]
Writing $A_c(z)=\Aph_c(z_c^{-1}(z))$ gives
\[
  \partial_zA_c=h_c\,\partial_\theta\Aph_c,
  \qquad
  \partial_z^2A_c=h_c^2\,\partial_\theta^2\Aph_c+h_ch_c'\,\partial_\theta\Aph_c .
\]
These identities prove the conjugacy and Eq.~\eqref{eq:physical-operator}. The first one also shows that one boundary derivative vanishes if and only if the other does.
\end{proof}

For the logarithmic clock ($h_c=1+\theta/\psi_c$, $h_c'=1/\psi_c$) in the single-curve case ($\Nc=1$, $Q=1$, $k=\mu^{-2}$), Eq.~\eqref{eq:physical-operator} reduces to $\Aph-\mu^{-2}[(1+\theta/\psi)^2\partial_\theta^2\Aph+\psi^{-1}(1+\theta/\psi)\partial_\theta\Aph]$, which is the operator of \citet{lecoz_bouchaud_2024} in the limit $\nu\to\infty$.

\section{Why naive coupling gives symmetric cross blocks}\label{app:naive}

\begin{proposition}[Common-grid limitation]\label{prop:common-grid}
Let
\[
  M=Q\otimes I_m+K\otimes L,
  \qquad K=\diag(k_1,\ldots,k_{\Nc}),
\]
where $Q=Q^{\T}\succ0$ and $L=L^{\T}\succeq0$ is a maturity operator common to all countries. Suppose that the source covariance is block diagonal in the eigenbasis of $L$. Equivalently, its maturity part commutes with $L$. Then each country cross-covariance block of the stationary velocity field, and of its integrated increments over a common window, is symmetric under exchange of its maturity indices.
\end{proposition}

\begin{proof}
Let $Lu_q=\ell_qu_q$ with $\{u_q\}$ orthonormal. In this basis the drift separates maturity mode by maturity mode,
\[
  M_q=Q+\ell_qK.
\]
The commuting-source assumption prevents different modes $q$ from mixing. At any common observation horizon, the country-space covariance carried by mode $q$ is some symmetric matrix $\Gamma_q$. Transforming back to maturity coordinates gives the $(c,d)$ block
\[
  C_{cd}=\sum_q(\Gamma_q)_{cd}\,u_qu_q^{\T}.
\]
Every projector $u_qu_q^{\T}$ is symmetric, hence $C_{cd}=C_{cd}^{\T}$ as a maturity matrix. Independently, the symmetry of the global covariance gives $C_{dc}=C_{cd}^{\T}$.
\end{proof}

Correlations normalize these raw covariance blocks by the variances of the two countries. In the image representation of Appendix~\ref{app:closed-forms}, every kernel of the common grid has the form $R(|n-m|)+R(n+m)$, Eqs.~\eqref{eq:Dp}--\eqref{eq:residue-D1}, where $R(r)$ is a sum of powers $\zeta_j^{\,r}$ of roots inside the unit disk. The variance of country $c$ at node $n$ is therefore $R_{cc}(0)+R_{cc}(2n)$, which tends geometrically to $R_{cc}(0)$ as $n$ moves away from the short end. There, the ratio of the variances of two countries no longer depends on maturity, and the maturity-index symmetry of each covariance block carries over to the corresponding correlation block.

For the reflected finite-difference operator \eqref{eq:latent-grid-laplacian}, the symmetry assumption of Proposition~\ref{prop:common-grid} is imposed after the transformation
\begin{equation*}
  \widehat L
  =G_{\mathrm{lump}}^{1/2}
   L_{\mathrm{grid}}
   G_{\mathrm{lump}}^{-1/2},
\end{equation*}
where $G_{\mathrm{lump}}$ is defined in Eq.~\eqref{eq:modelA-mass-lumping}.
The transformed source covariance is
\begin{equation*}
  \begin{aligned}
    \widehat{\Bsrc}_{\mathrm{nat}}
      &=G_{\mathrm{lump}}^{1/2}
        J_{\mathrm{nat}}^2
        G_{\mathrm{lump}}^{1/2}\\
      &=\delta z\,\diag
        \left(2,1,\ldots,1,\tfrac12\right).
  \end{aligned}
\end{equation*}
For the reference source $\delta z\,I_m$, the proposition applies exactly, and transforming the covariance blocks back to nodal coordinates preserves their maturity-index symmetry. The actual source differs from this reference at both endpoints and can generate departures from that symmetry. These corrections originate at the boundaries but are not strictly confined to boundary nodes.

On equal physical tenors, the naive extension reaches $\Sigma=6.0\%$, with a bias of $-3.5$ \% and almost no ordered-block asymmetry. Adding a shared logarithmic clock improves the surface fit. In the symmetric-source reference, any remaining correlation asymmetry comes from maturity-dependent variance ratios, while the finite-grid source also permits the boundary-generated correction described above.
At the calibrated tensions and coupling, this variance-ratio channel is negligible (see Fig.~\ref{fig:channels}(b)). A freely refitted common-clock model can enlarge this channel only by spreading country tensions across nearly five orders of magnitude, incurring a bias three times larger than the model while still capturing less of the empirical France-Italy asymmetry map over the 231 cells of Sec.~\ref{subsec:motivation} ($76\%$ of its sum of squares, versus $90\%$ for the model). Country-specific clocks are therefore required to account for the ordered-block asymmetry.\\

In the modal expansion of Proposition~\ref{prop:common-grid}, the maturity-uniform mode ($\ell_q=0$, $u_q\propto\mathbf 1$) adds the same constant to each country's latent variance at all depths, so the latent variance ratio $\Lambda_{cd}$ varies only through the non-uniform modes. Along the anti-diagonal cut $\rho_{cd}(\theta_{\mathrm{cut}}+u,\theta_{\mathrm{cut}}-u)$, the two legs have latent separation $\Delta z(u)=z_c(\theta_{\mathrm{cut}}+u)-z_d(\theta_{\mathrm{cut}}-u)$. Near its peak, the latent cross-correlation is even in $\Delta z$, $\ln\varrho\simeq\ln\varrho^\ast-\Delta z^2/(2\varsigma_z^2)$, where $\varsigma_z$ denotes its latent width. The ridge peak sits at $\Delta z(u^\ast)=0$. To first order in $\Delta\psi=\psi_c-\psi_d$, with $\bar\psi=(\psi_c+\psi_d)/2$,
\begin{equation}
  u^\ast\simeq-\Delta\psi\,\frac{\bar\psi+\theta_{\mathrm{cut}}}{2\bar\psi}
  \Big[\ln\Big(1+\frac{\theta_{\mathrm{cut}}}{\bar\psi}\Big)-\frac{\theta_{\mathrm{cut}}}{\bar\psi+\theta_{\mathrm{cut}}}\Big],
  \label{eq:ridge-shift-app}
\end{equation}
which simplifies to $u^\ast\simeq-(\Delta\psi/2\bar\psi)\,\theta_{\mathrm{cut}}[\ln(\theta_{\mathrm{cut}}/\bar\psi)-1]$ for $\theta_{\mathrm{cut}}\gg\bar\psi$.
Expanding $\Delta z(u^\ast+\delta)=g_1\delta+\tfrac12g_2\delta^2$ around the equal-latent pair, where $g_1=z_c'+z_d'$ and $g_2=z_c''-z_d''$, the flank asymmetry $F(\delta)=\ln\rho(u^\ast+\delta)-\ln\rho(u^\ast-\delta)$ and the symmetric drop $S(\delta)=\ln\rho(u^\ast)-\tfrac12[\ln\rho(u^\ast+\delta)+\ln\rho(u^\ast-\delta)]$ satisfy
\begin{equation}
  F(\delta)/S(\delta)=-2\,(g_2/g_1)\,\delta,
\end{equation}
from which $\varsigma_z$ (and hence the tensions and coupling) cancels out. Because $z''=-z'/(\psi+\theta)$, $g_2<0$ whenever the ridge pairs a shorter tenor with a longer tenor, making the flank facing the main diagonal flatter than the outer flank. The tensions and coupling enter through the physical half-width of the cut, $\varsigma_{\mathrm{cut}}=\varsigma_z/g_1$. For $u>0$, the ordered-block asymmetry along the cut, $A(u)=\rho_{cd}(\theta_{\mathrm{cut}}-u,\theta_{\mathrm{cut}}+u)-\rho_{cd}(\theta_{\mathrm{cut}}+u,\theta_{\mathrm{cut}}-u)$, satisfies $A(u)\simeq-2\rho^\ast u\,u^\ast/\varsigma_{\mathrm{cut}}^2$. Thus, the clocks fix the ridge displacement $u^\ast$, while the tensions and coupling determine how much correlation asymmetry that displacement produces.

\section{Covariance over a finite window}\label{app:finite-tau}
The below proposition and derivations follow from standard literature of multivariate Ornstein-Uhlenbeck processes \citep{godreche_luck_2019}.

\begin{proposition}[Integrated Ornstein-Uhlenbeck covariance]\label{prop:finite-tau}
For the stable system of Eq.~\eqref{eq:generic-ou}, let $X$ solve $MX+XM^{\T}=\Bsrc$. The stationary unequal-time covariance is
\begin{equation}
  \Gamma(s)=\avg{a_{t+s}a_t^{\T}}=
  \begin{cases}
  \dfrac{2D}{\tau}e^{-sM/\tau}X, & s\geq0,\\[0.6em]
  \dfrac{2D}{\tau}Xe^{sM^{\T}/\tau}, & s<0,
  \end{cases}
  \label{eq:unequal-covariance}
\end{equation}
and the covariance of $\int_t^{t+\Delta t}a_u\,\dd u$ is given by Eq.~\eqref{eq:exact-covariance}.
\end{proposition}

\begin{proof}
The stationary mild solution is
\begin{equation}
  a_t=\frac{\sqrt{2D}}{\tau}\int_{-\infty}^{t}e^{-(t-u)M/\tau}J\,\dd W_u.
  \label{eq:mild-solution}
\end{equation}
It\^o's isometry gives $\Gamma=(2D/\tau)X$ with
\[
  X=\int_0^\infty e^{-vM}\Bsrc e^{-vM^{\T}}\,\dd v.
\]
Differentiating the integrand and using stability yields $MX+XM^{\T}=\Bsrc$, and Eq.~\eqref{eq:unequal-covariance} follows. Splitting the square integration region by the sign of the time difference gives
\[
  C_{\Delta t}=\frac{2D}{\tau}\int_0^{\Delta t}(\Delta t-s)
  \Bigl(e^{-sM/\tau}X+Xe^{-sM^{\T}/\tau}\Bigr)\dd s.
\]
Finally,
\[
  \int_0^{\Delta t}(\Delta t-s)\,e^{-sM/\tau}\,\dd s
  =\tau\,\Phi_{\Delta t,\tau}(M),
\]
which proves Eq.~\eqref{eq:exact-covariance}.
\end{proof}

Expanding $e^{-\Delta t\,M/\tau}$ in $\Phi_{\Delta t,\tau}(M)$ and using the Lyapunov equation $MX+XM^{\T}=\Bsrc$ yields the short-horizon expansion~\eqref{eq:short-limit}. At long horizons ($\Delta t\gg\tau/\lambda_{\min}$), $e^{-\Delta t\,M/\tau}$ vanishes and the identity $M^{-1}X+XM^{-\T}=M^{-1}(XM^{\T}+MX)M^{-\T}=M^{-1}\Bsrc M^{-\T}$ gives Eqs.~\eqref{eq:long-limit} and~\eqref{eq:plateau-approach}. These formulas apply to integrated raw velocity fields, from which the observation map and the factored amplitudes recover yield covariances.

\section{Closed forms and clock regimes}\label{app:closed-forms}

The Fourier and image formulas below describe the continuum, or translation-invariant bulk, forcing convention, with reflection encoded by an image factor. The exact finite-grid model uses the Lyapunov formula of Appendix~\ref{app:finite-tau} with $\Bsrc=J^2$. The image factor and the matrix $J$ must therefore not be applied together in the same calculation.

On a uniform latent half-lattice $z_n=n\,\delta z$, the positive nearest-neighbor Laplacian has the Fourier symbol
\begin{equation}
  \ell(\xi)=\frac{2(1-\cos\xi)}{(\delta z)^2},
  \qquad \widehat M(\xi)=Q+\ell(\xi)K.
  \label{eq:fourier-symbol}
\end{equation}
Neumann reflection is represented by the direct-plus-image factor
\begin{equation}
  \chi_{nm}(\xi)=e^{i\xi(n-m)}+e^{i\xi(n+m)}.
  \label{eq:image-factor}
\end{equation}
For translation-invariant bulk forcing, define, for $p=1,2$,
\begin{equation}
  \mathcal D_p(n,m)=\frac{1}{2\pi}\int_{-\pi}^{\pi}
  \chi_{nm}(\xi)\,\widehat M(\xi)^{-p}\,\dd\xi .
  \label{eq:Dp}
\end{equation}
On the full lattice $n\in\mathbb Z$ with reflection $R:n\mapsto-n$, bulk forcing with mirror-symmetric covariance $\Bsrc=I+R$ produces an even velocity field ($a_{-n}=a_n$) satisfying the discrete Neumann condition $a_{-1}=a_1$, and the Lyapunov equation $MX+XM=I+R$ is solved by $X=\tfrac12(I+R)M^{-1}$ (note that $\Bsrc_{00}=2$, which is why the image factor already doubles the boundary forcing variance and is not combined with $J$). Within this bulk image representation, the equal-time velocity-field covariance is $(D/\tau)\mathcal D_1$, and the integrated covariance is
\begin{multline}
  C_{\Delta t}(n,m)=\frac{2D}{2\pi}\int_{-\pi}^{\pi}\chi_{nm}(\xi)
  \Bigl\{\Delta t\,\widehat M(\xi)^{-2}\\
  +\tau\widehat M(\xi)^{-3}\bigl[e^{-\Delta t\,\widehat M(\xi)/\tau}-I\bigr]\Bigr\}\dd\xi.
  \label{eq:fourier-finite-tau}
\end{multline}
Thus $C_{\Delta t}\simeq(D(\Delta t)^2/\tau)\,\mathcal D_1$ at short horizons and $C_{\Delta t}\simeq2D\,\Delta t\,\mathcal D_2$ at long horizons.

Evaluating the Fourier integral on the unit circle via $\zeta=e^{i\xi}$, with
\begin{equation}
  \mathsf P(\zeta)=\zeta Q-\frac{K}{(\delta z)^2}(\zeta-1)^2,
  \qquad \widehat M=\zeta^{-1}\mathsf P(\zeta),
  \label{eq:residue-polynomial}
\end{equation}
the integrand is analytic at $\zeta=0$ for $n\geq0$ because $\mathsf P(0)=-K/(\delta z)^2$ is invertible. If $\zeta_j$ is a simple root of $\det\mathsf P$ inside the unit disk, with right and left null vectors $v_j$ and $w_j$, the residue theorem gives, for every integer $n\geq0$,
\begin{equation}
  \begin{gathered}
  R_1(n)=\sum_{|\zeta_j|<1}
  \zeta_j^{\,n}\,\frac{v_jw_j^{\T}}{w_j^{\T}\mathsf P'(\zeta_j)v_j},\\
  \mathcal D_1(n,m)=R_1(|n-m|)+R_1(n+m),
  \end{gathered}
  \label{eq:residue-D1}
\end{equation}
and $\mathcal D_2$ follows from the corresponding double-pole residues.

When all tensions equal $k$, diagonalize $Q$ with orthonormal eigenvectors $v_e$ and eigenvalues $\lambda^Q_e$, and set $\kappa_z=k/(\delta z)^2$. For country mode $e$, the root inside the unit disk and its discriminant are
\begin{equation}
  \Delta_e=\sqrt{\lambda^Q_e(\lambda^Q_e+4\kappa_z)},
  \qquad
  \beta_e=\frac{\lambda^Q_e+2\kappa_z-\Delta_e}{2\kappa_z},
  \label{eq:equal-tension-root}
\end{equation}
with $0<\beta_e<1$. Summing the single-pole residues over $e$ and using $(\lambda+k\ell)^{-2}=-\partial_\lambda(\lambda+k\ell)^{-1}$ for $\mathcal D_2$ yields
\begin{align}
  \mathcal D_1(n,m)
  &=\sum_e v_ev_e^{\T}\,
  \frac{\beta_e^{|n-m|}+\beta_e^{n+m}}{\Delta_e},
  \label{eq:D1-closed}\\
  \mathcal D_2(n,m)
  &=\sum_e \frac{v_ev_e^{\T}}{\Delta_e^3}
  \Bigl[\beta_e^{|n-m|}\bigl(\lambda^Q_e+2\kappa_z+|n-m|\Delta_e\bigr)\notag\\
  &\qquad\qquad+\beta_e^{n+m}\bigl(\lambda^Q_e+2\kappa_z+(n+m)\Delta_e\bigr)\Bigr].
  \label{eq:D2-closed}
\end{align}

In Model~A the latent lattice is uniform by construction, and the clocks enter only through the observation matrix $O_A(\psi)$. Equations~\eqref{eq:fourier-symbol}--\eqref{eq:D2-closed} therefore apply to the latent velocity field for any clocks, with the bulk image source. In Model~B, the regular physical grid maps onto nonuniform latent nodes. The closed forms are then useful only when $\psi_c$ is large relative to the quoted tenors, where the clock is nearly affine. The calibrated scales, between $0.31$ and $0.36$ years for Model~A (Table~\ref{tab:daily-params}) and between $0.33$ and $0.36$ years for Model~B (Appendix~\ref{app:models}), lie in the strongly nonlinear regime, where the exact spectral formula should be used instead. Let a stable, diagonalizable finite matrix $M$ have a complete biorthogonal eigenbasis,
\begin{equation}
  Mu_\alpha=\lambda_\alpha u_\alpha,
  \quad
  w_\alpha^{\T}M=\lambda_\alpha w_\alpha^{\T},
  \quad
  w_\alpha^{\T}u_\beta=\delta_{\alpha\beta}.
  \label{eq:biorthogonal-eigenvectors}
\end{equation}
With $b_{\alpha\beta}=w_\alpha^{\T}\Bsrc w_\beta$, solving $MX+XM^{\T}=\Bsrc$ in this basis gives
\begin{multline}
  C_{\Delta t}=2D\sum_{\alpha,\beta}
  \frac{b_{\alpha\beta}}{\lambda_\alpha+\lambda_\beta}\\
  \times\bigl[\Phi_{\Delta t,\tau}(\lambda_\alpha)
       +\Phi_{\Delta t,\tau}(\lambda_\beta)\bigr]
  u_\alpha u_\beta^{\T},
  \label{eq:spectral-covariance}
\end{multline}
where the scalar version of Eq.~\eqref{eq:phi} is
\[
  \Phi_{\Delta t,\tau}(\lambda)
  =\frac{\Delta t}{\lambda}
  +\frac{\tau}{\lambda^2}\bigl(e^{-\lambda\Delta t/\tau}-1\bigr).
\]
For the canonical Galerkin operator, mass whitening supplies a real eigenbasis of this kind (Appendix~\ref{app:galerkin}).

\section{Coordinate conjugacy and the native-white distinction}\label{app:native-white}

\begin{proposition}[Jacobian law for white noise]\label{prop:jacobian}
If $\eta_c^z$ is white in latent maturity,
\[
  \avg{\eta_c^z(t,z)\,\eta_d^z(s,z')}
  =2D\,\delta_{cd}\,\delta(t-s)\,\delta(z-z'),
\]
then its direct physical representation satisfies
\begin{multline}
  \bigl\langle\eta_c^z(t,z_c(\theta))\,\eta_d^z(s,z_d(\theta'))\bigr\rangle\\
  =2D\,\delta_{cd}\,\delta(t-s)\,h_c(\theta)\,\delta(\theta-\theta').
  \label{eq:latent-white-physical}
\end{multline}
Conversely, physical-white noise pulled back to $z$ has covariance
\begin{equation}
  2D\,\delta_{cd}\,\delta(t-s)\,
  \frac{\delta(z-z')}{h_c(z_c^{-1}(z))}.
  \label{eq:physical-white-latent}
\end{equation}
\end{proposition}

\begin{proof}
For a strictly increasing differentiable function $g$, $\delta(g(x)-g(y))=\delta(x-y)/g'(x)$. Apply this identity first to $g=z_c$, for which $z_c'=1/h_c$, and then to $g=z_c^{-1}$, for which $(z_c^{-1})'=h_c$.
\end{proof}

\begin{proposition}[Equality of complete stochastic laws]\label{prop:law-equality}
Suppose that the initial laws and the drifts are identified by the pullback $U$. Latent-native Model~A with common amplitude $D_A$ and physical-native Model~B with common amplitude $D_B$ have the same Gaussian transition laws if and only if their diffusion covariances agree under the same pullback. In physical coordinates this requires
\begin{equation}
  D_A\,h_c(\theta)=D_B
  \label{eq:equivalence-condition}
\end{equation}
for every country and almost every tenor.
\end{proposition}

\begin{proof}
For $\dd a=-\mathsf F a\,\dd t+\mathsf G\,\dd W$, the transition covariance over a short interval $u$ is
\[
  \int_0^ue^{-\mathsf Fv}\mathsf G\mathsf G^{\T}e^{-\mathsf F^{\T}v}\,\dd v,
\]
whose derivative at zero is $\mathsf G\mathsf G^{\T}$. Equality of all transition laws therefore requires, and for equal drifts and initial laws is implied by, equality of the instantaneous diffusion covariances. Equation~\eqref{eq:equivalence-condition} follows from Eq.~\eqref{eq:latent-white-physical}.
\end{proof}

With one common amplitude in each model, Eq.~\eqref{eq:equivalence-condition} forces every $h_c$ to be the same constant, that is, every clock to be the same linear map. A possible outlook would be to consider country-wise different source amplitudes. Such constants could absorb country-specific affine Jacobians, but no finite set of constants absorbs a maturity-dependent $h_c(\theta)$.

\section{Galerkin transfer and convergence}\label{app:galerkin}

This appendix develops the finite-element construction behind Sec.~\ref{subsec:discretization}: the transfer through which, in Model~B, country $d$ is read at the latent maturities of country $c$, and the weight matrix $G_c$ of the neutral factors of Eq.~\eqref{eq:neutral-factor}. The calculations below are standard finite-element practices that, while technical, properly define the discretization of Model~B. Let $V_{c,h}\subset H^1(I)$ be the space of continuous piecewise-linear functions on the latent nodes of country $c$, with hat basis $\{\phi_{c,n}\}$. Define the mass, cross-mass, and stiffness matrices
\begin{equation}
  \begin{gathered}
  (G_{cd})_{nm}\triangleq\int_I\phi_{c,n}\phi_{d,m}\,\dd z,
  \qquad G_c\triangleq G_{cc},\\
  (L_c)_{nm}\triangleq\int_I\phi'_{c,n}\phi'_{c,m}\,\dd z .
  \end{gathered}
  \label{eq:fem-matrices}
\end{equation}
In Model~A all countries share a uniform latent grid, so both $G_c$ and $L_c$ are independent of country. The finite-difference operator in Eq.~\eqref{eq:latent-grid-laplacian} uses the lumped mass matrix
\begin{equation}
  \begin{aligned}
    G_{\mathrm{lump}}
      &\triangleq \diag(G_c\bm 1)\\
      &=\delta z\,\diag
        \left(\tfrac12,1,\ldots,1,\tfrac12\right),\\
    L_{\mathrm{grid}}
      &=G_{\mathrm{lump}}^{-1}L_c .
  \end{aligned}
  \label{eq:modelA-mass-lumping}
\end{equation}
Model~B instead uses the consistent mass matrices $G_c$ in Eq.~\eqref{eq:galerkin-drift}. Although $L_{\mathrm{grid}}$ is not symmetric in Euclidean coordinates, it satisfies $G_{\mathrm{lump}}L_{\mathrm{grid}} =L_{\mathrm{grid}}^{\T}G_{\mathrm{lump}}=L_c$. Consequently, the drift $M_z$ is self-adjoint in the weight $I_{\Nc}\otimes G_{\mathrm{lump}}$ and has positive eigenvalues when $Q\succ0$ and $k_c>0$.

Mass lumping also preserves the integration weights $\bm 1^{\T}G_c=\bm 1^{\T}G_{\mathrm{lump}}$. Since $L_c\bm 1=0$, it follows that $\bm 1^{\T}G_cL_{\mathrm{grid}}=0$. Thus the neutral average in Eq.~\eqref{eq:neutral-factor} eliminates the tension term exactly for Model~A as well. In Model~A all countries share the latent nodes, and $G_c$ is the same matrix for every country.

On each native regular half-grid, following the half-line convention of \citet{lecoz_bouchaud_2024}, the reflected source is implemented by loading shocks onto the nodes and doubling the weight of the first node at the short-end reflecting boundary $\theta=0$ ($z=0$), while leaving the far boundary at $\zmax$ in the compressed logarithmic tail unweighted:
\begin{equation}
  J_{\mathrm{nat}}=\diag(2,1,\ldots,1),
  \qquad
  \Bsrc_{\mathrm{nat}}=J_{\mathrm{nat}}J_{\mathrm{nat}}^{\T}=J_{\mathrm{nat}}^2 .
  \label{eq:J-native}
\end{equation}
This endpoint weighting specifies the source covariance separately from the Neumann conditions on the drift. In particular, reflection at $z=\zmax$ does not require doubling the last diagonal entry of $J_{\mathrm{nat}}$.
The country-level source maps are
\begin{equation}
  J_z=I_{\Nc}\otimes J_{\mathrm{nat}},
  \qquad
  J_\theta=\bdiag(J_{\mathrm{nat}},\ldots,J_{\mathrm{nat}}),
  \label{eq:J-country-blocks}
\end{equation}
for Models~A and~B, respectively. In Model~B the blocks may have different sizes. These matrices specify the canonical finite-grid source normalization. They are not combined with the bulk image factor of Appendix~\ref{app:closed-forms}. For continuum refinement, let $\delta$ denote the common
grid spacing in the native coordinate. The amplitude $D$
in the finite-grid dynamics is understood as
\begin{equation}
    \label{eq:conv_amp_source}
    D_\delta = \frac{D_{\mathrm{cont}}}{\delta},
\end{equation}
with $D_{\mathrm{cont}}$ held fixed as $\delta\to0$.
Thus the spatial white-noise intensity is preserved under
refinement. At any fixed resolution, this common amplitude
is factored out of the velocity-field correlations.

\begin{proposition}[Transfer identities]\label{prop:transfer}
The $L^2(I)$ Galerkin transfer $P_{d\to c}\triangleq G_c^{-1}G_{cd}$ satisfies
\begin{align}
  P_{d\to c}\bm 1_d&=\bm 1_c,
  \label{eq:P-constant}\\
  G_cP_{d\to c}&=P_{c\to d}^{\T}G_d,
  \label{eq:P-reciprocity}\\
  \|P_{d\to c}a_d\|_{G_c}&\leq\|a_d\|_{G_d},
  \label{eq:P-contraction}
\end{align}
where $\|a\|_{G}^2=a^{\T}Ga$. If
\begin{equation}
  (M_\theta)_{cd}=Q_{cd}P_{d\to c}+\delta_{cd}\,k_cG_c^{-1}L_c,
  \label{eq:galerkin-drift}
\end{equation}
then $M_\theta=G^{-1}H$, with $G=\bdiag(G_1,\ldots,G_{\Nc})$ and $H$ the symmetric positive-definite matrix with blocks $H_{cd}=Q_{cd}G_{cd}+\delta_{cd}k_cL_c$. In particular, $M_\theta$ is self-adjoint in $G$, $GM_\theta=M_\theta^{\T}G$.
\end{proposition}

\begin{proof}
Testing the relaxation equation of country $c$ against $\phi_{c,n}$ and integrating the tension term by parts under the Neumann conditions gives $\tau G_c\dot a_c=-\sum_dQ_{cd}G_{cd}a_d-k_cL_ca_c+\text{noise}$, which yields the blocks of Eq.~\eqref{eq:galerkin-drift}. The Galerkin projection $v_c$ of $v_d$ is characterized by $(\phi_{c,n},v_c)_{L^2}=(\phi_{c,n},v_d)_{L^2}$, giving $G_ca_c=G_{cd}a_d$. Both hat bases form partitions of unity, so $G_{cd}\bm 1_d=G_c\bm 1_c$, which proves Eq.~\eqref{eq:P-constant}. Since $G_{cd}=G_{dc}^{\T}$, Eq.~\eqref{eq:P-reciprocity} follows. Orthogonal projection in $L^2(I)$ is contractive, giving Eq.~\eqref{eq:P-contraction}. Finally, $H$ is the matrix of the coercive form~\eqref{eq:bilinear-form} on $\bigoplus_cV_{c,h}$, hence symmetric positive definite, and $GM_\theta=H$ block by block.
\end{proof}

Because $G^{1/2}M_\theta G^{-1/2}=G^{-1/2}HG^{-1/2}$ is symmetric positive definite, the nodal drift matrix $M_\theta$, though generally nonsymmetric in Euclidean coordinates, has a real positive spectrum and a $G$-orthogonal eigenbasis after mass whitening.

\begin{proposition}[Continuum convergence]\label{prop:convergence}
Under the source amplitude convention of Eq.~\eqref{eq:conv_amp_source}, assume $Q\succ0$, $k_c\geq k_0>0$, uniformly regular $C^2$ clocks, shape-regular meshes, and finite-element spaces dense in $H^1(I)$. Then the reconstructed Model~A and Galerkin Model~B drift semigroups converge strongly, uniformly on compact positive time intervals, to the coordinate-conjugate continuum semigroups. If both discretizations use the same pushed-forward source covariance, their bounded observed finite-$\tau$ covariances converge to the same limit. Native-white sources instead converge to the distinct limits of Appendix~\ref{app:native-white}.
\end{proposition}

\begin{proof}[Proof sketch]
For Model~B, the cross-mass construction represents the coupling term of the coercive form \eqref{eq:bilinear-form}, and standard Galerkin approximation gives strong resolvent convergence \citep{thomee_2006}. For Model~A, Eq.~\eqref{eq:modelA-mass-lumping} identifies the reflected finite-difference scheme with the mass-lumped approximation on a uniform grid. The lumped and consistent mass norms are uniformly equivalent, and their quadrature difference vanishes under refinement, giving the same resolvent limit. Uniform stability then yields convergence of the drift semigroups on compact positive time intervals. In one spatial dimension, parabolic smoothing, together with the source normalization \eqref{eq:conv_amp_source}, permits passage to the limit in the observed covariance integrals. Matching source covariances therefore gives the same covariance limit after the coordinate change, whereas the native-white conventions retain the distinct Jacobian factors of Appendix~\ref{app:native-white}.
\end{proof}

\section{Neutral modes and calibration}\label{app:neutral}

Define the continuum neutral factor
\begin{equation}
  r_c(t)\triangleq\frac{1}{\zmax}\int_IA_c(t,z)\,\dd z.
  \label{eq:neutral-continuum}
\end{equation}
For Model~A, the Neumann conditions remove the tension term after integration, and equal-latent coupling gives
\begin{equation}
  \tau\,\dd r_t=-Qr_t\,\dd t+\sqrt{2D}\,J_r\,\dd W_t,
  \label{eq:neutral-ou}
\end{equation}
with $V_A=J_rJ_r^{\T}=v_AI_{\Nc}$. For Model~A, the identity $\bm 1^{\T}G_cL_{\mathrm{grid}}=0$ established in Appendix~\ref{app:galerkin} removes the tension term from the discrete average exactly. For Model~B, multiplying the country equation by $\bm 1_c^{\T}G_c$ gives the same closure because $\bm 1_c^{\T}L_c=0$ and Eqs.~\eqref{eq:P-constant}--\eqref{eq:P-reciprocity} imply $\bm 1_c^{\T}G_{cd}=\bm 1_d^{\T}G_d$, with $\bm 1_d^{\T}G_d\bm 1_d=\zmax$. For Model~A, the common latent grid and identical reflected-source pattern make the projected covariance country-isotropic ($V_A=v_AI_{\Nc}$, where $v_A$ can be absorbed into $D$). For Model~B, Eq.~\eqref{eq:physical-white-latent} gives
\begin{equation}
  V_B(\psi)=\diag_c\left[
  \frac{1}{\zmax^2}\int_I\frac{\dd z}{h_c(z_c^{-1}(z))}
  \right],
  \label{eq:VB}
\end{equation}
or the corresponding finite-grid projection with the fixed native source. At long horizons, Eq.~\eqref{eq:long-limit} applied to Eq.~\eqref{eq:neutral-ou} gives the covariance of the time-integrated raw neutral factors over a long window,
\begin{equation}
  C_r\propto Q^{-1}VQ^{-\T}.
  \label{eq:neutral-long}
\end{equation}
When $V\propto I_{\Nc}$ and $Q$ is symmetric, this yields Eq.~\eqref{eq:Q-update-A}. In the orthonormal eigenbasis $Q=E\,\diag(\varpi_1,\ldots,\varpi_{\Nc})\,E^{\T}$, the projected source remains isotropic ($E^{\T}V_AE=v_AI_{\Nc}$), so the modal projections $\tilde r=E^{\T}r$ decouple into the $\Nc$ independent scalar Ornstein-Uhlenbeck processes of Eq.~\eqref{eq:spread-modes}. In Model~B, where $V_B(\psi)$ is diagonal but clock-dependent, $E^{\T}V_BE$ is generally non-diagonal. Solving $QC_rQ\propto V_B(\psi)$ via the symmetric positive-definite matrix $\Psi=C_r^{1/2}QC_r^{1/2}$ then yields the unique positive-definite update
\begin{equation}
  Q\propto C_r^{-1/2}
  \bigl[C_r^{1/2}V_B(\psi)C_r^{1/2}\bigr]^{1/2}
  C_r^{-1/2}.
  \label{eq:Q-update-B}
\end{equation}

Because the latent velocity field is not observed, empirical proxies for the neutral factors are constructed by interpolating the observed yield changes of each country in latent maturity. In this empirical construction, $\Delta Y_b$ denotes the yield change of bond $b$ divided by its estimated volatility $\sigma_b$. For a bond $b$ of country $c$, let $z_b\triangleq z_c(\theta_b)$ be its latent node, and let $\phi_b$ be the piecewise-linear hat function equal to one at $z_b$ and zero at the neighboring latent nodes of the same country:
\begin{equation}
  \widehat{\Delta Y}_c(t,z)\triangleq\sum_{b\in c}\Delta Y_b(t)\,\phi_b(z).
  \label{eq:latent-interpolation}
\end{equation}
Averaging over the latent interval covered by all countries,
\begin{equation}
  z_-\triangleq\max_c\,\min_{b\in c}z_b,
  \qquad
  z_+\triangleq\min_c\,\max_{b\in c}z_b ,
  \label{eq:common-support}
\end{equation}
defines the empirical neutral factor of country $c$,
\begin{equation}
  \begin{gathered}
  \hat r_c(t)\triangleq\frac{1}{z_+-z_-}\int_{z_-}^{z_+}\widehat{\Delta Y}_c(t,z)\,\dd z
  =\sum_{b\in c}w_b(\psi_c)\,\Delta Y_b(t),\\
  w_b(\psi_c)\triangleq\frac{1}{z_+-z_-}\int_{z_-}^{z_+}\phi_b(z)\,\dd z.
  \end{gathered}
  \label{eq:empirical-neutral}
\end{equation}
For the raw velocity field, averaging the latent operator~\eqref{eq:latent-operator} over $[z_-,z_+]$ leaves a boundary-gradient flux term $k_c[\partial_zA_c]_{z_-}^{z_+}/(z_+-z_-)$. Neglecting this endpoint flux closes the raw averages under $Q$ as in Eq.~\eqref{eq:neutral-ou}. The empirical average $\hat r_c$ is a proxy for these raw averages because its support is finite and the standardized change $\Delta Y_b$ carries the raw velocity-field increment of bond $b$ divided by the bondwise factor $\sigma_{A,b}$, which the weighted sum does not cancel. With the clock-dependent quadrature weights $w_b(\psi_c)$, the empirical covariance $\widehat C_r$ of $\hat r_c$ provides the initial estimate of $Q$,
\begin{equation}
  \widehat Q=\Nc\,\frac{\widehat C_r^{-1/2}}{\tr\widehat C_r^{-1/2}} .
  \label{eq:Q-estimate}
\end{equation}

Although $Q$ is positive definite and of full rank, the long-horizon covariance of the neutral factors in Model~A scales as $Q^{-2}$. Under the eigenvalues of Table~\ref{tab:spread-modes}, the common level carries $99.9\%$ of the long-horizon variance, the softest spread mode carries $0.08\%$, and each stiffer mode carries less than $0.02\%$. Because the long-horizon kernel weights mode $\alpha$ by $1/\varpi_\alpha^2$, daily correlations constrain the most persistent direction of $Q$ much more tightly than the stiffer spread directions. The modal weights in Sec.~\ref{subsec:spread-factors} should be interpreted with this in mind. At intraday horizons comparable to the relaxation times $\tau/\varpi_\alpha$ of the spread modes (which span $\sim 28\tau$ down to $0.30\tau$, versus $\sim 980\tau$ for the common level), the modal weighting shifts towards the equal-time covariance $\propto Q^{-1}$, resolving the spread modes more sharply.

\begin{figure*}[t]
  \centering
  \includegraphics[width=\textwidth]{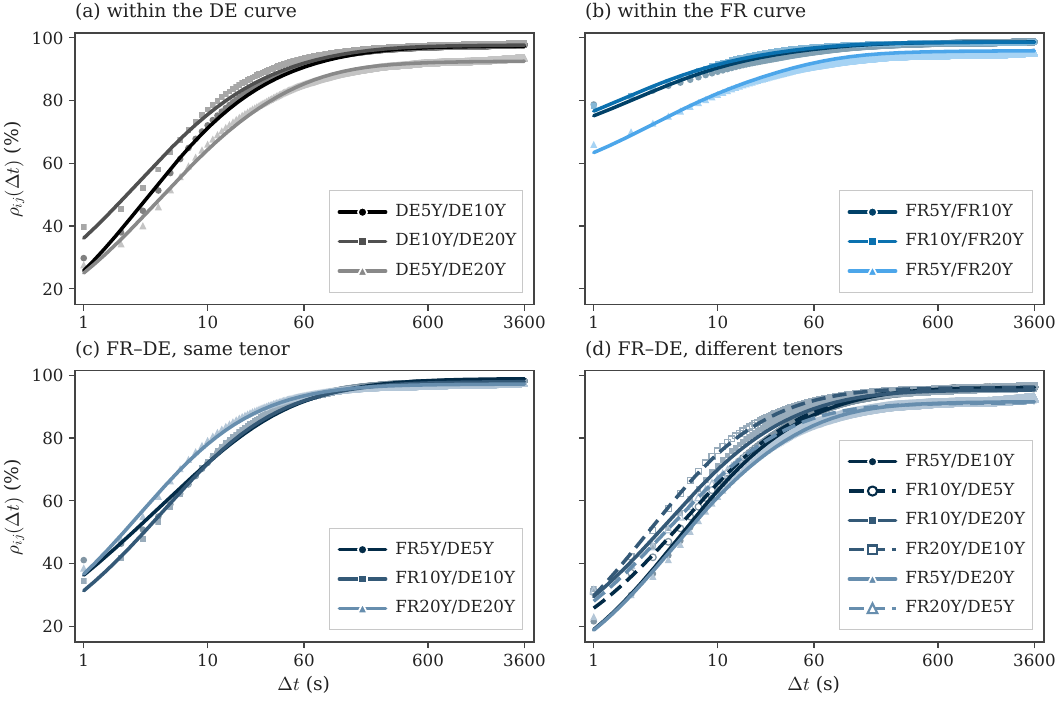}
  \caption{Conditional pairwise Epps recalibrations across the France-Germany box at 5, 10, and 20 years (intraday sample): (a) within DE, (b) within FR, (c) FR-DE at the same tenor, and (d) FR-DE at different tenors (solid: FR at the shorter tenor, dashed: DE at the shorter tenor). Markers show empirical correlations, and curves show conditional pairwise recalibrations with $\widetilde Q$ and $k$ fixed at the secondary, spectrally adjusted geometry of Appendix~\ref{app:block-fits} (Eq.~\eqref{eq:Q-common-mode}, Table~\ref{tab:block-fits}) and $(\tau_{ij},\epsilon_i,\epsilon_j)$ fitted separately to each pair as in Fig.~\ref{fig:epps}.}
  \label{fig:pairwise}
\end{figure*}

Fitting the coupling of a single pair of countries in isolation attributes to that single link all the co-movement that the full network distributes across the $\Nc$ issuers. An $\Nc\times\Nc$ matrix with unit diagonal and uniform off-diagonal $q$, $(1-q)I+q\bm 1\bm 1^{\T}$, is positive definite only for $-1/(\Nc-1)<q<1$ ($q>-1/8$ for $\Nc=9$). To see this algebraically, suppose the neutral factors have unit variance and uniform correlation $\rho_r$, $C_r=(1-\rho_r)I+\rho_r\bm 1\bm 1^{\T}$. In Model~A, $Q\propto C_r^{-1/2}=(1-\rho_r)^{-1/2}(I-\Pi)+[1+(\Nc-1)\rho_r]^{-1/2}\Pi$ with $\Pi=\bm 1\bm 1^{\T}/\Nc$, whose normalized off-diagonal entry is
\[
  q_{\Nc}=\frac{[1+(\Nc-1)\rho_r]^{-1/2}-(1-\rho_r)^{-1/2}}{(\Nc-1)(1-\rho_r)^{-1/2}+[1+(\Nc-1)\rho_r]^{-1/2}}.
\]
For $\rho_r=0.98$, a pair fitted in isolation ($\Nc=2$) yields $q_2\simeq-0.82$, whereas the same pair embedded in the nine-country graph gives $q_9\simeq-0.12$, safely inside $-1/8$. The joint identification~\eqref{eq:Q-update-A} avoids this confound by estimating all entries of $Q$ simultaneously, while conditional pairwise recalibrations on the spectrally adjusted geometry of Appendix~\ref{app:block-fits} illustrate individual pair curves across the France-Germany box (Fig.~\ref{fig:pairwise}).

For the daily surface, the calibration proceeds in three steps. First, I initialize the clock and tension parameters and determine the common latent support $[z_-,z_+]$. Second, I compute the empirical neutral-factor proxies~\eqref{eq:empirical-neutral} and the resulting initial estimate of $Q$ via Eq.~\eqref{eq:Q-estimate} (or Eq.~\eqref{eq:Q-update-B} in Model~B). Third, I minimize the typical error $\Sigma$ jointly over the clocks, the tensions, and $Q$ (parameterized by its Cholesky factor under $\tr Q=\Nc$), rebuilding $O_A(\psi)$ or $P_{d\to c}$ at each clock update. At intraday horizons ($\Delta t\sim\tau$), the relation between $C_r$ and $Q$ is nonlinear. I use the neutral-factor estimate to initialize $Q$ and optimize it numerically alongside all remaining parameters, as described in Sec.~\ref{subsec:intraday-fit}.

\section{Microstructure lemma}\label{app:epps-proof}

\begin{lemma}[Competing $\Delta t$ and $(\Delta t)^2/\tau$ terms]\label{lem:epps}
Let $A(t)$ be stationary with lag covariance $\Gamma(s)$ continuous at zero, and let
\begin{multline*}
  Y_{\Delta t}=\int_t^{t+\Delta t}A(u)\,\dd u\\
  +\sqrt{2D}\,\diag(\epsilon)^{1/2}\bigl[\widetilde W(t+\Delta t)-\widetilde W(t)\bigr],
\end{multline*}
where $\widetilde W$ has independent components and is independent of $A$. Then
\begin{equation}
  \Cov(Y_{\Delta t})
  =2D\,\Delta t\,\diag(\epsilon)+\Gamma(0)\,(\Delta t)^2+o\bigl((\Delta t)^2\bigr).
  \label{eq:epps-lemma}
\end{equation}
For the relaxation equation~\eqref{eq:generic-ou}, $\Gamma(0)=2DX/\tau=O(D/\tau)$.
\end{lemma}

\begin{proof}
Independence makes the two covariance terms additive. For the velocity field,
\begin{align*}
  \Cov\!\left(\int_t^{t+\Delta t}A(u)\,\dd u\right)
  &=\int_0^{\Delta t}\!\!\int_0^{\Delta t}\Gamma(u-v)\,\dd u\,\dd v\\
  &=\Gamma(0)(\Delta t)^2+o\bigl((\Delta t)^2\bigr)
\end{align*}
by continuity. The covariance of the Brownian increment is $2D\,\Delta t\,\diag(\epsilon)$.
\end{proof}

As in Eq.~\eqref{eq:micro-observation}, $Y_{\Delta t}$ is expressed in model units, and the velocity field and the noise of each bond share the omitted positive amplitude $\sigma_i/\sigma_{A,i}$, which cancels from $\rho_{ij}$. To leading order at short horizons, $\Var Y_i=2D\Delta t\,[\epsilon_i+\Delta tX_{ii}/\tau]$ and $\Cov(Y_i,Y_j)=2D(\Delta t)^2X_{ij}/\tau$ for $i\neq j$, so that
\[
  \rho_{ij}(\Delta t)=\frac{(\Delta t/\tau)\,X_{ij}}
  {\sqrt{(\epsilon_i+\Delta tX_{ii}/\tau)(\epsilon_j+\Delta tX_{jj}/\tau)}}.
\]
Microstructure noise dominates the variance of bond $i$ below the crossover time $\tau_i^{\star}=\tau\epsilon_i/X_{ii}$ of Eq.~\eqref{eq:crossover}. When both $\epsilon_i,\epsilon_j>0$, $\rho_{ij}(\Delta t)$ vanishes linearly as $(\Delta t/\tau)X_{ij}/\sqrt{\epsilon_i\epsilon_j}$. When only $\epsilon_i$ is positive ($\epsilon_i>0=\epsilon_j$), $\rho_{ij}(\Delta t)$ vanishes as $(\Delta t/\tau)^{1/2}X_{ij}/\sqrt{\epsilon_iX_{jj}}$. When neither is positive ($\epsilon_i=\epsilon_j=0$), $\rho_{ij}(\Delta t)$ converges to the instantaneous velocity-field correlation $X_{ij}/\sqrt{X_{ii}X_{jj}}$. At long horizons ($\Delta t\gg\tau$), both the velocity-field covariance $2D\Delta t\,\mathcal K$ (with $\mathcal K=M^{-1}\Bsrc M^{-\T}$) and the microstructure covariance grow linearly in $\Delta t$. Weak noise intensities therefore lower the long-horizon correlation plateau only by the relative amount $\epsilon_i/(2\mathcal K_{ii})+\epsilon_j/(2\mathcal K_{jj})$ noted in Sec.~\ref{subsec:microstructure}, with $X$ and $\mathcal K$ replaced by $OXO^{\T}$ and $O\mathcal KO^{\T}$ in observed bond coordinates.

\section{Numerical implementation and calibrated parameters}\label{app:numerics}

\subsection{Discretization, daily coupling, and joint intraday fit}\label{app:numerics-joint}

Model~A uses a regular latent grid covering every quoted $z_c(\theta)$, with $O_A(\psi)$ evaluating the velocity field at the observed bond tenors. Model~B assembles $G_c$, $L_c$, and $G_{cd}$ by interval quadrature and uses $P_{d\to c}=G_c^{-1}G_{cd}$. A selector $O_B$ extracts the observed bonds from any finer internal grid.

For each trial, $Q$ is parameterized through a Cholesky factor with the normalization $\tr Q=\Nc$, and the positive clock scales and tensions logarithmically. After assembling a positively stable $M$, one solves $MX+XM^{\T}=\Bsrc$, evaluates Eq.~\eqref{eq:exact-covariance}, reads it at the observed bonds through Eq.~\eqref{eq:observed-covariance}, normalizes to correlations, and computes Eq.~\eqref{eq:fit-error} on $\mathcal I$. The source is the native block pattern of Eq.~\eqref{eq:J-native}, without an additional finite-element mass factor.

\begin{figure*}[t]
  \centering
  \includegraphics[width=0.75\textwidth]{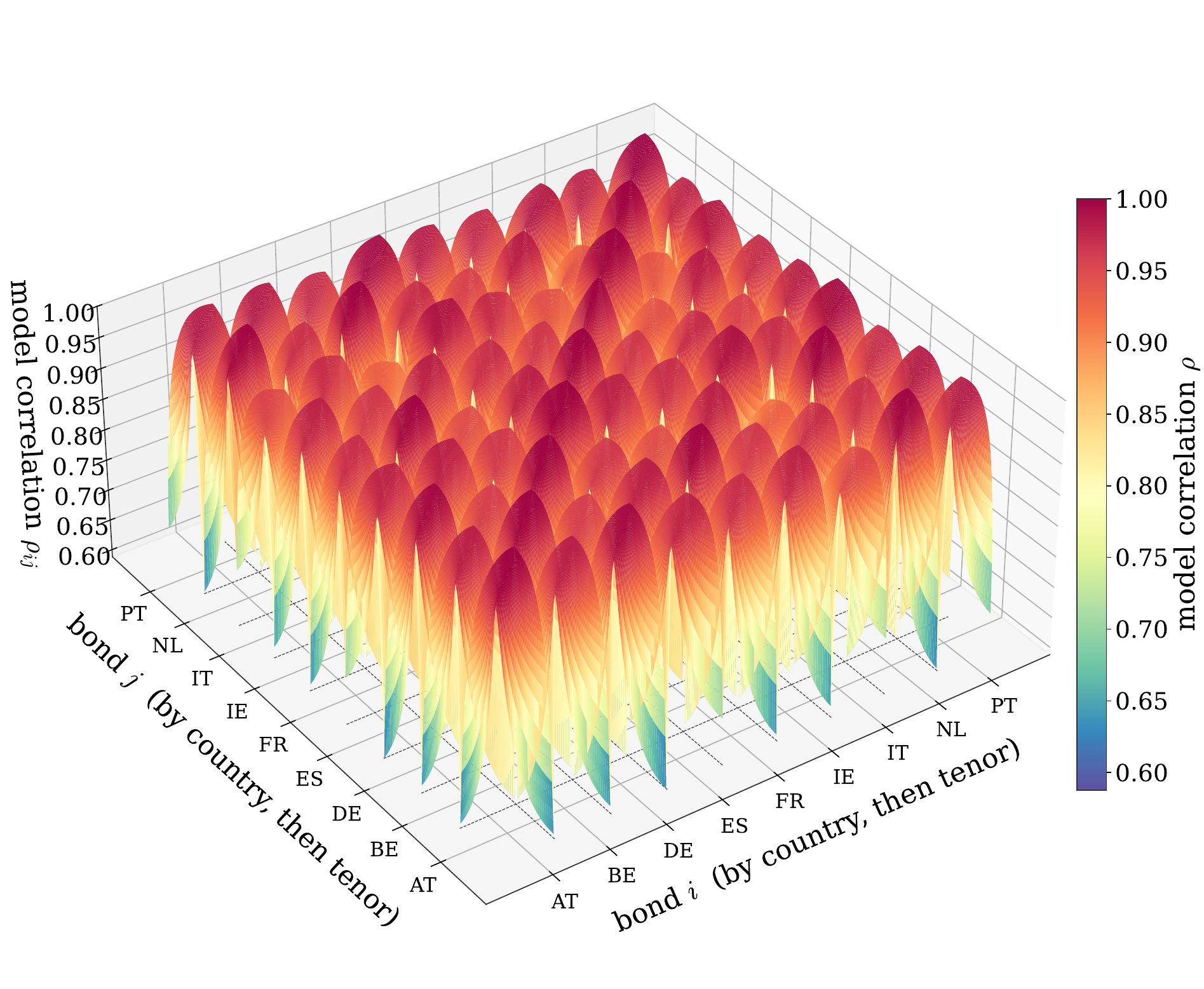}
  \caption{Model correlation surface for the nine issuers (daily calibration), drawn as in Fig.~\ref{fig:frit-3d}. Bonds are ordered by country, in alphabetical order, and, within each country, by tenor. The surface comprises the nine diagonal blocks and all cross blocks.}
  \label{fig:all9-3d}
\end{figure*}

The daily calibration of Model~A uses a latent grid of 96 nodes (with refinement to 128 nodes leaving $\Sigma$ unchanged at the reported precision) and reaches $\Sigma=1.7\%$. Its clock scales and tensions are given in Table~\ref{tab:daily-params}, and its coupling matrix in Table~\ref{tab:Q}. The matrix is positive definite, with eigenvalues between $1.0\times10^{-3}$ and $3.30$ (Table~\ref{tab:spread-modes}). Figure~\ref{fig:all9-3d} displays the model correlation surface of the nine issuers.

\begin{table*}
\caption{Country coupling $Q$ from the daily calibration of Model~A, normalized by $\tr Q=\Nc$.}
\label{tab:Q}
\begin{ruledtabular}
\begin{tabular}{lrrrrrrrrr}
 & AT & BE & DE & ES & FR & IE & IT & NL & PT\\
\hline
AT & $+$0.057 & $-$0.109 & $-$0.095 & $-$0.059 & $-$0.089 & $+$0.057 & $-$0.086 & $-$0.101 & $-$0.036\\
BE & $-$0.109 & $+$1.302 & $+$0.209 & $+$0.152 & $-$0.541 & $+$0.002 & $+$0.213 & $+$0.040 & $+$0.044\\
DE & $-$0.095 & $+$0.209 & $+$0.448 & $+$0.034 & $+$0.228 & $-$0.139 & $+$0.253 & $-$0.237 & $-$0.010\\
ES & $-$0.059 & $+$0.152 & $+$0.034 & $+$3.238 & $+$0.044 & $-$0.089 & $-$0.284 & $+$0.182 & $-$0.168\\
FR & $-$0.089 & $-$0.541 & $+$0.228 & $+$0.044 & $+$0.759 & $-$0.236 & $+$0.108 & $+$0.130 & $+$0.012\\
IE & $+$0.057 & $+$0.002 & $-$0.139 & $-$0.089 & $-$0.236 & $+$0.208 & $-$0.113 & $-$0.115 & $-$0.045\\
IT & $-$0.086 & $+$0.213 & $+$0.253 & $-$0.284 & $+$0.108 & $-$0.113 & $+$0.539 & $+$0.011 & $-$0.353\\
NL & $-$0.101 & $+$0.040 & $-$0.237 & $+$0.182 & $+$0.130 & $-$0.115 & $+$0.011 & $+$0.810 & $+$0.120\\
PT & $-$0.036 & $+$0.044 & $-$0.010 & $-$0.168 & $+$0.012 & $-$0.045 & $-$0.353 & $+$0.120 & $+$1.638\\
\end{tabular}
\end{ruledtabular}
\end{table*}

\begin{figure*}[p]
  \centering
  \includegraphics[width=\textwidth,height=0.85\textheight,keepaspectratio]{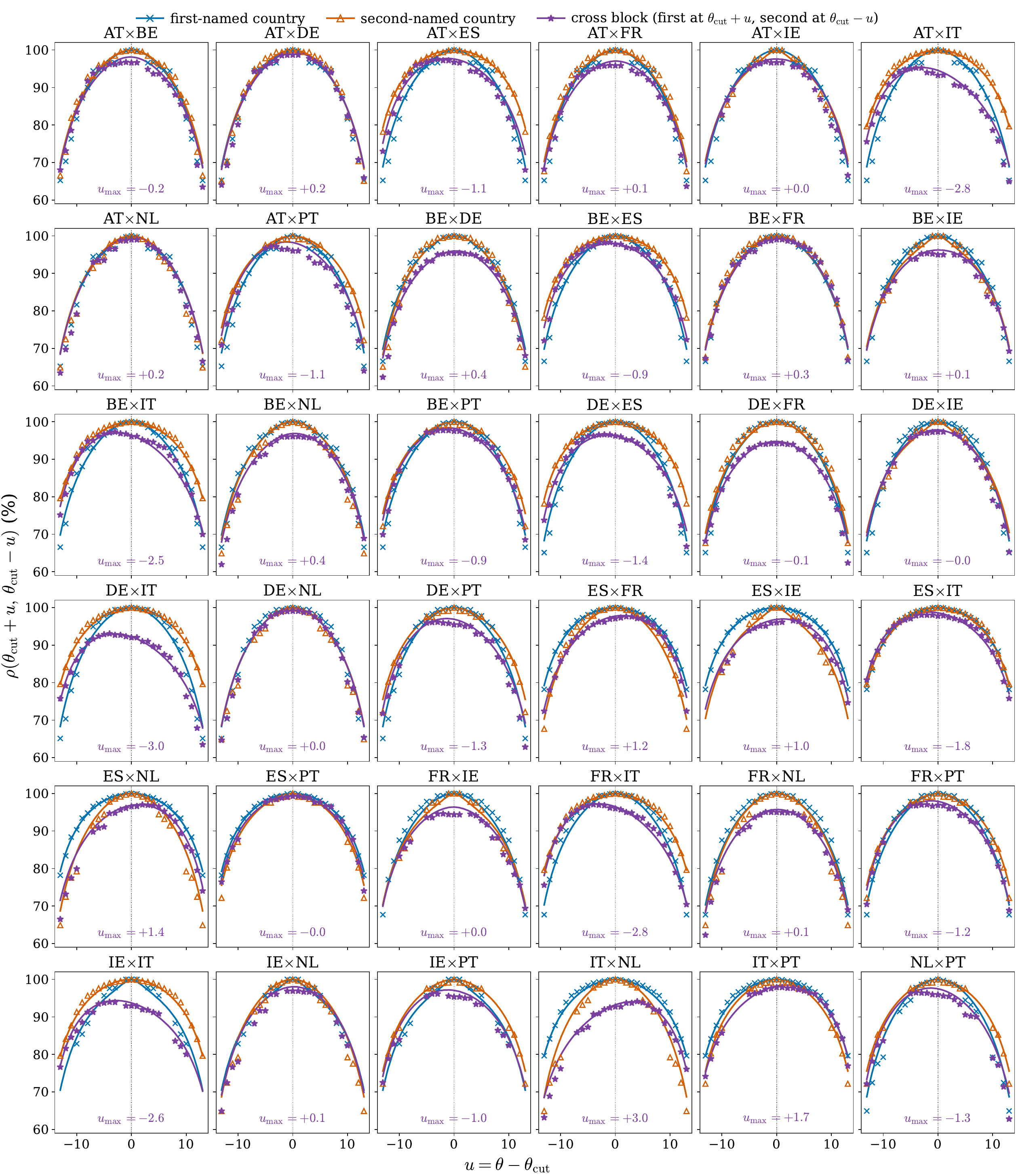}
  \caption{Cuts perpendicular to the diagonal through $\theta_{\mathrm{cut}}=15$ years, as in Fig.~\ref{fig:antidiag}, for all 36 country pairs (daily sample): $\rho(\theta_{\mathrm{cut}}+u,\theta_{\mathrm{cut}}-u)$ within the first-named country ($\times$), within the second-named country ($\triangle$), and in the ordered cross block, with the first-named country at $\theta_{\mathrm{cut}}+u$ and the second at $\theta_{\mathrm{cut}}-u$ ($\star$). Markers show the daily sample, lines show the model, and $u_{\max}$ reports the position, in years, of the maximum of the model cross-block cut.}
  \label{fig:allpairs-cuts}
\end{figure*}

Figure~\ref{fig:allpairs-cuts} repeats the cut of Fig.~\ref{fig:antidiag} for every pair of countries. The model tracks the observed cuts across the entire graph, and the position $u_{\max}$ of the cross-block maximum follows the clock difference of each pair (Table~\ref{tab:daily-params}). Specifically, $u_{\max}$ lies within $0.4$ years of the diagonal for Spain-Portugal and for all pairs among Austria, Belgium, Germany, France, Ireland, and the Netherlands, whose clock scales lie within $0.005$ years of one another. It reaches $0.9$ to $1.4$ years for Spain or Portugal paired with those six issuers, $1.7$ to $1.8$ years for Italy paired with Spain or Portugal, and $2.5$ to $3.0$ years for Italy paired with the other six issuers. Wherever the displacement is discernible, it shifts towards a shorter tenor of the issuer with the larger clock scale, as in the France-Italy block.

In the baseline intraday calibration, I initialize the clocks at the daily geometry estimated over the calendar window of the intraday sample and initialize the coupling matrix $Q$ from the empirical neutral factors of the intraday sample via Eq.~\eqref{eq:Q-update-A}. I then optimize the clocks, $Q$, the tensions, the bond-specific microstructure intensities $\epsilon_i$, and the single global relaxation parameter $\tau$ jointly against the exact finite-horizon covariance of Eq.~\eqref{eq:exact-covariance} (reaching Fisher-transformed correlation mean squared error $14.7$ for Model~A; Appendix~\ref{app:models}), using the Nelder-Mead simplex algorithm started from multiple initial points. The global parameter of Eq.~\eqref{eq:tau-value} is reproduced across starts and lies well away from the boundaries of the search domain. Any bond intensity that varies across optimizer starts is treated as unidentified and left unreported (hatched grey cells of Fig.~\ref{fig:eps}). Separately from this baseline joint fit, Figs.~\ref{fig:epps}, \ref{fig:pairwise}, \ref{fig:epps-blocks}, and Table~\ref{tab:block-fits} report conditional pairwise and block-wise recalibrations carried out on the secondary, spectrally adjusted geometry described below.

\subsection{Conditional block-wise and pairwise recalibrations on a spectrally adjusted geometry}\label{app:block-fits}

\begin{figure*}[!p]
  \centering
  \includegraphics[width=\textwidth,height=0.82\textheight,keepaspectratio]{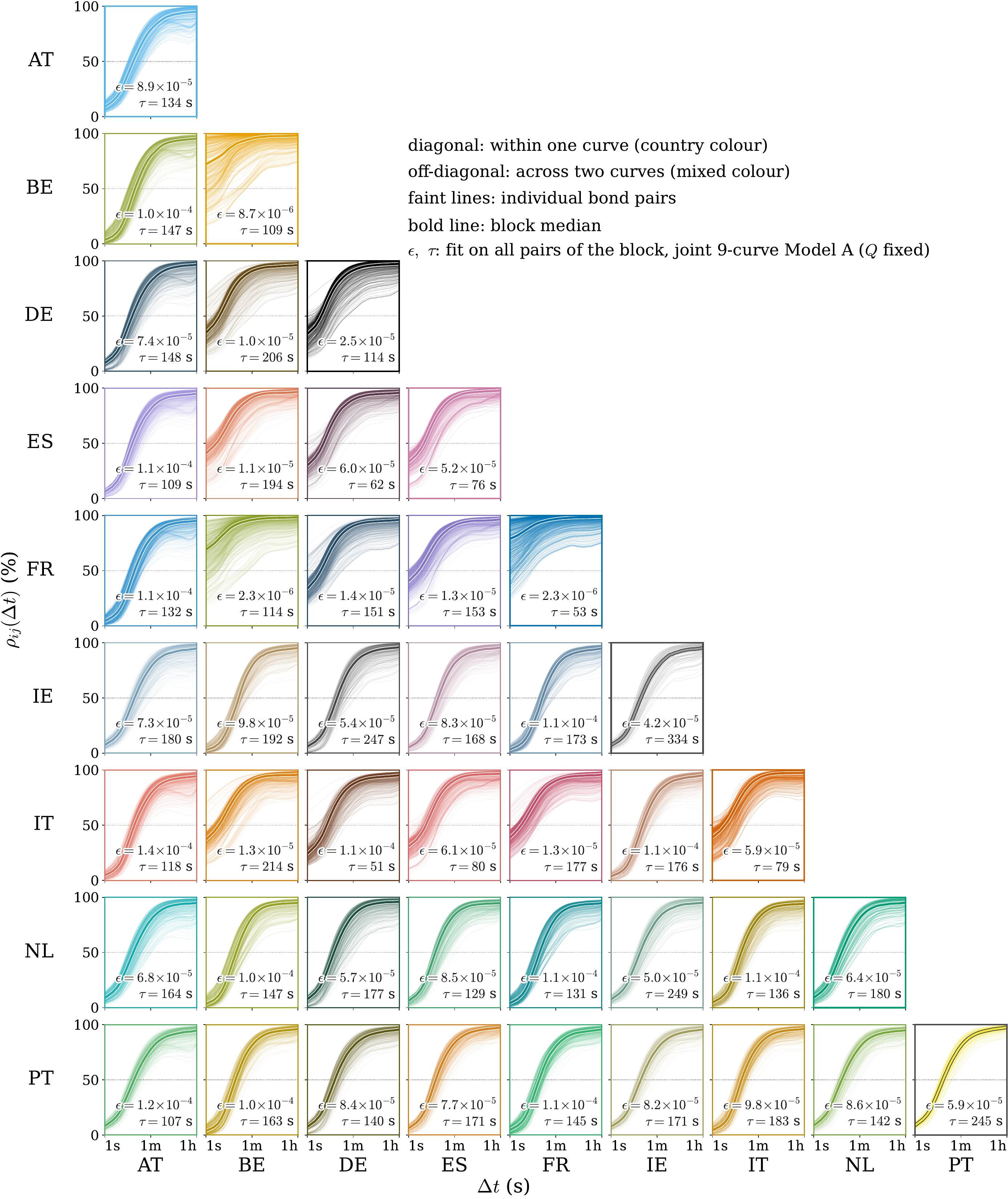}
  \caption{Complete set of empirical Epps curves by country block (intraday sample): Pearson correlation $\rho_{ij}(\Delta t)$ from one second to one hour across all within-country blocks (diagonal) and cross-country blocks (lower triangle), covering all $22{,}155$ bond pairs ($105$ to $960$ per block). Faint lines show individual bond pairs and bold lines show the block median. Each panel reports the parameters $\tau_{cd}$ and $\epsilon_{cd}$ of a conditional block-wise recalibration with the spectrally adjusted coupling $\widetilde Q$ of Eq.~\eqref{eq:Q-common-mode} and adjusted tensions $k_c$ held fixed (full parameter table and accuracies of these conditional fits in Table~\ref{tab:block-fits}).}
  \label{fig:epps-blocks}
\end{figure*}

Figure~\ref{fig:epps-blocks} displays, for each pair of countries $(c,d)$, the empirical Epps curves $\rho_{ij}(\Delta t)$ of all bond pairs $i\in c$, $j\in d$ (diagonal panels show pairs within one curve and off-diagonal panels show pairs across two curves), from $\Delta t=1$~s to $1$~h. Each panel reports the parameters $\tau_{cd}$ and $\epsilon_{cd}$ of a conditional block-wise recalibration fitted on the curves of that block ($22{,}155$ pairs in total, $105$ to $960$ per block) under the adjusted operator defined below.

In the eigenbasis $\{\lambda_m\}$ of the drift operator $M$, the covariance of observed yield changes in model units over a horizon $\Delta t$ (Eqs.~\eqref{eq:observed-covariance} and~\eqref{eq:micro-covariance}, up to the common amplitude $2D$) reads
\begin{equation}
  \begin{aligned}
  C_{ij}(\Delta t)&=\sum_m c_{ij}^{(m)}\varphi_m(\Delta t)+\epsilon_i\,\Delta t\,\delta_{ij},\\
  \varphi_m(\Delta t)&\triangleq\frac{\Delta t}{\lambda_m}-\frac{\tau}{\lambda_m^2}\bigl(1-e^{-\lambda_m\Delta t/\tau}\bigr),
  \end{aligned}
  \label{eq:block-modal-cov}
\end{equation}
and $\rho_{ij}(\Delta t)=C_{ij}(\Delta t)/[C_{ii}(\Delta t)C_{jj}(\Delta t)]^{1/2}$. The modal weights $c_{ij}^{(m)}$ are fixed by the geometry (clocks $\psi_c$, tensions $k_c$, coupling $Q$). The relaxation parameter $\tau$ sets the time scale of the build-up ($\varphi_m$ is quadratic in $\Delta t$ for $\Delta t\ll\tau$ and linear for $\Delta t\gg\tau$), and $\epsilon_i$ depresses the correlation at the shortest horizons.

As shown in Appendix~\ref{app:neutral}, the coupling $Q$ cannot be estimated pair of countries by pair of countries, because the two-country problem is strongly collinear, $q_{cd}=Q_{cd}/\sqrt{Q_{cc}Q_{dd}}$ trades against $\tau$ and $\epsilon$ along an almost flat direction, and a single pair can yield $q_{cd}$ close to $-1$ where the joint estimate is moderate. To examine variation across pairs and country blocks without running into this collinearity, I construct a secondary, spectrally adjusted geometry from the joint intraday calibration and hold it fixed while fitting only the time and noise parameters separately within each pair (Figs.~\ref{fig:epps} and~\ref{fig:pairwise}) or block (Fig.~\ref{fig:epps-blocks} and Table~\ref{tab:block-fits}).

In the joint intraday fit, $Q$ has a nearly uniform in-phase eigenvector, with lowest eigenvalue $\varpi_1=0.024$ under $\tr Q=\Nc$, $14$ times below the next level $\varpi_2$. Using the spectral notation of Sec.~\ref{subsec:spread-factors} and Appendix~\ref{app:neutral}, I set the in-phase eigenvalue to one and scale the spread eigenvalues through the adjusted coupling matrix
\begin{equation}
  \widetilde Q\triangleq E\,\diag\bigl(1,s\varpi_2/\varpi_1,\ldots,s\varpi_9/\varpi_1\bigr)\,E^{\T},
  \label{eq:Q-common-mode}
\end{equation}
with $E=[e_1,\ldots,e_9]$ the orthonormal eigenvector matrix of the joint intraday $Q$ and $\varpi_1\leq\varpi_2\leq\cdots\leq\varpi_9$ its eigenvalues. Retaining the eigenvectors $E$ and baseline eigenvalue ratios $\varpi_\alpha/\varpi_1$, the spread-mode scale factor $s$ and the nine tensions $k_c$ are fitted once jointly across the $45$ blocks with the block parameters $(\tau_{cd},\epsilon_{cd},\beta_{cd})$ profiled out, giving $s=3.16$ and $k_c$ ranging from $1.69$ (BE) to $5.00$ (PT) (caption of Table~\ref{tab:block-fits}). Because $s=3.16\neq 1$, Eq.~\eqref{eq:Q-common-mode} alters the relative stiffness of the spread modes relative to the baseline joint fit: it is a spectral adjustment rather than a pure gauge rescaling of Sec.~\ref{subsubsec:scale-invar} (which would correspond to $s=1$), and it does not preserve the baseline correlation geometry. Conditional fits on individual pairs (Figs.~\ref{fig:epps} and~\ref{fig:pairwise}) or blocks (Fig.~\ref{fig:epps-blocks} and Table~\ref{tab:block-fits}) are therefore reported separately from the baseline joint reconstruction, and independent pair fits need not assign the same relaxation parameter or microstructure intensity to a bond across pairs.

In block $(c,d)$, the free parameters are $\tau_{cd}$, $\epsilon_{cd}$, and a tenor exponent $\beta_{cd}$:
\begin{equation}
  \epsilon_i=\epsilon_{cd}\,\frac{\theta_i^{-\beta_{cd}}}{\avg{\theta^{-\beta_{cd}}}_{cd}},
  \label{eq:block-eps-shape}
\end{equation}
where $\avg{\theta^{-\beta_{cd}}}_{cd}$ is the average over the bonds of block $(c,d)$, so that $\epsilon_{cd}$ is the block-mean intensity and $\beta_{cd}$ lets the noise concentrate at the short end. The loss
\begin{equation}
  \mathcal L_{cd}\triangleq\frac{1}{P_{cd}}\sum_{(i,j)\in cd}\sum_h w_h\bigl[\rho^{\mathrm{model}}_{ij}(\Delta t_h)-\rho^{\mathrm{emp}}_{ij}(\Delta t_h)\bigr]^2,
  \label{eq:block-loss}
\end{equation}
with $P_{cd}$ the number of bond pairs in block $(c,d)$, runs over $41$ log-spaced horizons from $1$~s to $1$~h, with weights $w_h$ inversely proportional to the number of horizons per third of a decade, as in the joint calibration. It is minimized by a grid scan over $(\tau_{cd},\epsilon_{cd},\beta_{cd})$ followed by L-BFGS-B from the best cells.

Table~\ref{tab:block-fits} reports the fitted parameters across all $45$ blocks. Within a curve, $\tau_{cc}$ ranges from $53$~s (FR) to $334$~s (IE), with a median of $114$~s. Across curves, $\tau_{cd}$ ranges from $51$ to $249$~s, with a median of $152$~s. The block-mean intensity $\epsilon_{cd}$ lies between $4\times10^{-5}$ and $1.4\times10^{-4}$ for most blocks, around $10^{-5}$ when BE or FR meets a low-noise partner, and between $2\times10^{-6}$ and $9\times10^{-6}$ for BE, FR, and BE-FR, the most liquid segments. The weighted root-mean-square error ($\mathrm{RMSE}=\mathcal L_{cd}^{1/2}$) is $2.9$ to $5.7\%$ within curves (median $4.0\%$) and $3.4$ to $6.0\%$ across curves (median $4.8\%$).

In every block, the profile of the loss in $\tau_{cd}$ has a single interior minimum. The RMSE rises by $3$ to $13$ points at the ends of the admissible range ($2$~s and $4$~h), which excludes the branch $\tau\gg1$~h where only the product $\tau\epsilon$ would be determined (Eq.~\eqref{eq:epps-zero}). Setting $\epsilon_{cd}=0$ raises the RMSE by $3$ to $26$ points in all blocks except three (BE $+1.6$, BE-FR $+0.6$, and FR $+0.3$ points). A bootstrap over the pairs of each block ($100$ resamples) gives $16$--$84\%$ bands of median half-width $4\%$ for $\tau_{cd}$ and $5\%$ for $\epsilon_{cd}$ (widest on IE, $\tau=253$--$474$~s). No estimate or bootstrap replicate reaches a bound of the parameter space. Multiple starts agree within $4\%$ ($\tau_{cd}$) and $6\%$ ($\epsilon_{cd}$).

Because $\tau_{cd}$ and $\tau_{ij}$ are the in-phase relaxation parameters of the spectrally adjusted operator~\eqref{eq:Q-common-mode}, they share a common convention across the conditional block and pair fits (Figs.~\ref{fig:epps}, \ref{fig:pairwise}, \ref{fig:epps-blocks}, and Table~\ref{tab:block-fits}), whereas the baseline global parameter $\tau\approx222$~s of Eq.~\eqref{eq:tau-value} belongs to the unadjusted joint operator under $\tr Q=\Nc$.  In the latter case, individual modes relax over $\tau/\lambda_\alpha(M)$ and the observed correlation build-up reflects their combined contributions after variance normalization. Because Eq.~\eqref{eq:Q-common-mode} alters the spread-to-common stiffness ratio ($s=3.16$), the two parameter sets are not related solely by the scale invariance of Sec.~\ref{subsubsec:scale-invar}. The conditional fit errors, that are reported in Table~\ref{tab:block-fits}, measure those conditional recalibrations and do not substitute for the baseline joint calibration's accuracy (Fisher-transformed correlation mean squared error of $14.7$ for Model~A; Appendix~\ref{app:numerics-joint} and Appendix~\ref{app:models}). Likewise, $\epsilon_{cd}$ and the conditional pair intensities refer to the adjusted geometry~\eqref{eq:Q-common-mode} rather than the baseline intensities of Sec.~\ref{subsec:intraday-fit}. Meanwhile, the crossover times $\tau_i^{\star}$ are invariant under pure rescaling and display the same country ordering (sub-second for French and Belgian bonds, and longest for AT, IE, NL, and PT) across both the baseline joint fit and the conditional block fits.

\begin{table*}[!tbp]
\caption{Conditional block-wise Epps recalibrations across the $45$ country blocks (diagonal blocks first, then cross-country blocks). Here $\tau_{cd}$ is the in-phase relaxation parameter of the spectrally adjusted operator~\eqref{eq:Q-common-mode}, $\epsilon_{cd}$ the block-mean microstructure intensity in that adjusted geometry, and $\beta_{cd}$ the tenor exponent of $\epsilon_i$ in Eq.~\eqref{eq:block-eps-shape}. Uncertainty bands report the $16$--$84\%$ range across $100$ pair-level bootstrap resamples, and RMSE is the horizon-weighted root-mean-square error $\mathcal L_{cd}^{1/2}$ (in $\%$) across all pairs and horizons of the block, describing these conditional block fits rather than the baseline joint calibration. Jointly selected parameters of the adjusted geometry~\eqref{eq:Q-common-mode} are $s=3.16$ and $k_c$ equal to AT $2.59$, BE $1.69$, DE $1.81$, ES $2.61$, FR $1.97$, IE $2.79$, IT $2.45$, NL $3.03$, and PT $5.00$.}
\label{tab:block-fits}
\footnotesize
\renewcommand{\arraystretch}{0.88}
\begin{ruledtabular}
\begin{tabular}{lrrcccrr}
Block & Pairs & $\tau_{cd}$ (s) & $\tau_{cd}$ 16--84\% & $\epsilon_{cd}$ & $\epsilon_{cd}$ 16--84\% & $\beta_{cd}$ & RMSE (\%)\\
\hline
AT & 253 & 134 & 124--143 & $8.9\times10^{-5}$ & $8.4\times10^{-5}$--$9.7\times10^{-5}$ & $+$0.04 & 3.92\\
BE & 351 & 109 & 105--113 & $8.7\times10^{-6}$ & $7.7\times10^{-6}$--$9.8\times10^{-6}$ & $+$5.22 & 4.00\\
DE & 276 & 114 & 104--126 & $2.5\times10^{-5}$ & $2.3\times10^{-5}$--$2.8\times10^{-5}$ & $+$0.21 & 4.03\\
ES & 325 & 76 & 70--82 & $5.2\times10^{-5}$ & $4.8\times10^{-5}$--$5.7\times10^{-5}$ & $-$0.15 & 5.68\\
FR & 496 & 53 & 51--54 & $2.3\times10^{-6}$ & $2.2\times10^{-6}$--$2.4\times10^{-6}$ & $+$4.29 & 3.12\\
IE & 105 & 334 & 253--474 & $4.2\times10^{-5}$ & $2.9\times10^{-5}$--$5.6\times10^{-5}$ & $+$0.11 & 4.14\\
IT & 435 & 79 & 75--84 & $5.9\times10^{-5}$ & $5.6\times10^{-5}$--$6.3\times10^{-5}$ & $+$0.12 & 4.40\\
NL & 153 & 180 & 169--195 & $6.4\times10^{-5}$ & $5.9\times10^{-5}$--$6.9\times10^{-5}$ & $+$0.28 & 2.96\\
PT & 120 & 245 & 225--268 & $5.9\times10^{-5}$ & $5.4\times10^{-5}$--$6.4\times10^{-5}$ & $-$0.14 & 2.85\\
\hline
AT-BE & 621 & 147 & 144--152 & $1.0\times10^{-4}$ & $1.0\times10^{-4}$--$1.1\times10^{-4}$ & $+$0.32 & 5.19\\
AT-DE & 552 & 148 & 144--155 & $7.4\times10^{-5}$ & $7.0\times10^{-5}$--$7.6\times10^{-5}$ & $+$0.12 & 4.92\\
AT-ES & 598 & 109 & 107--113 & $1.1\times10^{-4}$ & $1.0\times10^{-4}$--$1.1\times10^{-4}$ & $-$0.00 & 5.19\\
AT-FR & 736 & 132 & 129--135 & $1.1\times10^{-4}$ & $1.1\times10^{-4}$--$1.2\times10^{-4}$ & $+$0.27 & 4.85\\
AT-IE & 345 & 180 & 168--194 & $7.3\times10^{-5}$ & $6.7\times10^{-5}$--$7.9\times10^{-5}$ & $+$0.06 & 4.11\\
AT-IT & 690 & 118 & 116--121 & $1.4\times10^{-4}$ & $1.3\times10^{-4}$--$1.4\times10^{-4}$ & $-$0.02 & 5.33\\
AT-NL & 414 & 164 & 158--174 & $6.8\times10^{-5}$ & $6.3\times10^{-5}$--$7.1\times10^{-5}$ & $+$0.20 & 3.44\\
AT-PT & 368 & 107 & 102--114 & $1.2\times10^{-4}$ & $1.1\times10^{-4}$--$1.3\times10^{-4}$ & $-$0.03 & 3.85\\
BE-DE & 648 & 206 & 196--217 & $1.0\times10^{-5}$ & $9.5\times10^{-6}$--$1.1\times10^{-5}$ & $+$0.38 & 4.70\\
BE-ES & 702 & 194 & 182--205 & $1.1\times10^{-5}$ & $1.0\times10^{-5}$--$1.2\times10^{-5}$ & $+$0.26 & 5.79\\
BE-FR & 864 & 114 & 110--117 & $2.3\times10^{-6}$ & $1.8\times10^{-6}$--$2.9\times10^{-6}$ & $+$6.44 & 4.66\\
BE-IE & 405 & 192 & 186--198 & $9.8\times10^{-5}$ & $9.6\times10^{-5}$--$1.0\times10^{-4}$ & $+$0.28 & 5.15\\
BE-IT & 810 & 214 & 205--224 & $1.3\times10^{-5}$ & $1.2\times10^{-5}$--$1.3\times10^{-5}$ & $+$0.34 & 5.13\\
BE-NL & 486 & 147 & 145--152 & $1.0\times10^{-4}$ & $9.8\times10^{-5}$--$1.0\times10^{-4}$ & $+$0.40 & 4.66\\
BE-PT & 432 & 163 & 158--168 & $1.0\times10^{-4}$ & $9.9\times10^{-5}$--$1.0\times10^{-4}$ & $+$0.21 & 5.97\\
DE-ES & 624 & 62 & 56--73 & $6.0\times10^{-5}$ & $5.0\times10^{-5}$--$6.7\times10^{-5}$ & $+$0.12 & 5.09\\
DE-FR & 768 & 151 & 145--156 & $1.4\times10^{-5}$ & $1.3\times10^{-5}$--$1.5\times10^{-5}$ & $+$0.32 & 4.19\\
DE-IE & 360 & 247 & 230--261 & $5.4\times10^{-5}$ & $5.1\times10^{-5}$--$5.8\times10^{-5}$ & $+$0.01 & 4.83\\
DE-IT & 720 & 51 & 50--54 & $1.1\times10^{-4}$ & $1.0\times10^{-4}$--$1.1\times10^{-4}$ & $+$0.14 & 4.52\\
DE-NL & 432 & 177 & 169--184 & $5.7\times10^{-5}$ & $5.5\times10^{-5}$--$6.0\times10^{-5}$ & $+$0.22 & 4.55\\
DE-PT & 384 & 140 & 136--144 & $8.4\times10^{-5}$ & $8.1\times10^{-5}$--$8.6\times10^{-5}$ & $-$0.07 & 4.71\\
ES-FR & 832 & 153 & 147--160 & $1.3\times10^{-5}$ & $1.2\times10^{-5}$--$1.4\times10^{-5}$ & $+$0.13 & 5.23\\
ES-IE & 390 & 168 & 162--174 & $8.3\times10^{-5}$ & $7.9\times10^{-5}$--$8.5\times10^{-5}$ & $+$0.00 & 4.95\\
ES-IT & 780 & 80 & 76--86 & $6.1\times10^{-5}$ & $5.6\times10^{-5}$--$6.5\times10^{-5}$ & $-$0.11 & 5.13\\
ES-NL & 468 & 129 & 127--133 & $8.5\times10^{-5}$ & $8.2\times10^{-5}$--$8.7\times10^{-5}$ & $+$0.14 & 4.70\\
ES-PT & 416 & 171 & 165--179 & $7.7\times10^{-5}$ & $7.3\times10^{-5}$--$8.0\times10^{-5}$ & $-$0.09 & 5.10\\
FR-IE & 480 & 173 & 168--178 & $1.1\times10^{-4}$ & $1.1\times10^{-4}$--$1.1\times10^{-4}$ & $+$0.22 & 4.78\\
FR-IT & 960 & 177 & 174--183 & $1.3\times10^{-5}$ & $1.3\times10^{-5}$--$1.4\times10^{-5}$ & $+$0.23 & 4.60\\
FR-NL & 576 & 131 & 128--135 & $1.1\times10^{-4}$ & $1.1\times10^{-4}$--$1.1\times10^{-4}$ & $+$0.36 & 4.29\\
FR-PT & 512 & 145 & 141--149 & $1.1\times10^{-4}$ & $1.1\times10^{-4}$--$1.2\times10^{-4}$ & $+$0.11 & 5.42\\
IE-IT & 450 & 176 & 171--182 & $1.1\times10^{-4}$ & $1.0\times10^{-4}$--$1.1\times10^{-4}$ & $-$0.04 & 5.09\\
IE-NL & 270 & 249 & 229--282 & $5.0\times10^{-5}$ & $4.4\times10^{-5}$--$5.6\times10^{-5}$ & $+$0.15 & 3.69\\
IE-PT & 240 & 171 & 161--182 & $8.2\times10^{-5}$ & $7.7\times10^{-5}$--$8.8\times10^{-5}$ & $-$0.00 & 4.02\\
IT-NL & 540 & 136 & 133--138 & $1.1\times10^{-4}$ & $1.1\times10^{-4}$--$1.1\times10^{-4}$ & $+$0.10 & 4.94\\
IT-PT & 480 & 183 & 180--186 & $9.8\times10^{-5}$ & $9.6\times10^{-5}$--$1.0\times10^{-4}$ & $-$0.22 & 5.09\\
NL-PT & 288 & 142 & 137--148 & $8.6\times10^{-5}$ & $8.2\times10^{-5}$--$9.0\times10^{-5}$ & $+$0.12 & 3.65\\
\end{tabular}
\end{ruledtabular}
\end{table*}

\section{Model A and Model B}\label{app:models}

This appendix explains why the paper works with Model~A, compares the daily fits of the two realizations of Table~\ref{tab:models}, and asks whether the available data can discriminate between them.

\subsection{Why the implementations differ}\label{app:models-implementation}

\begin{table*}[t]
\caption{Daily accuracy of the two realizations (daily sample, 196 bonds): typical error $\Sigma$ of Eq.~\eqref{eq:fit-error}, root mean square of the residuals (RMSE), and mean residual (bias).}
\label{tab:AB}
\begin{ruledtabular}
\begin{tabular}{lcccccc}
 & \multicolumn{3}{c}{Model A} & \multicolumn{3}{c}{Model B}\\
 & $\Sigma$ & RMSE & bias & $\Sigma$ & RMSE & bias\\
\hline
in-sample & 1.73\% & 1.74\% & $+$0.18\% & 2.43\% & 3.83\% & $-$2.95\%\\
held-out bonds (21 of 196) & 1.59\% & 1.60\% & $+$0.14\% & 2.02\% & 3.62\% & $-$3.01\%\\
\end{tabular}
\end{ruledtabular}
\end{table*}

Model~A is simpler to implement than Model~B because one latent maturity grid is shared by all discretized curves. The state vector for model A has $\Nc$ times as many nodes as the latent grid ($9\times96=864$ in the baseline, with refinement to $9\times128=1{,}152$ leaving $\Sigma$ unchanged at the reported precision) regardless of the clock values. Model~B instead assigns a regular physical-tenor grid to each country with a common grid spacing across countries. A common spacing is required because the native source is unmassed and the amplitude $D$ is shared across issuers, so nodal shocks are white per unit tenor only when all cells have equal width. In addition, every country's grid must span the common latent support $[0,\zmax]$ for the neutral factors to close. Country $c$ must therefore extend to physical tenor $z_c^{-1}(\zmax)=\psi_c(e^{\zmax/\psi_c}-1)$, which grows exponentially as $\psi_c$ decreases. At the calibrated daily clocks, the state vector of Model~B has 979 nodes, but parameter regions visited during numerical optimization can push this dimension far higher, where a random $20\%$ perturbation of the clocks around their calibrated values yields a median of roughly $9{,}000$ nodes, a 90th percentile of $49{,}000$ nodes, and over one million nodes in extreme draws.

The clocks also enter Model~A solely through the observation matrix $O_A(\psi)$, whereas in Model~B the observation operator $O_B$ selects the quoted bonds from the internal grid, requiring every quoted tenor to coincide with a grid node. With integer-year tenors up to a 95-year Austrian bond, the common spacing must be one year (at least 96 nodes on the shortest curve). Cross-country coupling in Model~A reduces to the identity operator on the shared latent grid, whereas in Model~B it requires the mass-consistent Galerkin transfer $P_{d\to c}=G_c^{-1}G_{cd}$ of Appendix~\ref{app:galerkin} between distinct country grids, whose $\Nc(\Nc-1)/2$ cross blocks must be recomputed whenever the clocks change. In addition, in Model~B the number of grid nodes of each country is an integer-valued function of the clocks: when the clock that determines $\zmax$ shifts, the node counts of the other countries jump and the objective function changes in discrete steps, so numerical derivatives with respect to the clocks must be evaluated at a frozen node layout, whereas Model~A has a fixed grid dimension and smooth derivatives throughout.

At the analytical level, the projected source covariance of the neutral factors in Model~A is isotropic ($V_A\propto I$), which yields the closed-form relation $Q\propto C_r^{-1/2}$ (Eq.~\eqref{eq:Q-update-A}) and decouples the spread modes (Eq.~\eqref{eq:spread-modes}). In Model~B, the projected source $V_B(\psi)$ is diagonal but clock-dependent (Eq.~\eqref{eq:Q-update-B}), so the spread modes do not decouple (Appendix~\ref{app:neutral}). Finally, the spectral closed forms of Appendix~\ref{app:closed-forms} hold in Model~A for arbitrary clocks, whereas in Model~B they apply only when the clocks are nearly affine, which is far from the calibrated regime ($\psi_c$ between $0.33$ and $0.36$~years in Model~B).

\subsection{Comparison of the fits}\label{app:models-fits}

Both models are calibrated on the daily sample with the same algorithm, the same loss, and the same 196 bonds, Model~A on its latent grid of 96 nodes and Model~B on its physical grids with a common spacing of one year. Table~\ref{tab:AB} reports three measures of accuracy, in sample and on the held-out bonds of the test of Sec.~\ref{subsec:fit}: the typical error $\Sigma$ of Eq.~\eqref{eq:fit-error}, the root mean square of the residuals $\mathcal E_{ij}$ (RMSE), and their mean (bias).

\subsection{The bias of Model B}\label{app:models-bias}

Model~B underestimates the overall correlation level by about $3.0$ points, in sample and on held-out bonds alike, making its RMSE noticeably larger than its $\Sigma$ ($3.8\%$ against $2.4\%$), whereas Model~A has a mean residual of $+0.18\%$ in sample (Table~\ref{tab:AB}). Because the typical error $\Sigma$ subtracts the mean residual (Eq.~\eqref{eq:fit-error}), it is insensitive to a uniform shift of the correlation surface. Minimizing $\Sigma$ constrains the relative geometry of the surface without pinning down its average level. The latent-white calibration of Model~A reproduces the empirical level, whereas the physical-white calibration of Model~B settles about three points below it.

\begin{table*}
\caption{Realized volatility (bp per square-root year) at standard tenors (nearest bond within $\pm30\%$ of the tenor), with $95\%$ block-bootstrap intervals in brackets. A dash marks a tenor without such a bond.}
\label{tab:realized-vol}
\footnotesize
\begin{ruledtabular}
\begin{tabular}{lccccccc}
 & 1y & 2y & 5y & 10y & 20y & 30y & 50y\\
\hline
AT & 37 [26, 46] & 64 [47, 79] & 64 [50, 77] & 61 [50, 72] & 53 [45, 59] & 52 [45, 57] & 50 [44, 56]\\
BE & -- & 65 [46, 81] & 69 [53, 83] & 68 [55, 80] & 60 [51, 69] & 58 [51, 65] & 55 [50, 61]\\
DE & 42 [30, 53] & 63 [46, 78] & 61 [49, 72] & 56 [47, 65] & 52 [45, 58] & 50 [45, 56] & --\\
ES & 53 [38, 66] & 67 [49, 83] & 68 [52, 82] & 67 [53, 78] & 61 [50, 71] & 58 [49, 66] & 58 [49, 65]\\
FR & 42 [29, 52] & 64 [46, 81] & 71 [55, 86] & 71 [57, 83] & 64 [54, 73] & 62 [54, 69] & 58 [51, 65]\\
IE & 59 [42, 73] & 65 [47, 81] & 64 [51, 76] & 61 [49, 71] & 55 [46, 62] & 52 [45, 58] & --\\
IT & 48 [34, 59] & 74 [53, 93] & 82 [61, 102] & 82 [63, 99] & 73 [58, 87] & 69 [56, 81] & 57 [48, 65]\\
NL & 49 [35, 61] & 63 [46, 78] & 63 [50, 74] & 59 [49, 68] & 51 [45, 57] & 51 [45, 57] & --\\
PT & 58 [41, 72] & 63 [45, 79] & 67 [51, 81] & 66 [53, 78] & 60 [49, 70] & 57 [49, 66] & --\\
\end{tabular}
\end{ruledtabular}
\end{table*}

\subsection{Can the data discriminate between the two models?}\label{app:models-discrimination}

The long-horizon approximation in Eq.~\eqref{eq:long-limit} is controlled by $\lambda_{\min}(M)\Delta t/\tau$. Although the slowest mode need not have relaxed within one trading day, the empirical correlation plateau supports using
$M^{-1}BM^{-\top}$ to describe daily correlations, which offer limited material to discriminate between Model~A and Model~B. Intraday covariances instead provide additional information on the two realizations through their mode-dependent relaxation times $\tau/\lambda_\alpha$ and their different source covariances.

A latent Markov velocity field seen through a dimension-reducing observation is not Markov, in general, in observed coordinates. For the covariance $C_{\Delta t}(s)$ of yield changes over two windows of length $\Delta t$ separated by a lag $s$, the product $C_{\Delta t}(s+s')\,C_{\Delta t}(s)^{-1}$ depends on $s$, whereas an Ornstein-Uhlenbeck process whose state is the bonds themselves makes it independent of $s$. This property separates a latent model from a model whose state is the bonds themselves, while both Model~A and Model~B have more internal nodes than observed bonds. Structural tests on an estimated generator require an invertible observation map, which neither model has. The applicable tests are the multi-horizon equal-time moments (the Epps curves) and predictions for bonds withheld from the calibration.

Empirically, the equal-time daily structure favors Model~A, which achieves the lower $\Sigma$ in sample and on held-out bonds. Both models predict the signs of the France-Italy asymmetry identically across all 231 cells. On the intraday sample, the two joint calibrations reach similar Fisher-transformed correlation mean squared errors on the Epps curves ($14.7$ for Model~A, $13.3$ for Model~B). Removing a third of the bonds and refitting barely moves either model: over four random draws, the clocks and tensions change by less than $1\%$ and the coupling by at most $2.3\%$ (Model~A) and $0.9\%$ (Model~B) in Frobenius norm, while the removed bonds are predicted with typical errors of $1.6$ to $1.7\%$ (Model~A) and $2.2$ to $2.3\%$ (Model~B).

A formal test distinguishing their noise measures remains to be carried out but seems unreachable in the current setting with the available data. The daily-horizon fit favors Model~A, and at the resolution of the intraday data the two realizations are not distinguishable.

\section{Realized volatility of the daily sample}\label{app:vol}
\begin{figure}[!htbp]
  \centering
  \includegraphics[width=0.95\columnwidth]{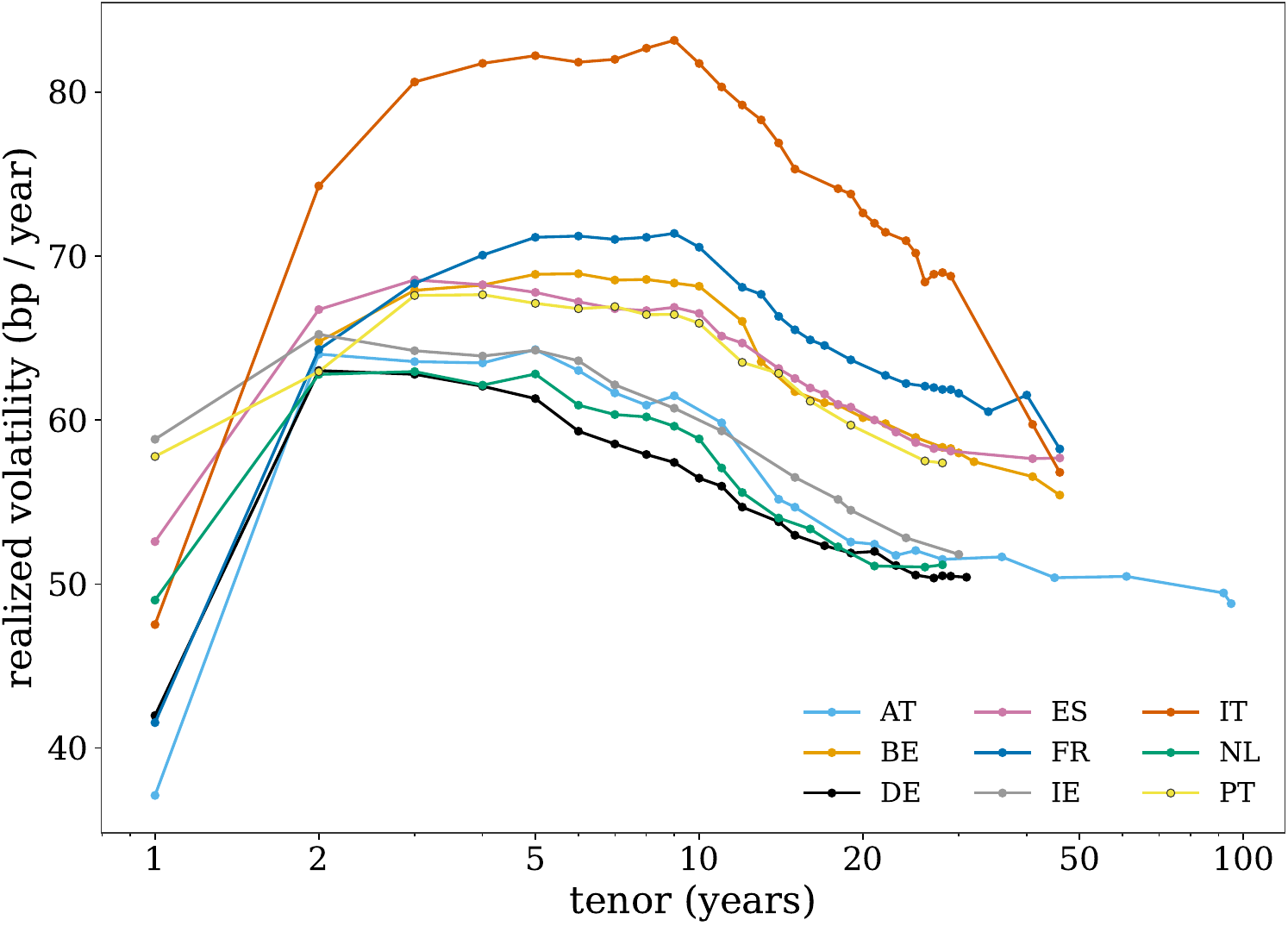}
  \caption{Realized volatility of daily yield changes by country and tenor (daily sample, 196 bonds): standard deviation of daily yield changes, annualized in basis points per square-root year.}
  \label{fig:realized-vol}
\end{figure}

Although the model is calibrated on correlations, the univariate volatility profile of the daily sample provides context for the yield-change amplitudes (Fig.~\ref{fig:realized-vol} and Table~\ref{tab:realized-vol}). Across all nine issuers, realized volatility is humped across maturity, with Italy highest at every tenor from two to thirty years, France second, and Germany, the Netherlands, Austria, and Ireland lowest. A ten-day block bootstrap yields $95\%$ confidence intervals with a median half-width of roughly $\pm11$~bp for individual bonds, resolving the higher volatility of Italy and France relative to the four core issuers while leaving differences among the remaining groups within sampling uncertainty. Volatility starts low at the short end, peaks between two and nine years, and declines beyond.

\end{document}